\documentclass[aps,twocolumn,superscriptaddress]{revtex4-2}

\usepackage{pstricks,pst-node}
\usepackage{graphicx}
\usepackage{dcolumn}
\usepackage{bm}

\usepackage[utf8]{inputenc}
\usepackage[T1]{fontenc}
\usepackage{mathptmx}
\usepackage{etoolbox}
\usepackage{xcolor}

\makeatletter
\newcommand*{\rom}[1]{\expandafter\@slowromancap\romannumeral #1@}
\def\@email#1#2{%
 \endgroup
 \patchcmd{\titleblock@produce}
  {\frontmatter@RRAPformat}
  {\frontmatter@RRAPformat{\produce@RRAP{*#1\href{mailto:#2}{#2}}}\frontmatter@RRAPformat}
  {}{}
}%
\makeatother
\usepackage{xcolor}
\begin{document}
\preprint{AIP/123-QED}

\title{Laser-induced magnonic band gap formation and control in YIG/GaAs heterostructure}

\author{K. Bublikov}
\email{konstantin.bublikov@savba.sk}
\affiliation{Institute of Electrical Engineering, Slovak Academy of Sciences, Dúbravská cesta 9, 841 04 Bratislava, Slovakia}

\author{M. Mruczkiewicz}
\affiliation{Institute of Electrical Engineering, Slovak Academy of Sciences, Dúbravská cesta 9, 841 04 Bratislava, Slovakia}
\affiliation{Centre For Advanced Materials Application CEMEA, Slovak Academy of Sciences, Dubravska cesta 9, 845 11 Bratislava, Slovakia}

\author{E.N. Beginin}
\affiliation{Laboratory "Magnetic Metamaterials", Saratov State University, Saratov 410012, Russia}

\author{M. Tapajna}
\affiliation{Institute of Electrical Engineering, Slovak Academy of Sciences, Dúbravská cesta 9, 841 04 Bratislava, Slovakia}

\author{D. Gregušová}
\affiliation{Institute of Electrical Engineering, Slovak Academy of Sciences, Dúbravská cesta 9, 841 04 Bratislava, Slovakia}

\author{M. Kučera}
\affiliation{Institute of Electrical Engineering, Slovak Academy of Sciences, Dúbravská cesta 9, 841 04 Bratislava, Slovakia}

\author{F. Gucmann}
\affiliation{Institute of Electrical Engineering, Slovak Academy of Sciences, Dúbravská cesta 9, 841 04 Bratislava, Slovakia}

\author{S. Krylov}
\affiliation{Institute of Electrical Engineering, Slovak Academy of Sciences, Dúbravská cesta 9, 841 04 Bratislava, Slovakia}

\author{A.I. Stognij}
\affiliation{Scientific-Practical Materials Research Center of National Academy of Sciences of Belarus, P. Brovki 19, 220072 Minsk, Belarus}

\author{S. Korchagin}
\affiliation{Financial University under the Government of Russian Federation, 4th Veshnyakovsky pr. 4, 111395 Moscow, Russia}

\author{S.A. Nikitov}
\affiliation{Laboratory "Magnetic Metamaterials", Saratov State University, Saratov 410012, Russia}
\affiliation{Kotel'nikov Institute of Radioengineering and Electronics, RAS, Moscow 125009, Russia}

\author{A.V. Sadovnikov}
\affiliation{Laboratory "Magnetic Metamaterials", Saratov State University, Saratov 410012, Russia}

\date{\today}

\begin{abstract}
We demonstrate the laser-induced control over spin-wave (SW) transport in the magnonic crystal (MC) waveguide formed from the semiconductor slab placed on the ferrite film. We considered bilayer MC with periodical grooves performed on the top of the \textit{n}-type gallium arsenide slab side that oriented to the yttrium iron garnet film. To observe the appearance of magnonic gap induced by laser radiation, the fabricated structure was studied by the use of microwave spectroscopy and Brillouin light-scattering. We perform detailed numerical studies of this structure. We showed that the optical control of the magnonic gaps (frequency width and position) is related to the variation of the charge carriers' concentration in GaAs. We attribute these to nonreciprocity of SW transport in the layered structure. Nonreciprocity was induced by the laser exposure of the GaAs slab due to SWs' induced electromagnetic field screening by the optically-generated charge carriers. We showed that SW dispersion, nonreciprocity, and magnonic band gap position and width in the ferrite-semiconductor magnonic crystal can be modified in a controlled manner by laser radiation. Our results show the possibility of the integration of magnonics and semiconductor electronics on the base of YIG/GaAs structures.  
\end{abstract}

\maketitle

\section{Introduction}

Currently, the intense research in the field of  dielectric magnonics~\cite{magn_roadmap, dielmagn,Nikitov_mag, demokritov2012mag,mag_blocks,kruglyak2010mag, gubbiotti_3Dmagnonics}  is focusing on the tasks of signal encoding, transport, and manipulation. This leads to the formation of the solid-state magnonics units, which in contrast with classical microelectronics, operate with magnons (quantum of spin wave) as information carriers \cite{Demokritov_magn, SWbook}. In analogy to the conventional complementary metal-oxide-semiconductor (CMOS)-based electronics \cite{asi2015itrs}, magnon-base components can act as the independent units in the magnonic networks \cite{Davies_networks, beginin_networks, sadovnikov_networks} with an aim to form functional devices. This is one of the concepts that may allow overcoming the almost achieved physical limits in CMOS-based electronics. In addition, semiconductor magnonics \cite{semicond_magn} might play the role of the bridge, allowing the combination of the magnonics and CMOS-based electronics elements. For this purpose, structures based on the bilayers of semiconductor-insulating magnetic material should be designed. Recently, a problem of incompatibility of the substrates used in insulating magnonics structures with the CMOS- components was investigated \cite{stognij_Si, Stognij_GaN, Lutsev_2DEgas, Sadovnikov_GaAs}. The perspective bilayers for semiconductor magnonics were grown, such as yttrium iron garnet (YIG) / gallium arsenide (GaAs) heterostructure. 

One of the structures  considered to be a magnonic component is the magnonic crystals (MCs) \cite{Krawczyk2014MC, chumak2017MC}. MCs are formed by the artificially created periodicity in the structure in the direction of SWs' propagation. This leads, due to the presence of Bragg resonances  in such  structures, to the formation of bandgaps in the spectrum of the propagating waves (also called magnonic band gaps). MCs are expected to be widely used in the magnonic data processing. 

The lattice of the early-designed MC was often fabricated as the periodic defects of the magnetic film surface (e.g., grooves or holes) or by growing the conductive stripes atop the magnetic layer \cite{Krawczyk2014MC}. Nowadays, an important task for magnonics is to design the reconfigurable MC in which periodic spatial variation of media properties will be induced by the physical effects due to external impact. Therefore, methods of manipulation of the lattice period in time are in demand. A few types of such MCs and the physical principles behind them will be highlighted next.

For instance, using  the current flow in the  electrode lattice atop of magnetic film layer allows  the creation of a spatial variation of the local magnetic field \cite{Chumak2009_current, nikitin2015_current}. MCs can also be performed based on the interaction of the SWs transport with the acoustic waves (which requires the excitation of acoustic waves) \cite{kryshtal2012surface}. Phenomena of these waves' interactions are studied in terms of Straintronics and Magnon Straintronics disciplines \cite{sadovnikov2019magnon, tikhonov2016brillouin, litvinenko2021tunable, litvinenko2018chaotic}.
More energy-efficient voltage tunable MCs (also controlled by the lattice of electrodes) may be created based on the periodic changes of the electric permittivity in the ferroelectric-magnetic insulator bilayers \cite{FE_Voltage_dynamic} or by the periodic variation of anisotropy in the nanoscale-size magnetic planar waveguide \cite{Voltage_anisotripy}.
Further, lattices formed in waveguides by the noncollinear magnetic states, e.g., domain walls \cite{MC_domain_walls, domain_walls_experiment} or skyrmions \cite{skyrmions_MC}, can be nucleated or annihilated by external stimuli.
Another interesting approach was proposed based on the superconductor-insulating magnetic heterostructure, where the periodic lattice of Abrikosov vortices induced a periodic local magnetic field modulation \cite{MC_fluxon}.
In the so-called moving MCs, a periodic lattice is formed by the strain-induced propagating acoustic waves \cite{moving_MC_dopler}. 
MCs based on the periodic variation of the magnetization in the magnetic film are also possible to realize by the optical means due to the heating of YIG film \cite{fetisov1996_heat, heatMc2012, heatMC2015}.

The concept of the dynamic MCs implies that tuning speed should be faster than the time of SWs' propagation through the structure. Since magnonics working frequencies are in the order of GHz (with possibilities to reach the THz range) \cite{magnonicsroadmap, serga2010yig}, to obtain a comparable rate of lattice manipulation, a physical mechanism that induces the lattice should be of the same range. In comparison with the SWs frequencies, the heating mechanisms and mechanisms of manipulation by the magnetization states are too inert in time to be applied for dynamic MCs. This constricts the number of possible methods to induce the lattice and thus limits reconfigurability methods in magnonics.

In this work, we propose to use semiconductor magnonics \cite{Sadovnikov_GaAs} in order to obtain the optically tunable MC based on YIG/GaAs heterostructure. The possibility of optical manipulation by the SWs' properties in the magnetic-semiconductor bilayers was theoretically and experimentally demonstrated in works \cite{seshadri1970surface, kawasaki1974interaction, stancil1986phenomenological, kindyak1995magnetostatic_screening, almeida1996eddy, fetisov1996optically_semicond, kindyak1999nonlinear, Kindyak, Sadovnikov_GaAs} (e.g., for YIG/GaAs heterostructure), and it was proved that this tuning was related to the variation of conductivity in the semiconductor layer due to the external light irradiation. The influence of the semiconductor screening layer conductivity on the propagating SW' may be compared to the mechanism of screening the SW's- induced electric field by a metal layer which was demonstrated in works \cite{Beginin_metal_MC, MC_stripes, mruczkiewicz2014nonreci_Me}, So it also induces the SWs' nonreciprocity \cite{MC_nonreciprocity} in the semiconductor-magnetic heterostructures \cite{Sadovnikov_GaAs}.  Thus, creating a periodic variation of conductivity in the semiconductor layer allows obtaining the MCs similar to the one with metal stripes lattice. At the same time, the rate of the SWs' tuning by the mechanism of optical injection of nonequilibrium charge carriers in semiconductor magnonics is limited by injection-recombination processes, which are fast enough to be used in the dynamical magnonics blocks. We point here that in general, the definition of the electrodynamic properties of lattices composed from the periodically arranged semiconductors is a complex task, and it was considered, e.g.,  in works \cite{korchagin2021mathematical, korchagin2021mathematical} with the author of this thesis as a collaborator. 

The possibility of both manipulation of the SWs' characteristics and inducement of a lattice cell for MCs' by the optical means in the insulator magnetic-semiconductor heterostructures opens a broad perspective for designing new tunable magnonics devices. The periodic spatial variation of conductivity can be obtained by forming specific semiconductor patterns (e.g., deposition of the semiconductor stripes atop magnetic film). In this research, in order to study the phenomena of the formation of the magnonic band gaps in the GaAs/YIG, we have combined the heterostructure of the semiconductor with a grooved surface faced to YIG. Applying an external laser light irradiation allows us to increase the contrast of the spatial variation of the charge carriers' concentration in the GaAs layer. Based on the measurements of the YIG/GaAs MC sample and their comparison with the results of numerical simulations, we demonstrated the processes of the light-induced switching of the magnonic gaps and their light-induced frequency tuning characteristic.

\section{\label{sec:level1}Sample fabrication and experimental methods}

\subsection{Fabrication of YIG/GaAs magnonic crystals structure}

\begin{figure}
\includegraphics{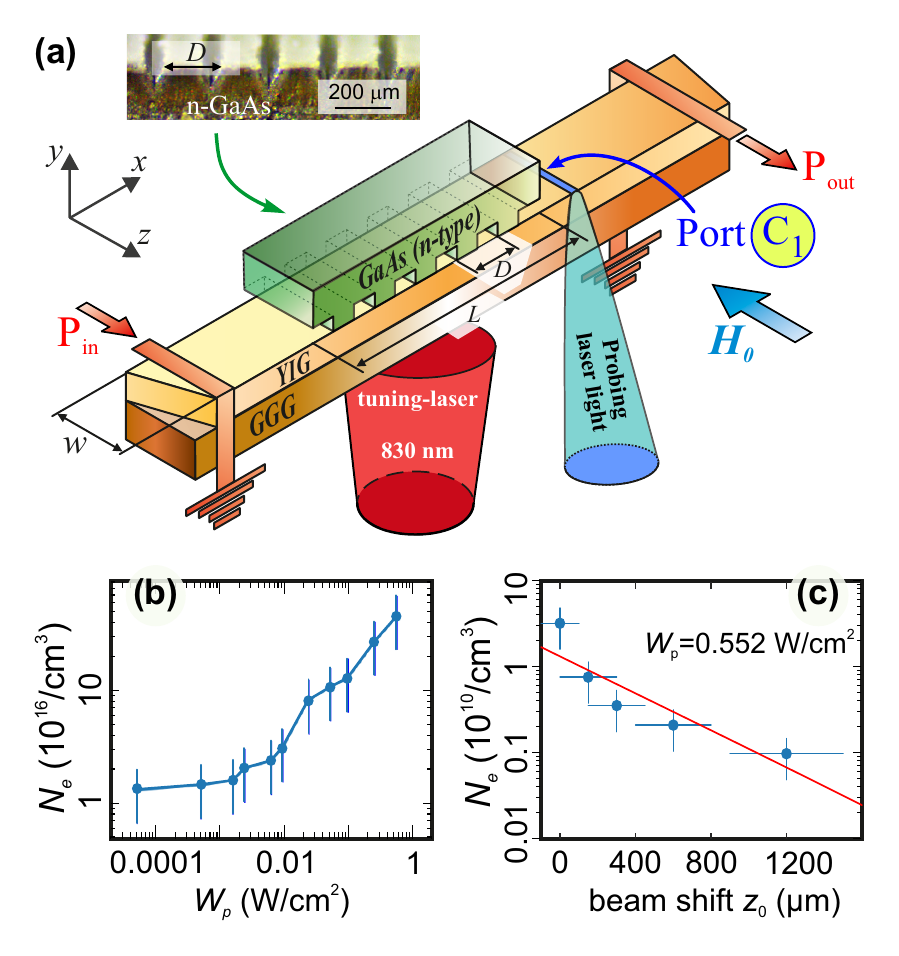}
\caption{\label{fig:structure_Sigm} 
\textbf{(a)} Sketch of the YIG/GaAs periodic structure and drawings of microwave (antennas P\textsubscript{in} and P\textsubscript{out}) and BLS (port C\textsubscript{1} and Probing laser light) experiments. Inset: Optical microscope side-view image of the laser-scribed GaAs slab used in the studied structure.
\textbf{(b)}~Dependence of the electron concentration on the laser radiation power density measured for the GaAs slab from the grooved side (doped surface). 
\textbf{(c)}~Dependence of the electron concentration versus the laser beam position shift from the Ohmic contacts region ($z_0$) measured for the GaAs slab at the not grooved side (semi-insulating surface). The laser power was fixed at the maximum level ($W_p=0.552$~W/cm$^2$). Obtained data were interpolated by analytic function (\ref{diffus}).
}
\end{figure}

The sketch of the fabricated multilayered waveguide structure is presented in Fig.~\ref{fig:structure_Sigm}~(\textbf{a}). As a material for fabrication of the magnetic planar waveguide, we used the commercial ferrimagnetic YIG film \cite{cherepanovYIG, serga2010yig} grown by high-temperature liquid phase epitaxy on the gadolinium gallium garnet ($Gd_{3}Ga_{5}O_{12}$ (111), GGG) substrate. YIG film had the following parameters \cite{serga2010yig,cherepanovYIG}: thickness $9~\mu$m, permittivity $\epsilon=9$, saturation magnetization $M_\textrm{S} = 139.26$~kA/m, gyromagnetic ratio $\gamma=175.93$~rad~GHz/T, ferromagnetic
resonance (FMR) linewidth $\mu_0\Delta H = 28$~GHz/T measured at the frequency 9.7~GHz. 

A waveguide of the width $w=1$~mm and the length of 20~mm  was etched from the YIG/GGG film with the use of the laser ablation method \cite{laserablationboock}. The ablation setup was based on fiber YAG:Nd laser with the high precision 2D scanning galvanometric module (Cambridge Technology 6240H) working in a pulse mode with a pulse length of 50~ns and a pulse power of 5~mJ. This method was adapted for processing with YIG films of thickness 0.1 - 10~${\mu}$m and earlier used in works \cite{beginin2013spatiotemporal, beginin2014multimode}.

The layered MC was composed by positioning a semiconductor slab atop this YIG waveguide. The slab of width, $w=1$~mm,  and of length, $L=5$~mm, was etched by laser ablation from the commercial epitaxially grown GaAs film (manufactured for the components of the CMOS transistors). According to the passport information of the sample, this commercial semiconductor had 1~$\mu$m of doped n-type GaAs layer atop of 500~$\mu$m thick semi-insulating n~GaAs. The expected electron concentration of semi-insulating layer was $\sim 10^{10}$~cm$^{-3}$ and of the highly doped layer $\sim 10^{17}$~cm$^{-3}$, which are in agreement with, e.g.,  \cite{semicond_empirical,GaAs_mu_T_depandence}.

We applied laser ablation method to obtain periodic grooves ($D=200\, \mu$m period) on one face of the GaAs slab (see inset in Fig.~\ref{fig:structure_Sigm}~(\textbf{a})). We note here that this spatial periodic modulation of the GaAs slab thickness was performed from the highly-doped face of the GaAs slab. Thus, the highly doped layer was split into a lattice of periodic stripes. GaAs slab was fixed atop the center of the magnetic  waveguide with grooves facing the YIG film, perpendicular to the direction of the propagating waves (see Fig.~\ref{fig:structure_Sigm}~(\textbf{a})). Further we will call this structure periodic GaAs/YIG or YIG/GaAs magnonic crystal.

The fabricated YIG/GaAs MC was placed on a holder with microstrip microwave antennas on its surface. Fixation of multilayer structure on the holder was done in a way that microstrip antennas were in contact with the YIG surface, perpendicular to the waveguide direction (see Fig.~\ref{fig:structure_Sigm}~(\textbf{a})). The holder design allowed for connecting the microstrip antennas to a microwave power source through the coaxial cables. Microstrip antennas were used to perform the SWs' excitation and detection during microwave measurements (see section "Microwave spectroscopy") and the SWs' excitation during BLS measurements (see section "Brillouin light scattering spectroscopy"). 

In order to induce the carriers in the semiconductor and, therefore, the screening of SWs-induced electric field, we used the tuning-laser (see section "Lasers for optical control over GaAs properties"). The self-constructed stand was used to fix together the holder and laser in a way that the beam irradiated all the surface of GaAs. Because of the sample holder configuration, the light beam was oriented to the multilayer sample from a GGG side (see  Fig.~\ref{fig:structure_Sigm}~(\textbf{a})). The photo-induction in GaAs was possible since GGG and YIG are optically transparent materials and the main absorber of optical power, in this case, was the highly doped GaAs layer.  

The stand which fixed the laser and holder was used to place the holder inside the electromagnetic coil in a way that the multilayer structure was magnetized tangentially (see Fig.~\ref{fig:structure_Sigm}~(\textbf{a})). It allowed the excitation of the surface spin waves (SSW) type (so-called  Damon–Eshbach configuration) \cite{damon1961, bajpai1985}. The orientation of the external magnetic field $\mu_0H_0= 0.09$~T was chosen in a way that propagating SSWs in the direction from input to output antenna had a maximum electromagnetic field localized on the surface faced to the GaAs side \cite{damon1961,SWbook, gurevichmelkov}. It was necessary for the effective interaction of SSWs with the semiconductor layer.

\subsection{Lasers for optical control over GaAs properties}

We used laser irradiation with the aim to vary the charge carriers' concentration of GaAs by photo injection. Two laser setups were applied for this purpose.

\textbf{First laser}. 632.8~nm wavelength HeNe laser was used to study the optical effect on the conductivity of the GaAs sample. 
This laser setup had a possibility of power control, and the calibrated maximum output power was 1.1~mW. We used the focus lenses system in order to focus this relatively small output laser beam power in the area of Ohmic contacts. It allowed us to obtain the maximum power density $W_p=0.552$~W/cm$^2$ in order to experimentally observe the photo impact on the charge carriers' concentration in the semiconductor sample.

\textbf{Second laser (tuning-laser)}. 830~nm wavelength fiber laser was used to irradiate the YIG/GaAs multilayer structure during the microwave spectroscopy measurements and the Brillouin light scattering measurements (see Fig.~\ref{fig:structure_Sigm}~(\textbf{a})). This laser setup had a possibility of power control, and the calibrated maximum output power was 450~mW. The focusing setup was absent; the laser spot on the sample surface had an elliptical shape with the sizes of 9~mm $\times$ 6~mm. Due to the elliptical shape of the beam, to decrease the structure heating and uniformly irradiate the GaAs surface, the laser beam was orientated in a way that the beam ellipse' major axis was parallel to the GaAs slab width, and the beam irradiated full surface of the slab. The maximum of the power density for this beam was equal 1.06~$\textrm{W/cm}^2$.

\subsection{GaAs slab electron density distribution and other parameters}

We resorted to the Ohmic contacts resistance measurements to check the values of the dark electron density on the GaAs slab faces and to obtain the optical variation of GaAs electron concentration. The details of this process are given in "Appendix". Based on the measurements, we established that dark electron concentrations are $N_e=1\cdot10^9$~cm$^{-3}$ for semi-insulating layer and $N_e=1.3\cdot10^{16}$~cm$^{-3}$ for highly-doped layer. The corresponding values of electron mobility (also see "Appendix") are $\mu_e = 8400$~cm$^2/\textrm{V\,s}$ for the semi-insulating side of GaAs sample and $\mu_e = 4000$~cm$^2/\textrm{V\,s}$ for the highly-doped GaAs side. Important to note that further in this work we assume these values of mobility to be constants (e.g., independent on light exposure). The electron effective mass for the GaAs we considered to be $m_{\textrm{eff}}=0.13 \cdot m_e$ of electron mass $m_e$ since the sample was highly doped \cite{blakemore1982semiconducting,raymond1979electron}. In addition, for the GaAs, we considered permeability $\mu=1$, and crystal lattice contribution to the GaAs permittivity $\epsilon _{g}=12.9$ \cite{blakemore1982semiconducting,raymond1979electron, GaAs_mu_T_depandence}.

The electron concentration measurements performed under the laser exposure ("First laser", $W_p=0.552$~W/cm$^2$) gave values of electron concentration $N_e=3.49\cdot10^{10}$~cm$^{-3}$ for semi-insulating layer and $N_e=4.54\cdot10^{17}$~cm$^{-3}$ for highly-doped layer. This means the variation of the magnitude of photo-induced concentration is around 1.5 order. Based on the results of resistance measurements, the dependencies of the electron concentration vs the optical power density (for highly doped layer, Fig.~\ref{fig:structure_Sigm}~(\textbf{b})) and the diffusion electron concentration (for semi-insulating layer Fig.~\ref{fig:structure_Sigm}~(\textbf{c})) were obtained. The distance-dependence of the diffusion electron concentration for the semi-insulating layer was approximated by the exponential function (\ref{diffus}) and it is plotted together with the measured data (see red line in Fig.~\ref{fig:structure_Sigm}~(\textbf{c})). Therefore, the electron diffusion length, $L_n$ (distance, at which diffusion concentration decreases $e$-times from the maximum value) was estimated as: $L_n=315~\mu$m.

Summarizing the above results, the electron concentration along the GaAs thickness is estimated as follows. We expect the thickness-uniform distribution in the 1~$\mu m$ thick highly doped layer with the electron concentration $N_e$ (controlled by the tuning-laser intensity). The electrons in the highly doped layer act as a source of the diffused electrons into the 500~$\mu m$ thick semi-insulating layer. Then, the concentration of diffused electrons vs. the coordinate along with the thickness decrease exponentially, obeying the relation (\ref{diffus}).

\subsection{Microwave spectroscopy}

To perform the microwave spectroscopy analysis of the periodic GaAs/YIG structure, a pair of 30~${\mu}$m width microwave transducers for the excitation and detection of the SW was attached to the YIG surface. These antennas had 50~Ohm impedance and the level of the input signal on the P$_{in}$ transducer -10~dBm. The input power of the microwave signal was 0.1~${\mu}$W in order to avoid the nonlinear effects \cite{gurevichmelkov}. Transmission and dispersion of SSW at different intensities of the tuning laser light were experimentally measured using PNA-X Keysight Vector Network Analyzer (VNA). Transmission response  were obtained as the frequency dependence of the absolute value of $S_{21}$ coefficient in the case when the excitation and detection of the signal were performed by microstrip antennas. Further in the text by $S_{21}$ we mean the absolute value of this coefficient.

\subsection{Brillouin light scattering spectroscopy}

BLS method, which is based on the effect of inelastic light scattering on coherently excited magnons \cite{demokritov2001, Demidov2008}, was used to measure SSWs' spectra-like signal. The  BLS setup was in the quasi-backscattering configuration, so the BLS measured signal was proportional to the squares of the dynamic magnetization components of YIG film surface $I_\textrm{BLS}\sim(m_x^2+m_y^2)$, where the probing laser beam was focused on. Probing laser light (single-frequency laser EXLSR-532-200-CDRH with a wavelength of 532~nm and power of 1~mW) had a 25~$\mu$m-diameter spot on the sample surface. 

The same orientation and value of the external magnetic field as during the microwave spectroscopy measurements were used. The input microwave antenna $P_\textrm{in}$  was used to excite SSW on the specific frequency, and the BLS scanning was performed along the waveguide structure transverse line (which can be imagined as a virtual port in Fig.~\ref{fig:structure_Sigm}~(\textbf{a})) with a 25~nm step. To obtain spectral dependence, the accumulated in time signal data was integrated through all the Port C$_1$ measured points for each of defined value of excitation frequency.

\section{Results and discussion}
\subsection{Experimental demonstration of the magnonic gap tuning}

\begin{figure}
\centering
\includegraphics{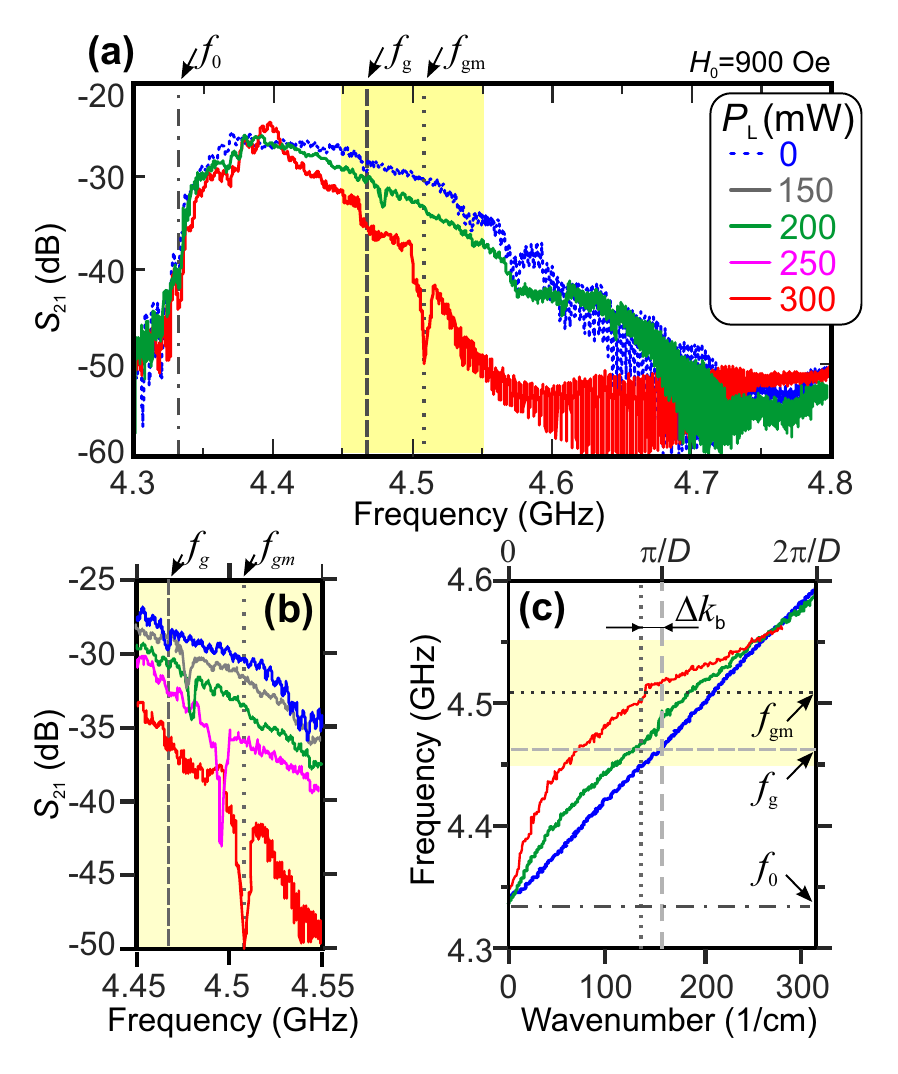}
\caption{
The results of measured frequency-dependant microwave transmission characteristics of the fabricated structure versus the power of the tuning-laser $P_\textrm{L}$ (marked in legend). $f_0$ marks the frequency of ferromagnetic resonance for in-plane magnetized ferrite film, $f_\textrm{g}$ - position of the first magnonic gap in a case of non-irradiated structure. \textbf{(a)} Transmission responce $S_{21}$ for the different $P_\textrm{L}$ values. Marked by the yellow fill sector is depicted in an enlarged scale on the panel \textbf{(b)}. 
\textbf{(c)} Dispersion relations of spin waves for cases of different $P_\textrm{L}$.  $\pi /D$ ratio marks the wavenumber value of the first Bragg resonance in the periodic structure with period $D$ without the presence of wave nonreciprocity effect. $\Delta k_b$ marks the value of Bragg resonance shift due to the appearance of nonreciprocity in the irradiated structure ($P_\textrm{L}$=275~mW). The yellow fill region corresponds to the yellow sector from panel (a).
}
\label{fig:S21_delay} 
\end{figure}

After preparation of the periodic GaAs/YIG multilayer sample and describing  its materials properties, microwave spectroscopy measurements were proceeded. Series of frequency-dependant transmission characteristics were obtained in dependence on the tuning-laser power $P_\textrm{L}$ (see Fig.~\ref{fig:S21_delay}~\textbf{(a)}). FMR frequency (the lowest frequency of the SSW transmission spectra) is marked on the transmission spectra by the $f_0$ label. We can conclude that in the range of $P_\textrm{L} \leq 275$~mW it stays constant: $f_0=4.332$~GHz. 

The dip $f_\textrm{g}=4.4673$~GHz on the transmission characteristics which corresponds to the first Bragg resonance is almost non-detectable at the low $P_\textrm{L}$ power (see Fig.~\ref{fig:S21_delay}~\textbf{(a)}, \textbf{(b)}). However, $f_\textrm{g}$ was possible to identify even at low $P_\textrm{L}$ power due to the sensitivity of the small dip to the variation of tuning-laser irradiation. The position of the first Bragg resonance dip, as well as the dip width and depth, were growing together with the increase of the tuning-laser power $P_\textrm{L}$ (see the enlarged scale in Fig.~\ref{fig:S21_delay}~\textbf{(b)}). The maximum frequency $f_{\textrm{gm}}=4.508$~GHz was obtained at $P_\textrm{L}=275$~mW. At the same moment, the growth of $P_\textrm{L}$ leads to the decrease of the whole $S_{21}$ level, which indicates the increase of the damping.

The growth of the dip width and depth under the light irradiation demonstrates the possibility of the magnonic bandgap formation in the GaAs/YIG bilayer by optical means. Taking into account the measured $N_e$ vs. $W_P$ characteristic of the GaAs sample  (see Fig.~\ref{fig:structure_Sigm}~(\textbf{b})) and the periodicity of the performed cavities in the GaAs slab (see Fig.~\ref{fig:structure_Sigm}~(\textbf{a})), we can relate the growth of the dip width and depth to the increase of the periodic structure contrast induced by the tuning-laser irradiation.

The depicted in Fig.~\ref{fig:S21_delay}~(\textbf{c}) experimentally measured dispersion curves, under the different $P_\textrm{L}$, corresponds to the SSW propagating through the studied structure. In the case of the non-irradiated structure, the dispersion curve in the plotted range is smooth and may be fitted by the linear function. Under the tuning-laser exposure, the dispersion curve experiences a jump that corresponds to the magnonic gap. At the same time, the growth of $P_\textrm{L}$ leads to the shift of the dispersion curve to the higher frequencies. In other words, we observe nonmonotonic growth of the frequency at the fixed wavenumber or decrease of the wavenumber at the specified frequency. This shift is more pronounced at the low wavenumber (and frequency).

Here we should describe the connection between the SSW dispersion and the magnonics bands in MCs. For MCs, it is known that bandgap formation may be explained by the Bragg diffraction between the propagating SW, $k^+$, and the scattered SW (formed due to a periodic lattice presence), $k^-$ \cite{Krawczyk2014MC}. 
The presence of  leads to the periodicity of the SSWs' wavevectors 
The intoduction of the reciprocal k-space and formation of Brillouin zones are usually introduced in the consoderation of the periodic lattice~\cite{Krawczyk2014MC}.

In general, when the frequencies of the propagating and scattered SSW are equal ($f(k^-)= f(k^+)$), and the wavevectors obey the so-called Bragg condition \cite{yeh1979electromagnetic, mruczkiewicz2013nonreciprocity}:

\begin{equation} \label{exchange_Bragg}
|k^-|+|k^+|= m \frac{ 2 \pi}{D},   
\end{equation}
formation of the magnonic band occurs. Here $m$ is an integer non negative number, with meaning of the order of the Bragg resonance, and $D$ is the structure period. We point here that $|k^-|$ does not need to be equal to $|k^+|$. In other words, the interference of the propagating and scattered waves results in the crossing of the dispersion branches of the propagating and scattered waves taking into account the nonreciprocal character of surface spin wave.

In the case of SSW transport in magnetic film  with the same boundary conditions at both faces, the SSWs' dispersion law is invariant to the reverse of the $\textbf{k}$ direction ($f(k)=f(-k)$). Such systems are called the reciprocal, and the regular Bragg diffraction condition is satisfied \cite{Beginin_metal_MC}:

\begin{equation} \label{Bragg_law}
k_b = m\frac{\pi}{D},    
\end{equation}
where $k_b$ is the wavelength of the Bragg resonance in the reciprocal system. The property of the Bragg diffraction in the reciprocal systems is that the length of the wavevectors of the propagating and scattered waves are equal. Thus, eq.(\ref{Bragg_law}) univocally connects the wavevectors of the magnonic bands to the lattice period, and the SSW's dispersion law defines the frequency of the magnonic bands. Here we can conclude that the appearance of SSW's nonreciprocity may be detected by a deviation of the $|k^-|+|k^+|$ from the $k_b$ value.

For the studied structure (see Fig.~\ref{fig:structure_Sigm}~(\textbf{a})), the SSW has the propagation direction from the excitation to the detection antenna, and the scattered SSW has the opposite direction. As we already mentioned, the orientation of the external magnetic field, $\textbf{H}_0$, provides the condition that propagating SSW has the electromagnetic field maximum localized at the YIG/GaAs interface, thus the scattered SSW has the electromagnetic field maximum localized at the YIG/GGG interface. Since the boundary conditions are different for the opposite sides of the YIG layer, we can expect the nonreciprocity of the propagating and scattered SSW due to the interaction of SSW with the semiconductor screening layer \cite{fetisov1996optically_semicond, semicond_magn}. And the nonreciprocity should depend on the electron concentration in the GaAs layer. The interaction of the SSW with the GaAs screening layer would depend on the penetration depth of the SSW- electromagnetic field into the GaAs. Waves with a longer wavelength should be stronger impacted by the semiconductor screening layer. 
Further, we will demonstrate the contribution of the semiconductor electron density variation on the nonreciprocal properties of SSW.

The bandgap frequencies $f_\textrm{g}$ (observed for the non-irradiated structure) and $f_{\textrm{gm}}$ (the maximum observed positive shift of bandgap in the structure under the variation of the laser exposure) are pointed in Fig.~\ref{fig:S21_delay}~(\textbf{b}) according to the dips on the corresponding $S_{21}$ dependencies. For the non-irradiated structure, the wavelength of the magnonic bandgap (defined on the measured dispersion characteristic by the $f_\textrm{g}$ value) is equal to the $k_b$ wavenumber followed from the eq.(\ref{Bragg_law}) at the $D=200 \mu$m, $m=1$: $k_b=157.08$~cm$^{-1}$. It means that for the non-irradiated structure, SSWs with $k < k_b$ are reciprocal, so the GaAs screening layer is not affecting the SSWs of these wavelengths.

Vertical dotted line in Fig.~\ref{fig:S21_delay}~(\textbf{b}) shows the $k$-position of the magnonic gap in the case of $P_\textrm{L}=275$~mW. The displacement of this position from the $k_b$ is $\Delta k_b=20.8$~cm$^{-1}$. The appearance of $\Delta k_b$ is related to the optically-induced growth of the GaAs screening layer influence on the SSW in the fabricated structure. Thus we can conclude that the SSW nonreciprocity was induced by optical means in the GaAs/YIG bilayer (for the waves with wavelengths at least below this magnonic gap), and this nonreciprocity manifested itself similarly to works \cite{mruczkiewicz2013nonreciprocity, mruczkiewicz2014nonreci_Me, MC_stripes}.

\begin{figure}
\centering
\includegraphics{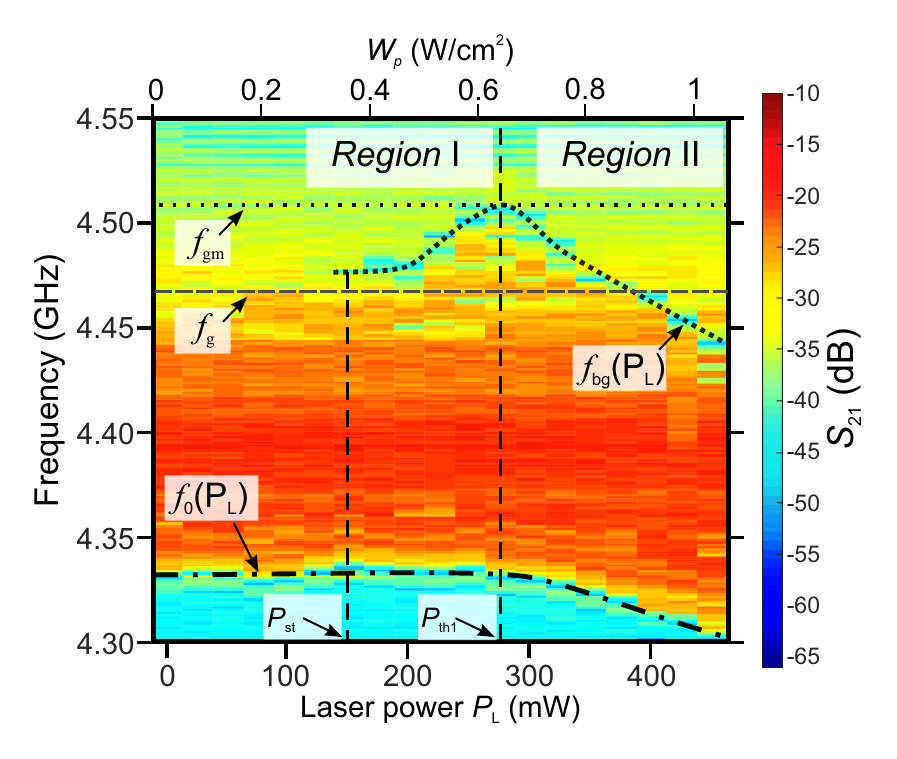}
\caption{
The color scale presents the frequency-dependent microwave transmission characteristics vs. the tuning-laser power $P_\textrm{L}$
measured for the studied structure.The power density scale is plotted along the upper axis according to the laser setup description given in Section "Lasers for optical control...". Dashed-dotted curve $f_0 (P_\textrm{L})$ marks the FMR frequency dependence vs $P_\textrm{L}$, dotted curve $f_{\textrm{bg}} (P_\textrm{L})$ marks the dependence of first magnonic gap vs tuning-laser power. This gap appears at $P_{\textrm{st}}$ and reach it maximum at $P_{\textrm{th1}}$. "Region~\rom{1}" and "Region~\rom{2}" corresponds to the negligible and well-presented laser-irradiation induced heating that impacted the characteristics of the studied microwave transport.
}
\label{fig:multispectra} 
\end{figure}

To analyze the magnonic gap frequency position dependence on the tuning-laser power ($f_{\textrm{bg}}(P_\textrm{L}$)), the series of transmission characteristics were measured with the $P_\textrm{L}$ step of 25~mW, and the result is plotted by color map in Fig.~\ref{fig:multispectra}. The dotted curve $f_{\textrm{bg}}(P_\textrm{L})$ is plotted in Fig.~\ref{fig:multispectra} with use of interpolation of the obtained at each $P_\textrm{L}$ step bandgap position. We defined the magnonic bandgap appearance threshold at power $P_{\textrm{st}}=150$~mW as the amplitude depth of the dip on the spectra curve $\leq2.5$~dB. We note here that $f_{\textrm{bg}}(P_\textrm{st}) > f_\textrm{g}$, which means that the frequency position of the Bragg resonance was already displaced due to the laser influence. At the same time, the analysis of the $S_{21}$ dip related to the Bragg resonance at the range below $P_\textrm{st}$ shows, that the dip frequency for $P_\textrm{L} < 100$mW is almost independent on $P_\textrm{L}$. This means that light control over the Bragg resonance also has a threshold behavior.

The $f_{\textrm{bg}}$ non-uniformly grows together with $P_\textrm{L}$ until its maximum value: $f_\textrm{g}(P_{\textrm{th1}}) = f_{\textrm{gm}}$, where $P_{\textrm{th1}}=275$~mW. The drop of $f_{\textrm{bg}}$ is observed above  $P_{\textrm{th1}}$. At the same time, $f_0(P_\textrm{L})$ stays approximately constant until $P_\textrm{L}<P_{\textrm{th1}}$ and then starts to decrease.

The reduction of $f_{\textrm{bg}}$ and $f_0(P_\textrm{L})$ in such a system may indicate the saturation magnetization decrease caused by the heating. Thus, $P_{\textrm{th}}$ divides the power range on the "Region~\rom{1}" and on "Region~\rom{2}". Inside the "Region~\rom{1}" the $f_{\textrm{bg}}$ grows along with $P_\textrm{L}$ and $f_0(P_\textrm{L})$ stays approximately constant (the laser-induced heating on the waveguide structure is insignificant). Along the growth of $P_{\textrm{th}}$ inside the "Region~\rom{2}",  $f_0(P_{\textrm{L}})$ decreases and the $f_{\textrm{bg}}(P_{\textrm{L}})$ also reduces (possible evidence of the YIG layer heating). Below in the text, we will describe the supposed influence of heating on the SSW transport through the studied structure in more detail.

\subsection{Numerical simulations of the YIG/GaAs bilayer and periodic GaAs/YIG structure}

In order to explain the experimentally observed results (in particular the $f_{\textrm{bg}}(P_{\textrm{L}})$ dependence), we used numerical simulations. The details of the simulation method are described in "Appendix".

\textbf{"Semi-infinite GaAs/YIG bilayer" model}

We started the analysis with the model of the in-plane magnetized infinite layers of YIG in direct contact with the GaAs with an aim to estimate the contribution of the GaAs- related parameters to the SSW characteristics.

For this task, we established the following parameters. The parameters of the YIG layer used in simulations correspond to the passport values of purchased YIG film (see section "Fabrication of YIG/GaAs magnonic crystals structure"). The orientation and value of the external magnetic field corresponded to the experimental one (see Fig.~\ref{fig:structure_Sigm}~(\textbf{a})). 

For this first simplified model, "semi-infinite GaAs/YIG bilayer" (see corresponding computation cell in Appendix, Fig.~\ref{fig:comp_cell}~(\textbf{a})), we set GaAs properties equal to one described in section "GaAs slab electron density distribution and other
parameters" with the following changes (and mark them as "Default Parameters"). We set GaAs uniform layer of 3~mm thickness in direct contact with the YIG surface. We should note that 3~mm thick GaAs layer was estimated to be enough (for the performed variation range of all the parameters) to consider the semiconductor layer as an infinite screening layer for the SSW. The electron concentration, $N_e$, was considered to be uniform within GaAs.

At first, for the "Default Parameters" and "semi-infinite GaAs/YIG bilayer" model, we relate the variation of the SSW frequencies with the electron concentration, $N_e$, in GaAs. We performed simulations for the wavelengths $kD/\pi=0.6$ and $kD/\pi=1$ as these wavelengths belong to the experimentally measured region. The range of the considered electron concentration in the GaAs layer ($N_e \in [10^{16}\,\textrm{cm}^{-3};\, 5\cdot10^{17}\,\textrm{cm}^{-3}]$) was comparable to the one expected for the optically-induced electron concentration in the GaAs slab doped layer.

\begin{figure}
\centering
\includegraphics{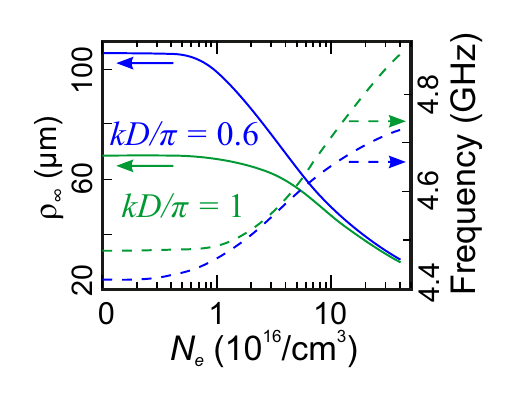}
\caption{
The numerical simulation results of the SSWs' frequency (dashed lines, right axis) and penetration depth, $\rho_ \infty$, (solid lines, left axes). Results were obtained for the wavelengths $kD/\pi=0.6$ and $kD/\pi=1$ (see color legend) in dependence of the electron concentration, $N_e$, in the structure defined by the "semi-infinite GaAs/YIG bilayer" model.  
}
\label{fig:screening} 
\end{figure}

The calculated $f(N_e)$ dependencies are plotted by the dashed lines in Fig.~\ref{fig:screening}. From these results, we can conclude that the GaAs layer is weakly impacting the frequency of SSW with wavelength $kD/\pi=1$ in the region $N_e < 1 \cdot 10^{16}$~cm$^{-3}$. At the same time, the frequency growth of the SSW with the value of wavelength $kD/\pi=0.6$ starts to be visible from the lower electron concentration: $N_e \approx 0.3\cdot 10^{16}$~cm$^{-3}$. Thus, here we confirm that for the case of the semi-infinite semiconductor screening layer, waves of the longer wavelengths are stronger affected by the semiconductor. At the same time, we can conclude that an increase of $N_e$ will lead to extend of the wavelengths interval $(0, kD/\pi)$ where the SSW's frequency is affected by GaAs. The further increase of the $N_e$ leads to the substantial growth of the SSW frequency. We note here that the SSWs frequencies limit at $N_e \rightarrow \infty$ is equal to the case of YIG film screened by the perfect electric conductor.

Next, we wanted to relate the influence of the GaAs layer on the SSWs' properties with the penetration depth (or skin depth) of the SSW's electromagnetic field. We defined the SSW's penetration depth, $\rho$, as a distance along with the $y$ axis, at which SSW's-induced $z$ component of the electric field, $E_z$, is reduced $e$-times, relative to the $E_z$ value on the YIG- facing GaAs edge. We point here that for the "semi-infinite GaAs/YIG bilayer" we mark such penetration depth as $\rho _ \infty$.

For the obtained dependencies of frequency vs. electron concentration depicted in Fig.~\ref{fig:screening}, we also plotted the corresponding dependencies of $\rho _ \infty (N_e)$ (continuous lines in Fig.~\ref{fig:screening}).  We can conclude that since SSW dependencies of $f(N_e)$ and $\rho _ \infty (N_e)$ are correlated to each other, the semiconductor screening layer influence the SSW mainly due to the SSW- electric field screening. Further, we will also call the attenuation of the SSW's- induced electric field due to the semiconductor screening layer as the SSW's screening.

Here we want to set the connection between the penetration depth and the SSWs' wavelengths. First, we can remind that for the structure depicted in Fig~\ref{fig:structure_Sigm}~(\textbf{a}), wavevector of propagating SSW's is $\textbf{k} \equiv \textbf{k}_x$ (since the numerical model considers the structure to be infinite along the $z$ direction, $\textbf{k}_z = 0$). Hereinafter $k=|\textbf{k}|$. The penetration depth of this wave along the $y$ direction outside the YIG film, $\rho=1/k_\textrm{S}$, where $k_\textrm{S}$ is the attenuation factor along the $y$ direction. In a case of pure YIG film (single magnetic layer in vacuum), $k_\textrm{S}=k$ and we mark the corresponding SSWs' skin depth by $\rho_0=1/k$. This SSW's attenuation we will call the intrinsic one.

Next, we can link to the phenomenological model of SSWs propagation in the structure magnetic insulator/metal layer, which was considered in works \cite{van1991magnetostatic, mruczkiewicz2014nonreci_Me}. In case of the direct contact of the YIG with the semi-infinite metal layer, the SSW's attenuation factor along the $y$ direction, $k_\textrm{S}$, can be written in a form:

\begin{equation} \label{SSW_GaAs_k}
k_\textrm{S}(\sigma,k) = \sqrt{k^2 + \frac{2 \, i}{\delta_\infty^2 (f(k), \sigma)}},
\end{equation}
where $\delta_\infty  (f(k), \sigma)$ is an intrinsic skin depth of a metal, $\sigma$ - electrical conductivity of the metal. 1st term of the eq.(\ref{SSW_GaAs_k}) is responsible for the impact of the intrinsic SSWs' attenuation into the attenuation factor of SSWs' in the semi-infinite metal layer. 2nd term describes the influence of the metal layer on the SSW's attenuation along the normal direction to the magnetic film (the SSW's screening). If this term grows, the SSW's penetration depth reduces. We point here that metal layer with thickness much larger then $1/k_\textrm{S}$ can be considered as semi-infinite layer.

Works \cite{van1991magnetostatic, mruczkiewicz2014nonreci_Me} introduce the analytical form of the $\delta_\infty  (f(k), \sigma)$. However, we do not know the analytical form of the eq.(\ref{SSW_GaAs_k}) in case of the magnetic layer contact with the semiconductor. Therefore we can not perform quantitative analysis. Nevertheless, we can extract the influence of the semiconductor layer on the SSW's screening from the $\rho$ obtained by the simulations. In analogy to the $\delta_\infty$ for the magnetic layer/semi-infinite metal multilayer, we introduce the $\rho_\infty$, the SSWs' penetration depth in the YIG/semi-infinite GaAs multilayer. The $\rho _ \infty$  would have contributions from the SSW's intrinsic decay, $\rho_0$, and from the intrinsic skin depth of the semi-infinite semiconductor screening layer. Thus, for the "semi-infinite GaAs/YIG bilayer" model by comparing the  $\rho_\infty$ with the $\rho_0$ it is possible to discuss the role of the GaAs screening layer on the SSWs screenings, which we will do next. Note here, that for the SSW's transport in the general YIG/GaAs multilayer structures, we have to compare $\rho$ with both  $\rho_0$ and $\rho_ \infty$. 

\textbf{"Default Parameters" variation}

As a second step of the numerical investigations of the SSW in the "semi-infinite GaAs/YIG bilayer" model (see corresponding computation cell in Appendix, Fig.~\ref{fig:comp_cell}~(\textbf{a})), we were focused on the influence of the "Default Parameters" on the SSWs' screening (through analysis of the penetration depth dependencies, $\rho$). We independently varied (while the other parameters were fixed and corresponded to the "Default Parameters") the electron diffusion length, $L_n$, the size of the air gap between the YIG and GaAs, $t_g$, and the GaAs electron collision frequency, $\nu$.

\begin{figure}[h!]
 \centering
 \includegraphics{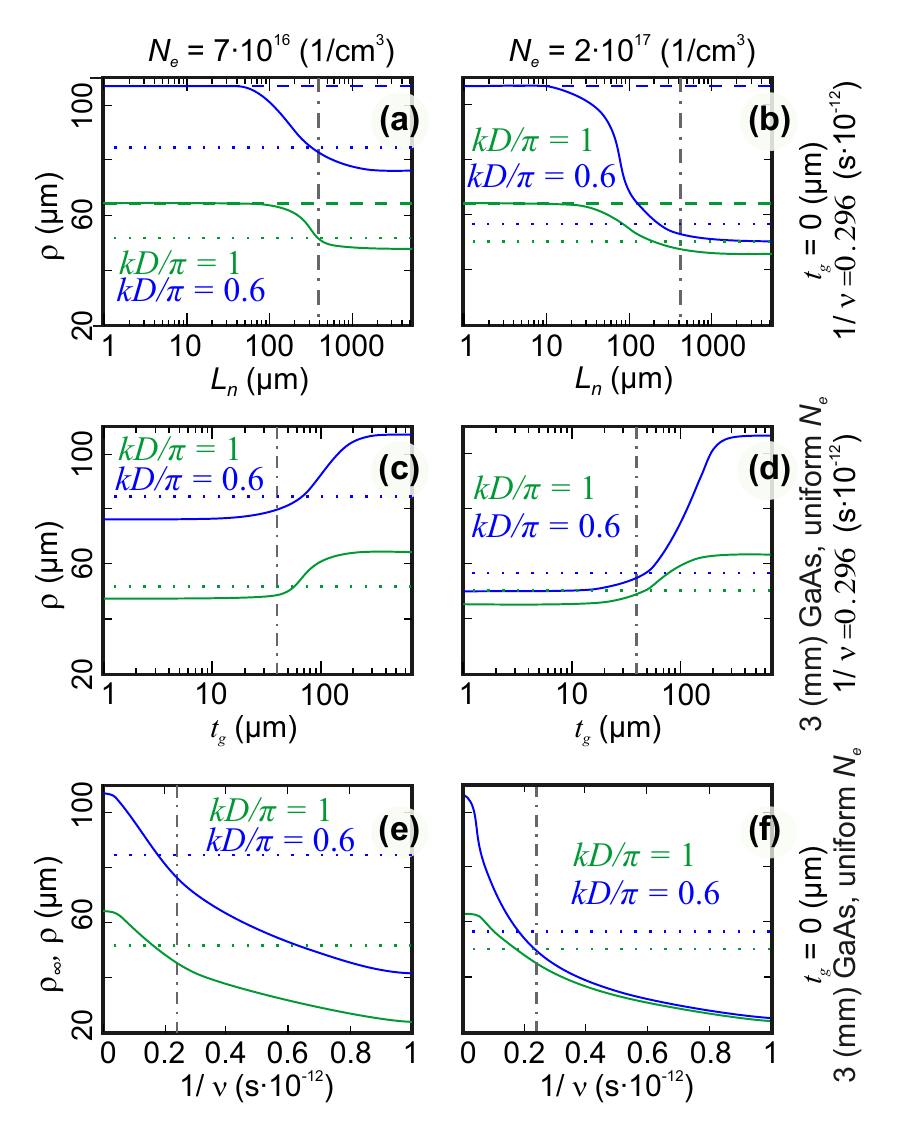}
 \caption{The dependencies of the SSWs' penetration depth, $\rho$, for the wavelengths $kD/\pi=0.6$ and $kD/\pi=1$, and the GaAs layer electron concentration $N_e=7\cdot 10^{16}$~cm$^{-3}$ (left column) and $N_e=2\cdot 10^{17}$~cm$^{-3}$ (right column). Results depicted by solid lines in panels (\textbf{a})-(\textbf{f}) were obtained in dependence on variation of the "semi-infinite GaAs/YIG bilayer" model parameters. Vertical dashed-dotted lines mark the "Experimental Parameters" values of the "experimental YIG/GaAs multilayer, and the horizontal dotted lines demonstrate SSW's $\rho$ obtained by this model. 
 (\textbf{a})-(\textbf{b}) SSWs' penetration depth were obtained vs. the electron diffusion length in the GaAs: $\rho(L_n)$. Horizontal dashed lines demonstrate the $\rho$ of  SSW's screened by the $1\,\mu$m thick layer of GaAs with the uniform electron concentration, $N_e$. (\textbf{c})-(\textbf{d}) SSWs' penetration depth
  were obtained vs. the air gap size, $t_g$ of the YIG/GaAs interface: $\rho(t_g)$. 
 (\textbf{e})-(\textbf{f}) SSWs' penetration depth, $\rho_ \infty$, were obtained vs. the GaAs mean collision frequency, $\nu$: $\rho_ \infty (\nu)$.}
\label{fig:rho}
\end{figure}

The penetration depth dependencies (see Fig.~\ref{fig:rho}) were performed for the same two wavelengths as above ($kD/\pi=0.6$ and $kD/\pi=1$) and for the particular cases of the electron concentration $N_e=7\cdot10^{16}~1/\textrm{cm}^3$ and $N_e=2\cdot10^{17}~1/\textrm{cm}^3$. These values of the GaAs $N_e$ correspond to the obtained concentrations of the experimentally-used sample (see Fig.~\ref{fig:structure_Sigm}~(\textbf{b})) at the weak and high power of the laser light irradiation.  Panels (\textbf{a}-\textbf{d}) in Fig.~(\ref{fig:rho}) show that electron diffusion length in the GaAs and size of the gap between YIG and GaAs layers influence the $\rho$ according to the "S"-shape dependence with the limits of free YIG film ($\rho _0(k)$) and saturation of the parameter impact on the SSW screening ($\rho_ \infty$). At the same time, the dependencies of $\rho_\infty(1/\nu)$ have a monotonous decrease from the limit of free YIG film ($\rho_\infty(\nu \rightarrow \infty)=\rho_0$ ). This means that the increase of the electron mean free time increases the efficiency of the SSW's screening by the GaAs layer and the limit skin depth of the SSW's in the "semi-infinite GaAs/YIG bilayer" model is a function of wavenumber, semiconductor electron concentration and semiconductor electron mean free time: $\rho_\infty (k, N_e, \nu)$. 

In Figs.~\ref{fig:rho}~(\textbf{a}-\textbf{b}) we present the numerically-obtained dependencies of the SSWs screening, $\rho$ vs the electron diffusion length, $L_n$, along the 3~mm thick- GaAs layer. Here we can consider that variation of the $L_n$ leads to the change of the effective thickness of the GaAs screening layer.
We can split the obtained dependence $\rho (L_n)$ into three regions (according to the $L_n$ range). If $L_n \ll \rho_\infty$, the GaAs layer influence on the SSW's is weak and ($\rho \approx \rho_0(k)$). If $L_n \gg \rho_\infty$, the GaAs layer influence on the SSW's is maximal and $L_n \approx \rho_\infty (k, N_e)$. The observed transition of $\rho$ (between the SSW's in the free YIG film and SSW's in the YIG/semi-infinite GaAs) happens, then $L_n \approx \rho_\infty$. We should point here, that difference between the $\rho_0(k)$ and $\rho_\infty (k, N_e)$ grows along with $N_e$ and wavenumber, defining the left and right limits of dependencies depicted in Figs.~\ref{fig:rho}~(\textbf{a} and \textbf{b}). Thus, the gradient of $\rho (L_n)$ in the transition range of $L_n$ also grows along with $N_e$ and $1/k$.

In addition, we studied the SSW's screening in the YIG/GaAs structure in the case of the GaAs layer of 1~$\mu$m thickness with the uniform $N_e$. This case corresponds to taking into account only the highly doped layer of the GaAs sample used in the fabricated structure. The dashed horizontal lines in Figs.~\ref{fig:rho}~(\textbf{a}-\textbf{b}) shows that 1 $\mu$m layer influence on the SSW's screening is negligible since for the considered $N_e$ values, the obtained $\rho(k) \approx \rho_0 (k)$. It means that for the considered wavelengths and the GaAs electron concentration, the role of the doped layer in the experimentally used GaAs sample is mainly a source of the diffused electrons to the undoped GaAs layer. The main screening of the SSWs' is by the diffused electrons to the undoped GaAs layer.

Next, we investigated the role of the air gap influence between the layers of the YIG/GaAs interface. The appearance of the air gap, $t_g$, leads to the weakening of the SSW's screening by the GaAs layer. The obtained dependencies of $\rho(t_g)$ depicted in Figs.~\ref{fig:rho}~(\textbf{c}-\textbf{d}) have a similar shape to the $\rho (L_n)$ dependencies presented in  Figs.~\ref{fig:rho}~(\textbf{a}-\textbf{b}). 
The $\rho(t_g)$ dependencies also can be split into three regions (according to the $t_g$ range). In the presence of nonzero $t_g$, the eq.(\ref{SSW_GaAs_k}) is not valid. We would characterize the $\rho(t_g)$ dependence by comparing the $t_g$ with $\rho_0$. If the $t_g \ll \rho _0$, the SSW's screening is similar to the one in the structure YIG/semi-infinite GaAs ($\rho \approx \rho _ \infty (N_e,k)$). If the air gap between the interfacial layers is big ($t_g \gg \rho _0$), the SSW's penetration depth is similar to the one in the YIG film without any screening layer ($\rho \approx \rho_ 0 (k)$). The transition between this states occurs then $t_g \approx \rho_0 (k, N_e)$. Again, due to difference between the $\rho_0(k)$ and $\rho_\infty (k, N_e)$ grows along with $N_e$, the gradient of $\rho (t_g)$ in the transition range of $t_g$ grows along with $N_e$ (see the difference between Figs.~\ref{fig:rho}~(\textbf{c} and \textbf{d}). 
 
In the next step, we wanted to study the influence of the electron mean collision frequency (semiconductor media damping), $\nu$, on the SSW's screening by the semiconductor layer. Point here that for these simulations, we varied only $\nu$ parameter in the "semi-infinite
GaAs/YIG bilayer" model, so the obtained SSWs' penetration depth was the $\rho _ \infty$ one.  The obtained $\rho _ \infty (\nu)$ dependencies are depicted in Figs.~\ref{fig:rho}~(\textbf{e}-\textbf{f}). We can see that the growth of the semiconductor electrons collision frequency, $\nu$, leads to the reduction of the SSW's screening by the semiconductor layer (value $\rho _ \infty (k,N_e)$ grows). Thus to obtain the reliable result of SSW's screening by the simulations, the choice of the $\nu$ parameter should be related to the range of the expected values valid for the experimental samples. The dependency between $\nu$ and $\mu$ (see eq.(\ref{e_collision}), Appendix) leads to the dependency of mean collision frequency on the variation of the semiconductor doping electron concentration \cite{semicond_empirical} and the semiconductor temperature \cite{GaAs_mu_T_depandence}. Thus, we can expect the $\nu$ growth along with the growth of the laser irradiation power, $P_\textrm{L}$.

As the result of analyzing the SW penetration depth vs the  "Default Parameters" variation, we can conclude about the importance of all parameters choice since the role of every parameter was found to be significant on the waves screening.

\textbf{"Experimental YIG/GaAs multilayer" model}

In order to make the correlation between the presented in Figs.~\ref{fig:rho}~(\textbf{a}-\textbf{f}) $\rho$ dependencies with the SSW's screening expected in the fabricated structure (see Fig.~\ref{fig:structure_Sigm}~(\textbf{a})), we introduced the advanced numerical model "experimental YIG/GaAs multilayer" with the following "Experimental Parameters". We established the 1~$\mu m$ GaAs layer with the defined concentration $N_e$ (the strongly n-doped layer) attached to the $500~\mu m$ thick layer with the diffusion length $L_n=315\,\mu$m (semi-insulating layer). We supposed the $t_g=40~\mu$m thick air gap in the YIG/GaAs interface in order to take into account the roughness of the YIG and GaAs film surfaces. The charge carriers collision frequency was set to be $1/ \nu = 0.296 \cdot 10^{-12}$~s. Thus, the difference between the "semi-infinite GaAs/YIG bilayer" and the "experimental YIG/GaAs multilayer" models is in the simultaneous presence of the diffusion length, the air gap in the YIG/GaAs interface, and the reduced thickness of the GaAs layer (see corresponding computation cell in Appendix, Fig.~\ref{fig:comp_cell}~(\textbf{a})).

Values of the "Experimental Parameters" parameters ($L_n$, $t_g$, $\nu$) were pointed in the panels of Figs.~\ref{fig:rho}~(\textbf{a}-\textbf{f}) by the vertical dashed-dotted lines. The crossing points of $\rho$ dependencies with these vertical dashed-dotted lines demonstrate that a small deviation of every parameter will lead to the change of the resulting $\rho$.  In other words, for all "Experimental Parameters" values $\rho$ dependencies are in the transition between the $\rho _ \infty (N_e,k)$ and $\rho _0(k)$. In conclusion, all the "Experimental Parameters" would significantly influence the SSWs' properties and their choice in the "experimental YIG/GaAs multilayer" model is important for the correspondence of the simulation results to the experiment. At the same time, simulations of the SSWs skin depth in dependence on a wide range of parameters may be useful for designing the waveguide structure based on YIG/GaAs interface. 

We calculated the values of SSW's $\rho$ with the use of the "experimental YIG/GaAs multilayer" model for the wavelengths $kD/\pi=0.6$ and $kD/\pi=1$ and for the electron concentration $N_e=7\cdot10^{16}~1/\textrm{cm}^3$ and $N_e=2\cdot10^{17}~1/\textrm{cm}^3$ (see the horizontal dotted lines in Fig.~\ref{fig:rho}). We point here that due to the impact of the presence of YIG/GaAs air gap and due to the reduction of the GaAs layer effective thickness ($500~\mu m$ thick layer with the diffusion length $L_n=315\,\mu$m), the SSW's screening in this simulated structure is weaker than the one for the "semi-infinite GaAs/YIG bilayer" with the same value of $\nu$.

\textbf{"Experimental YIG/GaAs multilayer" model with periodic GaAs groves}

In order to obtain the dispersion characteristics of the SSWs' in the periodic GaAs/YIG structure (see Fig.~\ref{fig:structure_Sigm}~(\textbf{a})) by numerical simulations, we extended the "experimental YIG/GaAs multilayer" model by taking into account the periodic grooves of the GaAs surface that faced the YIG (see corresponding computation cell in Appendix, Fig.~\ref{fig:comp_cell}~(\textbf{b})). Again, we considered the structure to be uniform (infinite) along the transverse direction ($z$- axis). The shape of the GaAs layer periodic grooves in simulations corresponded to the GaAs slab side- image (see insert in Fig.~\ref{fig:structure_Sigm}~(\textbf{a})). In the direction along the propagating SSWs ($x$- axis), we considered one period of the periodic GaAs/YIG structure.  The computation cell of the simulated structure contained this one period limited from the sides by the periodic boundary conditions.

\begin{figure}
\centering
\includegraphics{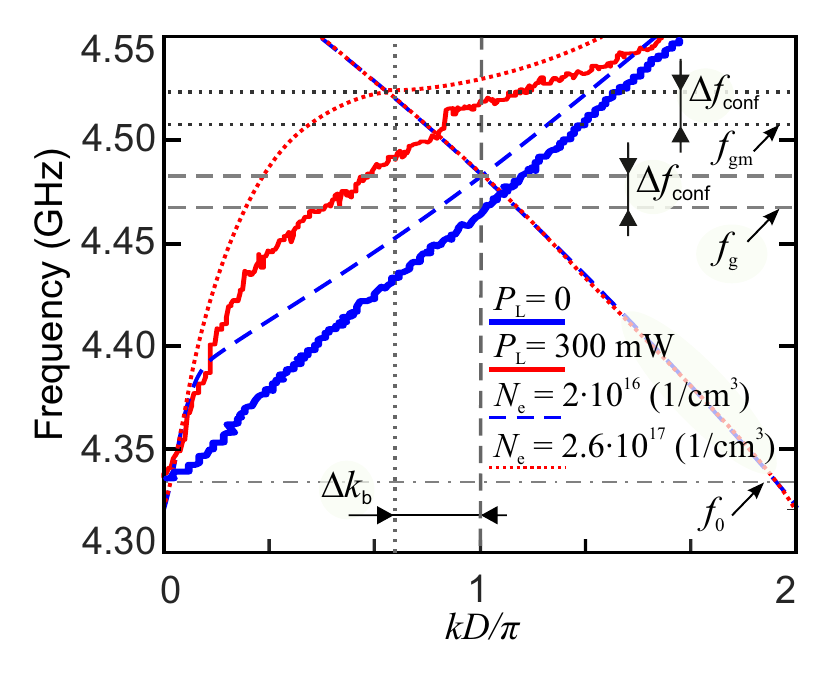}
\caption{
Dispersion relations of SSWs' in the periodic GaAs/YIG structure (depicted in Fig.~\ref{fig:structure_Sigm}~(\textbf{a})) obtained by microwave spectroscopy ($P_\textrm{L}=0$, blue solid curve; $P_\textrm{L}=275$~mW, red solid curve) and by FEM ($N_e=2\cdot10^{16}$~cm$^{-3}$, blue dashed curves; $N_e=2.6\cdot10^{17}$~cm$^{-3}$, red dotted curves). Horizontal lines marks the obtained from the measured $S_{21}$ dependencies: FMR frequency ($f_0$), measured magnonic gap position for the non-irradiated structure ($f_\textrm{g}$), and measured maximum frequency of the magnonic gap in the irradiated structure ($f_{\textrm{gm}}$). The $\Delta f_{\textrm{conf}}$ mark the frequency difference between the first Bragg resonance (when the SSWs are reciprocal) obtained by the microwave measurements and by numerical simulations, related to the experimental sample's finite size along the $z$ direction. 
}
\label{fig:comsol_disp} 
\end{figure}

The numerically obtained dispersion relations of the SSWs' in the periodic GaAs/YIG bilayer are depicted in Fig.~\ref{fig:comsol_disp} by the blue dashed and red dotted curves. We present the dispersion relations for the two electron concentrations: $N_e=2\cdot10^{16}~\textrm{cm}^{-3}$ (blue dashed curves) and $N_e=2.6\cdot10^{17}~\textrm{cm}^{-3}$ (red dotted curves). These concentrations belong to the experimental GaAs slab under the different power of the laser exposure (see Fig.~\ref{fig:structure_Sigm}~(\textbf{b}),(\textbf{c})). Due to the periodicity of the simulated structure, the dispersion relations of the SSWs in this structure are also periodic and within the Brillouin zone  contain both propagating SSW and scattered SSW. From the numerically-obtained results presented in Fig.~\ref{fig:comsol_disp}, we can conclude that scattered SSW is almost not influenced by the variation of the GaAs $N_e$. At the same time, we can see that simulated dispersions of the propagating SSW are influenced by $N_e$ in the same manner as in the experiment (see solid blue and red lines in Fig.~\ref{fig:comsol_disp}): dispersion curve shifts to the higher frequencies along with the growth of $N_e$. This confirms the appearance of the SSWs' nonreciprocity in the YIG/GaAs structure along with the growth of the semiconductor electron concentration, $N_e$.

The position of the crossings of propagating and scattered SSW dispersion curves within the Brillouin zone corresponds to the first Bragg resonance (see simulation results in  Fig.~\ref{fig:comsol_disp}). The SSWs' dispersion relation at the GaAs electron concentration $N_e=2\cdot10^{16}~\textrm{cm}^{-3}$ (blue dashed curves) demonstrate that propagating SSWs was affected by the semiconductor layer in the wavenumbers range lower then $kD/ \pi =1$. This means that Bragg resonance happens between the reciprocal SSWs (see eq.(\ref{Bragg_law})). At the same time, SSWs' dispersion relation at the $N_e=2.6\cdot10^{17}~\textrm{cm}^{-3}$ of the GaAs (red dotted curves) demonstrate the shift $\Delta k_b$ of the Bragg resonance to the lower wavenumber and higher frequency since the range of the propagating SSWs' wavelengths affected by the GaAs load increased along with the $N_e$ growth. For this case, Bragg resonance happens between the nonreciprocal waves (see eq.(\ref{exchange_Bragg})). At the same time, the threshold behavior over the Bragg resonance control is in agreement with the experimental results (see Fig.~\ref{fig:multispectra}).

\subsection{Comparison of the numerically and the experimentally obtained results}

In Fig.~\ref{fig:comsol_disp},
horizontal lines indicate the experimentally measured (see Fig.~\ref{fig:S21_delay}) frequency values of the FMR frequency, $f_0$, the magnonic gap for non-irradiated GaAs, $f_\textrm{g}$, and the magnonic gap for the structure under the  "tuning-laser" $P_{\textrm{th1}}$ exposure, $f_{\textrm{gm}}$.

As we mentioned above, the numerically obtained SSWs' dispersion relation at $N_e =2\cdot 10^{16}~\textrm{cm}^{-3}$ demonstrates the first Bragg resonance between the reciprocal waves (see blue dashed curves in Fig.~\ref{fig:comsol_disp}). At the same time, the frequency of this resonance is higher than the magnonic gap frequency, $f_\textrm{g}$, obtained by the microwave measurements of the experimental sample without the "tuning-laser" exposure (see $\Delta f_{\textrm{conf}} = 15$~MHz above the $f_\textrm{g}$ in Fig.~\ref{fig:comsol_disp}). We relate this difference with the finite transverse dimension (width) of the fabricated sample, which was not taken into account in the numerical model. According to the first theoretical work focused on the SSWs' transport in the transversely- confined magnetic slabs \cite{Okeeffe}, the reduction of the slab width leads to the shift of the SSWs' dispersion to the lower frequency in the range of short wavelengths.

The GaAs electron concentration value, $N_e = 2 \cdot 10^{16} \; \textrm{cm}^{-3}$, was fitted in a way that numerically obtained SSWs' dispersion dependency (see red dotted curves in Fig.~\ref{fig:comsol_disp}) would correspond to the Bragg resonance frequency $f_{\textrm{gm}}+\Delta f_{\textrm{conf}}$. Thus we roughly took into account the sample's width in the numerical model which would correspond to the microwave measurements of the magnonic gap in the irradiated sample at $P_{\textrm{th1}}=275$~mW.

In addition, we point that the quantitative difference between the dispersion curves obtained by microwave spectroscopy measurements with the numerically obtained dispersion relations depicted in Fig.~\ref{fig:comsol_disp} (for both values of $N_e$). We relate it to the finite transverse dimension of the fabricated sample \cite{Okeeffe}. In particular, for the electron concentration value $N_e=2.6\cdot10^{17}$~cm$^{-3}$ in the numerical model, the obtained shift of the Bragg resonance wavelength position, $\Delta k_b = 51.8\; \textrm{cm}^{-1}$, is bigger then the one observed in Fig.~\ref{fig:S21_delay}~(\textbf{c}). Also, the simulated dispersion of the periodic GaAs/YIG bilayer demonstrates a stronger influence of the GaAs screening layer on the SSW at the long wavelengths. One more manifestation of the difference between the model and experiment is in the lower limit frequency of the waves: for the numerically - obtained dispersion dependencies it lies below the experimentally measured $f_0$.

\begin{figure}
\centering
\includegraphics{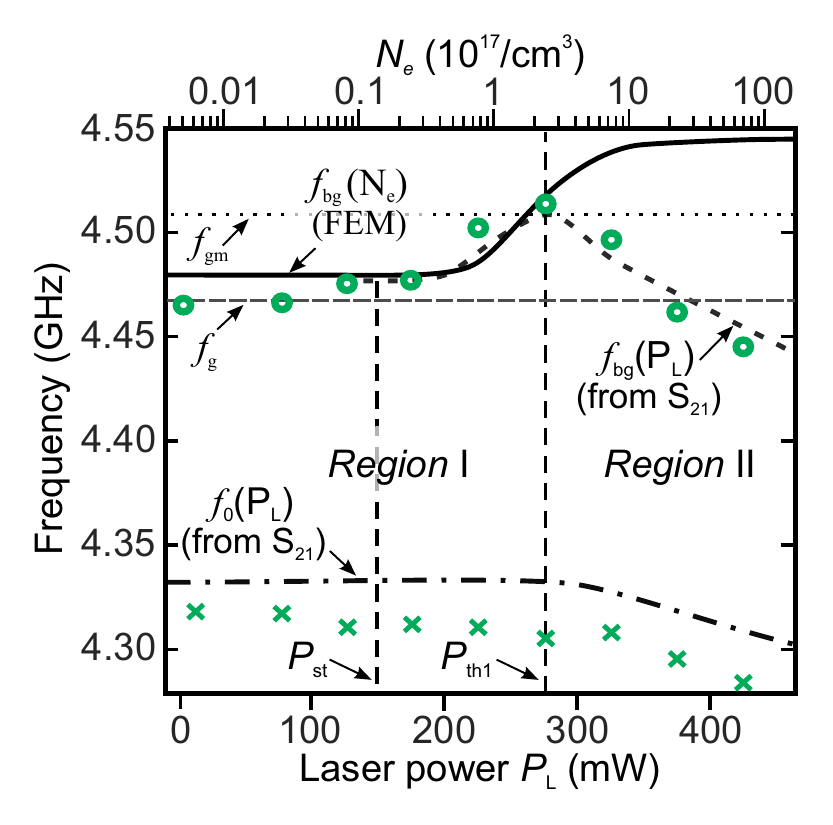}
\caption{
Data obtained from the measured $S_{21}$ dependencies (see Fig.~\ref{fig:multispectra}): $f_{\textrm{bg}}(P_\textrm{L})$ - the dependence of the first magnonic band gap frequency position versus the variation of the irradiation laser power, $P_{\textrm{st}}$ marks the threshold of the $f_{\textrm{bg}}(P_\textrm{L})$. $f_{\textrm{0}}(P_\textrm{L})$ - the dependence of the FMR frequency versus the variation of the irradiation laser power. "Region~\rom{1}" and "Region~\rom{2}" corresponds to the negligible and well-presented  impact of the heating (induced by the "tuning-laser") on the studied $S_{21}$ dependencies (according to FMR frequency). Data obtained by FEM- model: $f_{\textrm{bg}}(N_\textrm{e})$ - the dependence of the first magnonic gap frequency position versus variation of the GaAs electron concentration. Data obtained from BLS measurements: Green dots - the dependence of the first magnonic band gap frequency position versus the variation of the irradiation laser power. Green crosses - the dependence of the FMR frequency versus the variation of the irradiation laser power.
Horizontal lines marks the obtained from the measured $S_{21}$ dependencies: FMR frequency ($f_0$), measured magnonic gap position for the non-irradiated structure ($f_\textrm{g}$), and measured maximum frequency of the magnonic gap in the irradiated structure ($f_{\textrm{gm}}$). 
}
\label{fig:comparing}
\end{figure}

The numerically obtained dispersion relation for the SSWs in the layered structure demonstrates the opening near the crossing of the dispersion branches corresponding to the propagating and scattered waves ($f_{\textrm{bg}}$ point). However, the width of this opening is a few orders lower than the experimentally-obtained width of the magnonic bandgap dip (see Fig.~\ref{fig:S21_delay}~(\textbf{c}).  The partial differential equation for the eigenvalue problem was defined as $f(k)$. The imaginary part of wavenumber ($k"$) describing the spatially-related damping is not presented in the simulation output. Thus the numerical model is not provide the information about all types of losses related to the SSW interaction with the periodic semiconductor screening layer and the large bandgap observed in experiments.

From the previous analysis, we can expect that increasing the electron concentration in the GaAs periodic lattice increases the magnonic crystal contrast. This should lead to an increase in the interference between the propagating and scattered SSW, which results in an increase in the width of the band gap. So we can use the $f_{\textrm{bg}}$ increase as the parameter evidence of the contrast of periodic screening layer related to electrons concentration, $N_e$. This is confirmed by the experiment results as the magnonic bandgap forms and increases with the tuning-laser power (see Fig.~\ref{fig:S21_delay}).

The series of simulations with the varying of $N_e$ allowed to plot the $f_{\textrm{bg}}(N_e)$ dependence and compare it with the experimentally obtained $f_{\textrm{bg}}(P_\textrm{L})$ (see Fig.~\ref{fig:comparing}). The juxtaposed of $P_\textrm{L}$ and $N_e$ axis in Fig.~\ref{fig:comparing} was made, since the  $N_e(W_p)$ dependence can be fitted by a linear function (see Fig.~\ref{fig:structure_Sigm}~(\textbf{c})).

The tuning-laser irradiation of the studying structure  is expected to induce the heating of the sample. As the GGG and YIG are optically transparent materials, the tuning-laser beam directed to the structure from the GGG face should be mainly absorbed by the GaAs layer and thus heat it. However, the microwave SSW transport measurements of the studying multilayer structure can only indirectly detect the YIG temperature increase. It is  due to heat-induced  $M_\textrm{S}$ decrease, resulting in the  drop of the FMR frequency, $f_0$. According this analysis, the $f_0(P_\textrm{L})$ drop is observed only in the "Region~\rom{2}" in Fig.~(\ref{fig:multispectra}). Therefore, we expect an important contribution from the heating when the sample is irradiated with laser power exceeding the value $P=275$~mW.

We have to note here that the measurement setup configuration might mask the temperature-induced decrease of FMR frequency, $f_0$. We measure the transport of the SSW between two antennas separated by the distance, 9~mm, and the heat source (GaAs slab) covers only part of the YIG material (5~mm length). Therefore the transport of SSW between the antennas will be supported only if their frequency is above the threshold FMR frequency value,  $f_0$, in both YIG-heated and YIG-non-heated parts. The transmitted signal will be measured above the higher frequency of YIG-heated and YIG-non-heated parts. In our case, it is the material's properties that are close to the antennas (far from the heat source). 

Once the $f_0$ drop becomes pronounced on the transmission spectra, it means that the heating gradient reaches the antennas' region, which distance from the heating source is in the mm- range. Considering this range of heat transport, we can not expect that the magnonic gap may be influenced by the periodic temperature gradient in the YIG layer (caused by the GaAs periodicity) with thickness $t_s=9~\mu$m since the temperature profile is expected to be flattened in the YIG layer.

In the case of the uniform heating of the YIG layer, the approximately linear decrease of the dispersion relation frequencies should be observed, as the $M_\textrm{S}$ decrease uniformly in the whole magnetic film. Thus the decrease rate of the $f_{\textrm{bg}}$ and of $f_0$ should be similar (unlike we see in Fig.~\ref{fig:comparing}). Therefore, one can expect that the region of YIG under the irradiated GaAs should have a higher temperature. The position of the rejection band  $f_{\textrm{gb}}$ is sensitive only to the temperature of the sample covered by the periodic GaAs. On the other hand, the measured $f_0$ is sensitive to the area of the sample close to the antennas, as discussed above. Therefore the drop of $f_{\textrm{gb}}$ will be more significant than the drop of frequency of $f_0$ as the tuning-laser power increases above P= 275~mW.

Additionally, due to the YIG/GaAs air gap in the composed structure and the effective light absorption by the semiconductor, the GaAs temperature can be much higher than in YIG below the GaAs. We can expect that with an increase in temperature of the GaAs, the electron mobility will decrease, and thus, the collision frequency $\nu$ will increase \cite{GaAs_mu_T_depandence}. As it was shown in Fig.~\ref{fig:rho}~(\textbf{e})-(\textbf{f}), the $1/\nu$ decrease leads to the weakening of SSW screening by the GaAs and, therefore, contributes to further lowering of the $f_{\textrm{bg}}$.

\begin{figure}
\centering
\includegraphics{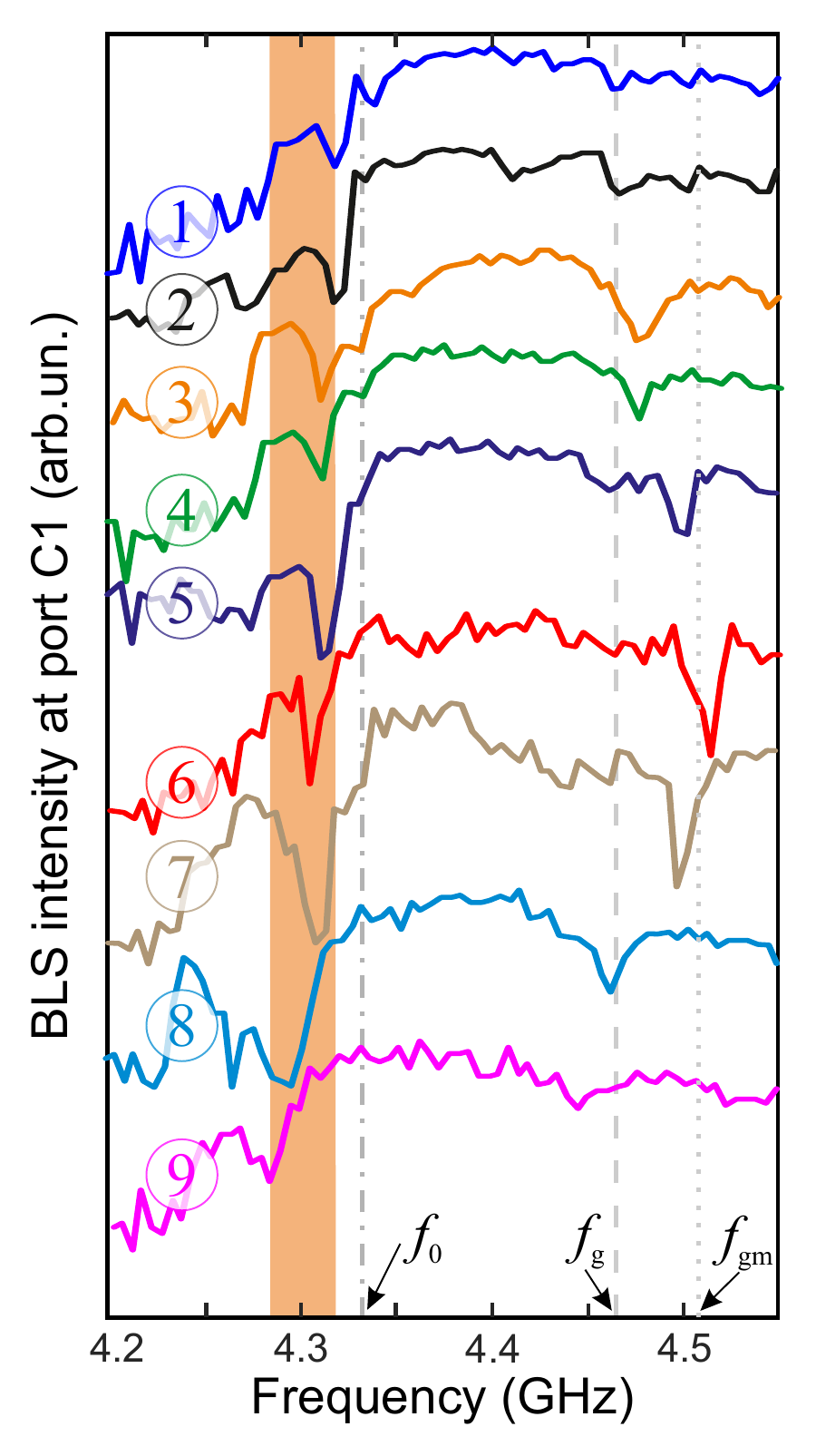}
\caption{
BLS spectra of SW propagating through the periodic GaAs/YIG structure. Curves were measured by the variation of the tuning-laser $P_\textrm{L}$: 1- 0~mW, 2- 75~mW, 3- 125~mW, 4- 175~mW, 5- 225~mW, 6- 275~mW, 7- 325~mW, 8- 375~mW, 9- 425~mW. $f_0$, $f_\textrm{g}$, and $f_{\textrm{gm}}$ correspondingly marks obtained by the microwave spectroscopy the FMR frequency, the position of the first magnonic gap in a case of non-irradiated structure, and the position of the magnonic gap at the laser power $P_\textrm{L}=275$~mW. Red fill marks the BLS-obtained range of the FMR thermal decrease. 
}
\label{fig:BLS} 
\end{figure}

The sample holder configuration did not allow the displacement of the excitation microwave antennas closer to the semiconductor slab edges. In order to study the influence of the regions not covered by GaAs on the microwave spectroscopy result (for instance, induced by the non-uniform heating), the BLS measurement setup was used instead.
The points where the BLS scanning was performed were closest possible to the semiconductor slab edge (see Fig.~\ref{fig:structure_Sigm}~(\textbf{c})). The frequency range during the BLS scanning was chosen to resolve the first Bragg resonance and the FMR frequency range. This was done in order to study the SSW screening, YIG layer heating, and the $f_{\textrm{bg}}$ formation in more detail.

The BLS spectra were measured with the mean step of  $P_\textrm{L} =50 \textrm{mW}$ in the interval from 0 to  425~mW. The plotted spectra are presented in Fig.~\ref{fig:BLS}.
Also, the BLS- resolved FMR frequency and first Bragg resonance frequencies in dependence on tuning-laser power are depicted in Fig.~\ref{fig:comparing}. The related to FMR dip on the BLS spectra at $P_\textrm{L} = 0 \textrm{mW}$ has the lower frequency value in comparison to obtained by the microwave spectroscopy $f_0$.  This may be linked to the specificity of the light-resolving of the magnetization dynamics of the multimode SSWs' transport \cite{demokritov2001brillouin}. In particular, the magnetization dynamics probe by BLS consists of accumulating the amplitude-frequency characteristics in discrete points along the Port $C_1$ line (see Fig.~\ref{fig:structure_Sigm}~(\textbf{a})) and then sum them up. In contrast, detecting the SSW transport by the probing antenna means obtaining the microwave current induced in the antenna by the SSW. Thus the resulting signal obtained by these two methods will be different if the SSWs spectra contains any spatially nonsymmetrical mods with respect to the central waveguide axis.  This modes may be especially significant near the FMR frequency since the spectra of long-wavelengths SSWs becomes strongly affected by the confinement of the structure \cite{Okeeffe,demokritov2001}.  We expect that the difference between the FMR frequency obtained by the BLS and microwave spectroscopy keeps constant for all the measured tuning-laser powers. 

We will compare the relative decrease of  $f_0$ in BLS and microwave transport measurements. The shift of this FMR-related dip obtained by the BLS scanning (in the interval $P_\textrm{L}~[0~\textrm{mW}-425~\textrm{mW}]$ is 0.0344~GHz (compared to 0.02~GHz in the same $P_L$ interval at the microwave spectroscopy). Unlike the results obtained by the microwave spectroscopy, this dip frequency decrease does not have the threshold $P_{\textrm{th1}}$ value before which FMR frequency stays constant (see Fig.~\ref{fig:comparing}).

This FMR dip behavior may be related to our assumption about the non-uniform heating of the structure along the $x$ direction. Thus the scanning closer to the GaAs slab (heating source) edge should better detect this non-uniform heating of the YIG layer. Thus, the comparison of $f_0(P_\textrm{L})$ dependencies obtained by the microwave spectroscopy and by the BLS prove the presence of the thermal nonuniformity within the YIG layer along the SSW propagation direction.

Fig.~\ref{fig:comparing} demonstrates the good correspondence between the band gap frequency dependence on the tuning-laser power obtained from the microwave spectroscopy (dashed line marked by $f_{\textrm{bg}}$ and from the BLS measurements (green circles). The probed by the BLS technique $f_{\textrm{bg}}(P_\textrm{L} = 0~\textrm{mW})$ value is equal to one probed by the microwave spectroscopy. The maximum frequency of bandgap obtained by both methods (with the accuracy of tining-laser power $P_L$ discrete step) was achieved at at $P_\textrm{L}=P_{\textrm{th1}}=275$~mW. At the same time, the maximum frequency of bandgap extracted from BLS measurements (4.514~GHz) is slightly bigger than the one extracted from $S_{21}$ dependencies. 

The presented BLS spectra demonstrate the formation of the magnonic gap. With the tuning-laser power increase, $P_\textrm{L}$, the dip depth and width of the magnonic gap increase. We can not directly compare the BLS spectra signal intensity with the microwave spectroscopy one. However, a pronounced dip is observed above $P_{\textrm{st}}=150$~mW in both techniques. At the same time, for the high values of $P_\textrm{L}$ (curves "8" and "9" in Fig.~\ref{fig:BLS}), the magnonic gap dip becomes broadened and flattens out. We can relate this process to the heating-caused decrease in electron mobility \cite{GaAs_mu_T_depandence} leading to the weakening of SSW screening by the GaAs (see Fig.~\ref{fig:rho}~(\textbf{e})-(\textbf{f})).

\subsection{CONCLUSION}

As a result of the research conducted on the periodic GaAs/YIG magnonic crystal, we can conclude the following. We have confirmed the possibility of changing the magnonic crystal contrast by the optical means in a predictable manner. The degree of this contrast change is enough to form and switch off the rejection band in the spectra of SSW transport in the periodic structure. We unambiguously related the change of MC's contrast with the optically-induced electron density variation in the GaAs.

We claim that the critical point of the possibility of optical manipulation over the magnonic bandgap formation is the design of the YIG/GaAs bilayer. This design should allow the working wavelength range to effectively vary the SSW screening within the possible GaAs electron concentration in the experiment. In our research, we showed that in addition to the importance of the semiconductor dark electron concentration value and the possibility of significantly varying the electron concentration by the photo impact, the important role plays the effective thickness of the semiconductor layer (which should be thicker than the skin layer for the working range of wavelengths), and the semiconductor electron means collision frequency.

The obtained results will allow the design of new class of  the optically reconfigurable YIG/semiconductor magnonics elements. 
The spatial limitation for the sizes of the optically-induced lattice will be due to the semiconductor charges diffusion length and the semiconductor layer thickness enough to screen the SW wavelength of the working range.

\subsection{ACKNOWLEDGMENTS}
The research has received funding from the Slovak Grant Agency APVV (Nos APVV-19-0311(RSWFA), and APVV-21-0365), and from VEGA (Project No. 2/0068/21). Furthermore, this study  was performed during the implementation of the project Building-up Centre for advanced materials application of the Slovak Academy of Sciences, ITMS project code 313021T081 supported by Research \& Innovation Operational Programme funded by the ERDF (15\%). Also, this work was supported by the Ministry of Education and Science of the Russian Federation as part of the state assignment (Project No. FSRR-2023-0008).

\section{APPENDIX}

\subsection{Estimation of the semiconductor sample charge carriers' concentration}

\begin{figure}[b]
\centering
\includegraphics{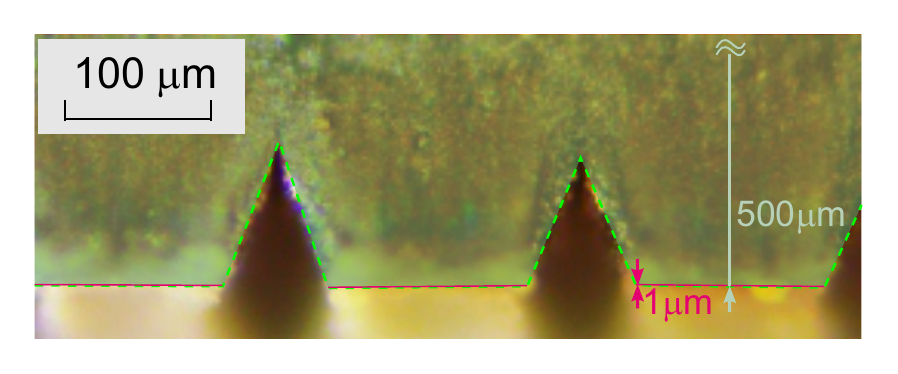}
\caption{Photo-microscope image of the side view of the GaAs slab. The Green dashed line schematically shows the edge of the slab and fabricated grooves on its surface. Pink fill schematically shows the highly n-doped GaAs layer (1~$\mu$m thickness). Green fill showed the semi-insulator n-doped GaAs substrate (500~$\mu$m thickness). 
}
\label{fig:GaAs_side}
\end{figure}

In this section, we are going to describe the sequence of steps we have used to obtain the parameters of the semiconductor sample. The slab of GaAs had fabricated periodic grooves on one of the side surfaces (see Fig.~\ref{fig:GaAs_side}) by the laser ablation method. This was done since this sample was prepared to serve as a periodic load for a multilayered waveguide structure. This shape of the sample defined the procedure that needed to be done to know the GaAs material parameters. 

According to the known facts about the GaAs slab we had (it was a commercial sample performed for the CMOS transistors fabrication), it consisted of the 500~$\mu$m thick substrate (expected electron concentration around $10^{10}$~cm$^{-3}$) and 1~$\mu$m thick strongly doped n-type layer (expected election concentration around $10^{17}$~cm$^{-3}$). The grooves were performed on the dopes side of the sample. However, these concentration values had to be examined in detail. Further, for the  purpose of the research on the YIG/GaAs multilayer,  the optical variation of GaAs electron concentration had to be checked.

For doped semiconductors, the charge transfer is mainly due to one type of charge carrier. Then, for n-type semiconductors, the dependence of the conductance $\sigma$ on the sample electron concentration $N_e$ may be written in a form \cite{sze2021physics}:

\begin{equation} \label{conduct)} 
\sigma = q N_e \mu_e, 
\end{equation}
where $q$ is the electron charge constant and $\mu_e$ - electron mobility. 

The standard approach to obtain the sample's electron concentration experimentally needs 4-Ohmic contacts. Then, the electrons concentration is calculated from Van der Pauw and Hall measurements that define the electron mobility and conductance \cite{pohorelec2020investigation}. Measured between the contacts, sheet resistance $R_S$ is related to the conductance $\sigma$ through the current flowed cross-sectional area $d$ between the Ohmic contacts:

\begin{equation} \label{resist)} 
\frac{1}{\sigma} = R_S d. 
\end{equation}

\begin{figure}
\centering
\includegraphics{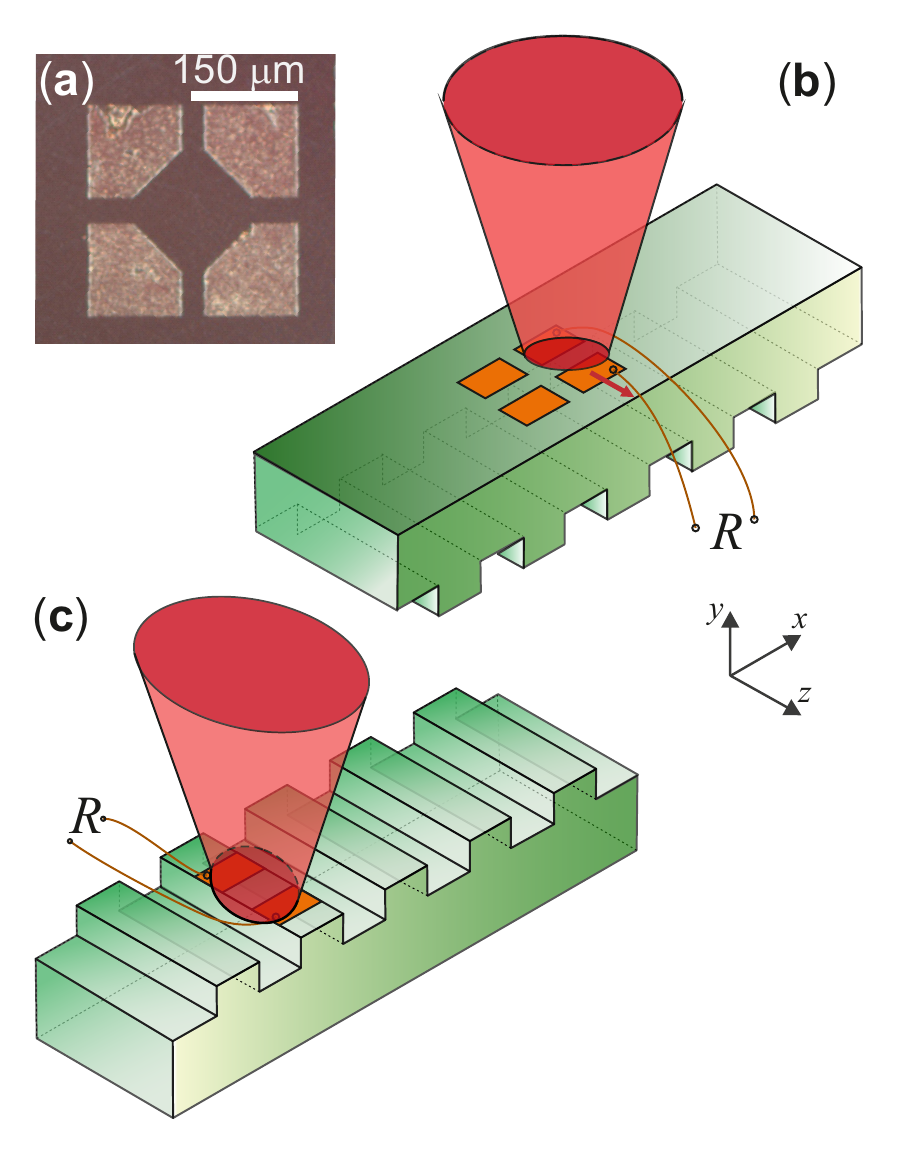}
\caption{(\textbf{a}) SEM image of 4-Ohmic contacts fabricated on a flat side of the GaAs slab. (\textbf{b}), (\textbf{c}) Panels show the experimental resistance measurements process schemes. The red arrow on panel (\textbf{b}) demonstrates the laser beam displacement performed for the diffusion characteristic obtaining (see attached directions of the Cartesian coordinates).}
\label{fig:Ohmic_contacts}
\end{figure}

However, fabricated periodic grooves on the doped face of the GaAs slab allowed to perform the deposition of only 2 Ohmic contacts on the stripe  between two grooves. On the other face of the slab (flat semi-insulating surface), 4 Ohmic contacts were deposited. Fabrication of the contacts consisted of a few steps. After the evaporation of multilayer (Ni(3~nm)/Al$_{0.88}$Ge$_{0.12}$(90~nm)/Ni(27~nm)) atop of GaAs slab, optical lithography was used to pattern contacts (two on a grooved side and four on the opposite side). Then the annealing was done (450°C, 1~min, N$_2$ ambient). Contacts were patterned on the distance 20~$\mu$m from each other (see Fig.~\ref{fig:Ohmic_contacts}~(\textbf{a})).

As a result, for a flat semi-insulating side of the GaAs sample, it was possible to make both resistance and electron mobility measurements. For the ribbed side of the highly doped side, it was possible to perform only resistance measurements. To obtain the electron concentration for the ribbed side, the results of resistance measurements were compiled with the known results of n-type GaAs room temperature electron mobility dependence vs. doping concentration  \cite{semicond_empirical}.

We obtained for the semi-insulating side of GaAs sample the electron mobility $\mu_e = 8400$~cm$^2/\textrm{V\,s}$ and the resulting dark concentration $N_e = 1\cdot 10^9$~cm$^{-3}$. The best matched pair of mobility (according the \cite{semicond_empirical}) and dark concentration for the highly-doped GaAs side is $\mu_e = 4000$~cm$^2/\textrm{V\,s}$ and $N_e = 1.3 \cdot 10^{16}$~cm$^{-3}$.

To obtain the laser power dependence of the GaAs electron concentration for both sides of the sample, we used the laser setup "First laser". Due to the technical issues, there was no possibility for simultaneous measurement of resistance and mobility under the optical irradiation. We measured the resistance relations vs. the optical power for both sides of the sample. At the same time, we had to ignore the mobility dependence on the photo-injected electrons. Moreover, we neglected the impact of the laser-induced variation of the holes concentration on the measured resistance (due to one-order lower holes mobility).  Thus, we univocally linked the laser-exposure impact on the resistance variation with an optical generation of pure electrons, and the eq.(\ref{conduct)}) keeps valid. 

For the highly-doped side, we were interested in the electron concentration variation vs. the laser irradiation's power density. For this purpose, the resistance between two Ohmic contacts was measured vs. the power density of the laser beam irradiating the contacts area. Consider the mobility of the highly doped side $\mu_e = 4000$~cm$^2/\textrm{V\,s}$ non-depend on the photo injected electrons, the dependence of electron concentration $N_e$ vs the  laser radiation power
density was obtained (see Fig.~\ref{fig:structure_Sigm}~(\textbf{b})). The maximum obtained electron concentration were $N_e=4.54 \cdot 10^{17}$~$1/$cm$^3$ (under the maximum available laser exposure power density $W_p=0.552$~W/cm$^2$). 

For the semi-insulating side, we were interested in the diffusion of the optically injected electrons. The reason for this is that the main conductance of the sample belongs to the highly-doped layer, which we used as a load for the SW screening. The thickness-dependence concentration of the diffused electrons from the doped layer to the semi-insulating layer can be approximated by the exponential function defined by the diffusion length, $L_n$, of the optically-induced electrons of the semi-insulating layer. To obtain it, the resistance measurements were done in dependence on the laser beam center displacement (beam shift $z_0$) out of the area of the Ohmic contacts. Consider the mobility of the semi-insulating side $\mu_e = 8400$~cm$^2/\textrm{V\,s}$ non-depend on the photo injected electrons, the diffusion dependence of electron concentration $N_e$ vs the distancing from the laser beam center $z_0$ was obtained (see the points in Fig.~\ref{fig:structure_Sigm}~(\textbf{c})). During this measurement, the power density of the laser beam was fixed on the maximum available value $W_p=0.552$~W/cm$^2$. The maximum obtained concentration (when the laser beam was irradiating the Ohmic contacts area) was $N_e= 3.49 \cdot 10^{10}$~$1/$cm$^3$. Since we know that usually, the spatial distribution of the concentration of diffusion carriers in semiconductors obeys an exponential law, we interpolated the measured data by function: 

\begin{equation} \label{diffus} 
N_e=1.7 \cdot 10^{10}e^{-0.003 y}, \; y(\textrm{nm}). 
\end{equation}
This function was plotted together with the measured data (see red line in Fig.~\ref{fig:structure_Sigm}~(\textbf{c})). Relation (\ref{diffus}) allowed us to choose the diffusion length: $L_n=315\,\mu$m, which is typical for the n- GaAs samples with similar doping \cite{GaAs_mu_T_depandence,semicond_empirical}.

In addition, we should point here that semiconductor electron mobility allows to express through the electron effective mass, $m^*$, the electron mean collision frequency, $\nu$:

\begin{equation} \label{e_collision}
    \mu_e = \frac{q}{m^* \nu}.
\end{equation}
The electron mean collision frequency is an important semiconductor electrodynamic parameter with a meaning of damping, which in terms of this work is used in the numerical simulations. Consider for the n- GaAs the table value of the electron effective mass $m^* = 0.13 \, m_\textrm{e}$ of the electron mass constant \cite{blakemore1982semiconducting,raymond1979electron}, $m_\textrm{e}$, we obtained $1/ \nu = 2.96 \cdot 10^{-13}$~s for the doped layer. We note here that since the electron mean collision frequency depends on the electron mobility, thus it strongly depends on the variation of the semiconductor doping electron concentration \cite{semicond_empirical} and on the semiconductor temperature \cite{GaAs_mu_T_depandence}.

\subsection{Numerical model and semiconductor describing approach}
\label{sec:fem} 

The numerical simulations of the YIG-GaAs periodic structure was based on the solution of the Helmholtz wave equation for the electric field vector \textbf{E} \cite{RFguide}: 
\begin{equation} \label{helmholtzeq}
 \nabla \times  (\mu^{-1} \nabla\times \textbf{E}) -
 \frac{(2 \pi f)^2}{c_0^2} 
 (\epsilon - \frac{i \sigma}{2 \pi f \epsilon_0})\textbf{E} = 0.
\end{equation}
This equation follows from the Maxwell system and set the conformity between the electric field vector \textbf{E} on the wave eigenmode frequency wave $f$ and coordinate-dependant material properties: $\mu$, the relative permeability, $\epsilon$, the relative permittivity, and $\sigma$, the electrical conductivity. $\epsilon_0$ is a vacuum permittivity, $c_0$ is a vacuum light speed constant. 

\textbf{YIG magnetic permeability tensor}

YIG properties were substituted to eq.(\ref{helmholtzeq}) as the constant-value permittivity, $\epsilon$, and gyrotropic frequency dependent permeability tensor, $\hat{\mu}$. $\hat{\mu}$ tensor can be obtained from the linearized damping-less Landau-Lifshitz equation without exchange interaction \cite{gurevichmelkov,SWbook}: 
\begin{equation} \label{mu_tensor} 
\hat{\mu} = \mu_0 \left[\begin{array}{ccc} 
{\mu}&{-i\mu_a}&{0}\\
{i\mu_a}&{\mu}&{0}\\
{0}&{0}&{1}
\end{array}\right],
\end{equation}
where diagonal and non-diagonal components are:

\begin{equation}
\label{mu_mua}
    \mu = 
    \frac{f_H(f_H+f_M)-f^2}
    {f_H^2-f^2}, \;
    \mu_a = 
    \frac{f_M \, f}
    {f_H^2-f^2}
\end{equation}
and  $2 \pi f_H = \gamma \mu_0 H_0$, $2 \pi f_M = \gamma \mu_0 M_\textrm{S}.$
The approach of describing the magnetic properties of solids through the permeability tensor holds for the samples with quasi-uniform magnetization (e.g., magnetized thin films and layers) without possible reverse of magnetization \cite{gurevichmelkov,SWbook,mruczkiewicz2014nonreci_Me,mruczkiewicz2013nonreciprocity}.

Thus, in our model YIG material properties were fully determined through the input parameters: gyromagnetic ratio, $\gamma$, saturation magnetization, $M_\textrm{S}$, and external magnetic field, $H_0$. This approach was widely used for numerical simulations of the similar waveguide structures, e.g., in works \cite{mruczkiewicz2014nonreci_Me,mruczkiewicz2013nonreciprocity, sadovnikov2015brillouin,sadovnikov2015nonreciprocal}.

\textbf{GaAs permittivity tensor}

GaAs properties were substituted to eq.(\ref{helmholtzeq}) as the permeability $\mu=1$, and frequency-dependant permittivity tensor, $\hat{\epsilon}$, which describes semiconductor gyroelectric properties. Such approach to describe the interaction of wave-induced electromagnetic field with the charges of semiconductor layer through the permittivity tensor was raised in \cite{gurevichmelkov, gurevichRussian, bass1997kinetic}.

The charges motion within the semiconductor may be described by the Boltzmann kinetic equation written in the hydrodynamic approximation \cite{gurevichmelkov}: 

\begin{equation} \label{chargedynamichydro}
    \frac{\partial \textbf{V}}{\partial t} + (\textbf{V} \cdot \nabla)\textbf{V} = \frac{q}{m_{\textrm{eff}}}(\textbf{E}+\textbf{V}\times\textbf{B})-\nu \textbf{V}, 
\end{equation}
where $\textbf{B}$ - magnetic field inside semiconductor, $\textbf{E}$ - electric field inside the semiconductor, $\textbf{V}$ - mean charge carriers drift speed, $q$ - elementary charge of charge carrier, $m_{\textrm{eff}}$ - charge effective mass, $\nu$ - charges mean collision frequency, followed from the charges mobility.

It is possible to get velocities of charge carriers as the solution of linearized eq.(\ref{chargedynamichydro}), and current density definition allows to transform these velocities to the conductivity tensor, $\hat{\sigma}$. We assumed only the electron conductance, and $N_e$ - electron concentration, $m_e$ - electron effective mass, $q$ - electron charge constant, $\nu$ follows from the electron mobility defined by eq.(\ref{e_collision}). Introducing the cyclotron and plasma frequencies: 

\[\omega_\textrm{c}=\frac{q}{m_e}\mu_0 H_0, \; \omega_p = \sqrt{\frac{N_e q^2}{m_e \epsilon_0}},\] it is possible to formulate the frequency-dependant permittivity tensor, $\hat{\epsilon}$. In case of no macroscopic drift of the electrons (external electric field $\textbf{E}$ is absent), only small oscillations of $\textbf{V}$ and $\textbf{E}$ are assumed, and for the Cartesian coordinates and external magnetic field orientations defined in Fig.~\ref{fig:structure_Sigm}~(\textbf{a}), 

\begin{equation} \label{epstensorV} 
\hat{\varepsilon} _{r} =\left|\begin{array}{ccc} 
{\epsilon_g-a_{11} } & {a_{12}} & {0} 
\\{a_{21}} & {\epsilon_g-a_{22}} & {0} 
\\{0} & {a_{32}} & {\epsilon_g+a_{33} } 
\end{array}\right|, 
\end{equation} 

\[a_{33} =\frac{i\omega _{p} ^{2} }{{\rm \omega }\left(-i{\rm \omega }-{\rm }\omega\right)} ,\]
\[a_{11} =a_{22} \left(f\right)=\left(-i \omega-\nu\right)\epsilon _{part} ,\]
\[a_{12} =-a_{21} \left(f\right)=\omega _{c} \epsilon _{part} ,\]
\[\epsilon _{part} =\frac{i\omega_{p} ^{2} }{{\rm \omega }\left({\rm \omega }^{2} +2\left(-i{\rm \nu }\right){\rm \omega }-{\rm }\omega^{2} -\omega_{c} ^{2} \right)} .\] 
Here $\epsilon _{g}$ is a crystal lattice contribution to the semiconductor's permittivity.

Thus, in our model GaAs material properties were fully determined through the input parameters: electron concentration, $N_e$, electron effective mass, $m_e$, electron collision frequency, $\nu$, crystal lattice contribution to the semiconductor's permittivity, $\epsilon _{g}$, external magnetic field $H_0$.  This approach in different variations was used to describe the semiconductor influence on the SW transport, e.g. in works \cite{Kindyak,kindyak1995magnetostatic_screening,kindyak1999nonlinear,kindyak2002surface, eliseeva2005electromagnetic, eliseeva2008dispersion, eliseeva2010anisotropy}.

\begin{figure}
\includegraphics{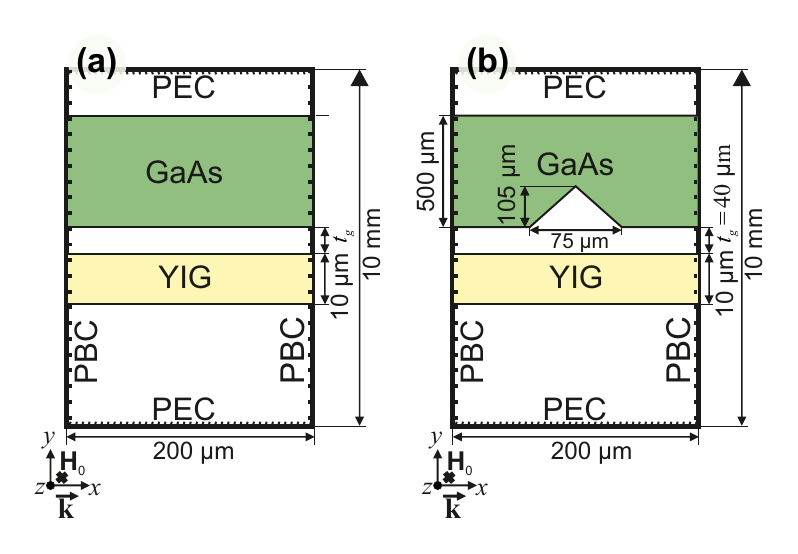}
\caption{The computation cells used in numerical simulations of YIG/GaAs multilayers by the finite element method. The boundary conditions forming the cells are periodic boundary condition (PBC) and perfect electric conditions (PEC). Directions of Cartesian coordinates, external magnetic field and wavevectors correspond to one depicted in Fig.~\ref{fig:structure_Sigm}~(\textbf{a}).  The computation cells was used for simulations of the YIG/GaAs multilayers: (\textbf{a}) infinite and uniform along $x$ and $y$ directions, (\textbf{b}) infinite and uniform along $y$ direction and periodic and infinite along $x$ direction.}
\label{fig:comp_cell}
\end{figure}

\textbf{Computation cell}

Considering the wave processes in waveguide structures (e.g., investigated structure, Fig.~\ref{fig:structure_Sigm}~\textbf{(a))}), equation (\ref{helmholtzeq}) has coefficients that are constants or periodic functions along the wave propagation direction $y$, (period $D$), thus the Bloch theorem can be used:
\begin{equation} \label{Bloch}
\textbf{E}(y,z) = \textbf{E} ^{'}(y,z)e^{ik_x x}.    
\end{equation}
$\textbf{E}^{'}(y,z)= \textbf{E}^{'}(y+a,y+a)$ - periodic function with period $a$ and $k_x \equiv k$ as we considering only waves propagating along $x$ direction. To obtain eigenvalues of \textbf{k} for considered system of equations (\ref{Bloch}) and (\ref{helmholtzeq}) supplemented with the YIG and GaAs material properties parameters $\mu$ and $\epsilon$, they were supplemented with boundary conditions. It allowed to form the computation cell (see Fig.~\ref{fig:comp_cell}). We used periodic boundary conditions (PBC) along the wave propagation direction ($x-$ axis) and perfect electric conductor (PEC) along the structure normal direction ($y-$ axis). In the approximation of structure to be infinite along the transverse direction ($z-$ axis) we could reduce the numerical task problem to the 2D. The partial differential equation for the eigenvalue task was formulated from the (\ref{helmholtzeq}) as $f(k)$. This defined task was solved by use of the
Comsol MULTIPHYSICS commercial software by the finite element method \cite{RFguide}.

Periodic boundary conditions allowed us to simulate multilayer infinite and uniform along the SW propagation direction (Fig.~\ref{fig:comp_cell}~(\textbf{a})) or multilayer infinitely periodic along the SW propagation direction (Fig.~\ref{fig:comp_cell}~(\textbf{b}), one period of structure is considered, GaAs shape corresponds to the experimental sample image from Fig.~\ref{fig:structure_Sigm}~(\textbf{a})).

After input of the materials parameters and sizes, for defined $k$ value the certain amount of eigensolutions where obtained.  The result of each solution is the simulated electromagnetic field spatial distribution within the computation region with the corresponding eigenfrequency.  This defined amount of eigensolutions where calculated around defined initial frequency value (which was chosen according the surface magnetostatic wave approximation \citep{damon1961}). This gives an effective tool for studying the magnonics structures, e.g., it is possible to investigate the spatial distribution of the electromagnetic fields within the studied structures, a reflection of electromagnetic wave (EM) on its boundaries, standing waves, and waves propagating processes. As a result, dispersion relations and information about wave attenuation can be obtained.

We mention, that computational cell depicted in Fig.~\ref{fig:comp_cell}~(\textbf{a}) was used for the "Semi-infinite GaAs/YIG bilayer" model, "Default Parameters" variation, and "Experimental YIG/GaAs multilayer" model. The computational cell depicted in Fig.~\ref{fig:comp_cell}~(\textbf{b}) was used for the "Experimental YIG/GaAs multilayer" model with periodic GaAs groves. The similar introduction of the boundary conditions for the system of eq.(\ref{helmholtzeq}) and eq.(\ref{Bloch}) was used to form computation cell for the waveguides with gyrotropic materials in works, e.g., \cite{sadovnikov2015brillouin, sadovnikov2015nonreciprocal, mruczkiewicz2014nonreci_Me, MC_nonreciprocity}.

\bibliography{1aipsamp}

\providecommand{\noopsort}[1]{}\providecommand{\singleletter}[1]{#1}%
\begin{thebibliography}{81}%
\makeatletter
\providecommand \@ifxundefined [1]{%
 \@ifx{#1\undefined}
}%
\providecommand \@ifnum [1]{%
 \ifnum #1\expandafter \@firstoftwo
 \else \expandafter \@secondoftwo
 \fi
}%
\providecommand \@ifx [1]{%
 \ifx #1\expandafter \@firstoftwo
 \else \expandafter \@secondoftwo
 \fi
}%
\providecommand \natexlab [1]{#1}%
\providecommand \enquote  [1]{``#1''}%
\providecommand \bibnamefont  [1]{#1}%
\providecommand \bibfnamefont [1]{#1}%
\providecommand \citenamefont [1]{#1}%
\providecommand \href@noop [0]{\@secondoftwo}%
\providecommand \href [0]{\begingroup \@sanitize@url \@href}%
\providecommand \@href[1]{\@@startlink{#1}\@@href}%
\providecommand \@@href[1]{\endgroup#1\@@endlink}%
\providecommand \@sanitize@url [0]{\catcode `\\12\catcode `\$12\catcode
  `\&12\catcode `\#12\catcode `\^12\catcode `\_12\catcode `\%12\relax}%
\providecommand \@@startlink[1]{}%
\providecommand \@@endlink[0]{}%
\providecommand \url  [0]{\begingroup\@sanitize@url \@url }%
\providecommand \@url [1]{\endgroup\@href {#1}{\urlprefix }}%
\providecommand \urlprefix  [0]{URL }%
\providecommand \Eprint [0]{\href }%
\providecommand \doibase [0]{http://dx.doi.org/}%
\providecommand \selectlanguage [0]{\@gobble}%
\providecommand \bibinfo  [0]{\@secondoftwo}%
\providecommand \bibfield  [0]{\@secondoftwo}%
\providecommand \translation [1]{[#1]}%
\providecommand \BibitemOpen [0]{}%
\providecommand \bibitemStop [0]{}%
\providecommand \bibitemNoStop [0]{.\EOS\space}%
\providecommand \EOS [0]{\spacefactor3000\relax}%
\providecommand \BibitemShut  [1]{\csname bibitem#1\endcsname}%
\let\auto@bib@innerbib\@empty
\bibitem [{\citenamefont {Barman}\ \emph
  {et~al.}(2021{\natexlab{a}})\citenamefont {Barman}, \citenamefont
  {Gubbiotti}, \citenamefont {Ladak}, \citenamefont {Adeyeye}, \citenamefont
  {Krawczyk}, \citenamefont {Gr{\"a}fe}, \citenamefont {Adelmann},
  \citenamefont {Cotofana}, \citenamefont {Naeemi}, \citenamefont {Vasyuchka}
  \emph {et~al.}}]{magn_roadmap}%
  \BibitemOpen
  \bibfield  {author} {\bibinfo {author} {\bibfnamefont {A.}~\bibnamefont
  {Barman}}, \bibinfo {author} {\bibfnamefont {G.}~\bibnamefont {Gubbiotti}},
  \bibinfo {author} {\bibfnamefont {S.}~\bibnamefont {Ladak}}, \bibinfo
  {author} {\bibfnamefont {A.~O.}\ \bibnamefont {Adeyeye}}, \bibinfo {author}
  {\bibfnamefont {M.}~\bibnamefont {Krawczyk}}, \bibinfo {author}
  {\bibfnamefont {J.}~\bibnamefont {Gr{\"a}fe}}, \bibinfo {author}
  {\bibfnamefont {C.}~\bibnamefont {Adelmann}}, \bibinfo {author}
  {\bibfnamefont {S.}~\bibnamefont {Cotofana}}, \bibinfo {author}
  {\bibfnamefont {A.}~\bibnamefont {Naeemi}}, \bibinfo {author} {\bibfnamefont
  {V.~I.}\ \bibnamefont {Vasyuchka}},  \emph {et~al.},\ }\href@noop {}
  {\bibfield  {journal} {\bibinfo  {journal} {Journal of Physics: Condensed
  Matter}\ } (\bibinfo {year} {2021}{\natexlab{a}})}\BibitemShut {NoStop}%
\bibitem [{\citenamefont {Nikitov}\ \emph {et~al.}(2020)\citenamefont
  {Nikitov}, \citenamefont {Safin}, \citenamefont {Kalyabin}, \citenamefont
  {Sadovnikov}, \citenamefont {Beginin}, \citenamefont {Logunov}, \citenamefont
  {Morozova}, \citenamefont {Odintsov}, \citenamefont {Osokin}, \citenamefont
  {Sharaevskaya}, \citenamefont {Sharaevsky},\ and\ \citenamefont
  {Kirilyuk}}]{dielmagn}%
  \BibitemOpen
  \bibfield  {author} {\bibinfo {author} {\bibfnamefont {S.~A.}\ \bibnamefont
  {Nikitov}}, \bibinfo {author} {\bibfnamefont {A.~R.}\ \bibnamefont {Safin}},
  \bibinfo {author} {\bibfnamefont {D.~V.}\ \bibnamefont {Kalyabin}}, \bibinfo
  {author} {\bibfnamefont {A.~V.}\ \bibnamefont {Sadovnikov}}, \bibinfo
  {author} {\bibfnamefont {E.~N.}\ \bibnamefont {Beginin}}, \bibinfo {author}
  {\bibfnamefont {M.~V.}\ \bibnamefont {Logunov}}, \bibinfo {author}
  {\bibfnamefont {M.~A.}\ \bibnamefont {Morozova}}, \bibinfo {author}
  {\bibfnamefont {S.~A.}\ \bibnamefont {Odintsov}}, \bibinfo {author}
  {\bibfnamefont {S.~A.}\ \bibnamefont {Osokin}}, \bibinfo {author}
  {\bibfnamefont {A.~Y.}\ \bibnamefont {Sharaevskaya}}, \bibinfo {author}
  {\bibfnamefont {Y.~P.}\ \bibnamefont {Sharaevsky}}, \ and\ \bibinfo {author}
  {\bibfnamefont {A.~I.}\ \bibnamefont {Kirilyuk}},\ }\href {\doibase
  10.3367/ufne.2019.07.038609} {\bibfield  {journal} {\bibinfo  {journal}
  {Physics-Uspekhi}\ }\textbf {\bibinfo {volume} {63}},\ \bibinfo {pages} {945}
  (\bibinfo {year} {2020})}\BibitemShut {NoStop}%
\bibitem [{\citenamefont {Nikitov}\ \emph {et~al.}(2015)\citenamefont
  {Nikitov}, \citenamefont {Kalyabin}, \citenamefont {Lisenkov}, \citenamefont
  {Slavin}, \citenamefont {Barabanenkov}, \citenamefont {Osokin}, \citenamefont
  {Sadovnikov}, \citenamefont {Beginin}, \citenamefont {Morozova},
  \citenamefont {Filimonov}, \citenamefont {Khivintsev}, \citenamefont
  {Vysotsky}, \citenamefont {Sakharov},\ and\ \citenamefont
  {Pavlov}}]{Nikitov_mag}%
  \BibitemOpen
  \bibfield  {author} {\bibinfo {author} {\bibfnamefont {S.~A.}\ \bibnamefont
  {Nikitov}}, \bibinfo {author} {\bibfnamefont {D.~V.}\ \bibnamefont
  {Kalyabin}}, \bibinfo {author} {\bibfnamefont {I.~V.}\ \bibnamefont
  {Lisenkov}}, \bibinfo {author} {\bibfnamefont {A.}~\bibnamefont {Slavin}},
  \bibinfo {author} {\bibfnamefont {Y.~N.}\ \bibnamefont {Barabanenkov}},
  \bibinfo {author} {\bibfnamefont {S.~A.}\ \bibnamefont {Osokin}}, \bibinfo
  {author} {\bibfnamefont {A.~V.}\ \bibnamefont {Sadovnikov}}, \bibinfo
  {author} {\bibfnamefont {E.~N.}\ \bibnamefont {Beginin}}, \bibinfo {author}
  {\bibfnamefont {M.~A.}\ \bibnamefont {Morozova}}, \bibinfo {author}
  {\bibfnamefont {Y.~A.}\ \bibnamefont {Filimonov}}, \bibinfo {author}
  {\bibfnamefont {Y.~V.}\ \bibnamefont {Khivintsev}}, \bibinfo {author}
  {\bibfnamefont {S.~L.}\ \bibnamefont {Vysotsky}}, \bibinfo {author}
  {\bibfnamefont {V.~K.}\ \bibnamefont {Sakharov}}, \ and\ \bibinfo {author}
  {\bibfnamefont {E.~S.}\ \bibnamefont {Pavlov}},\ }\href {\doibase
  10.3367/ufne.0185.201510m.1099} {\bibfield  {journal} {\bibinfo  {journal}
  {Physics-Uspekhi}\ }\textbf {\bibinfo {volume} {58}},\ \bibinfo {pages}
  {1002} (\bibinfo {year} {2015})}\BibitemShut {NoStop}%
\bibitem [{\citenamefont {Demokritov}\ and\ \citenamefont
  {Slavin}(2012)}]{demokritov2012mag}%
  \BibitemOpen
  \bibfield  {author} {\bibinfo {author} {\bibfnamefont {S.}~\bibnamefont
  {Demokritov}}\ and\ \bibinfo {author} {\bibfnamefont {A.}~\bibnamefont
  {Slavin}},\ }\href {https://books.google.sk/books?id=CK-3S9Xa-z4C} {\emph
  {\bibinfo {title} {Magnonics: From Fundamentals to Applications}}},\ Topics
  in Applied Physics\ (\bibinfo  {publisher} {Springer Berlin Heidelberg},\
  \bibinfo {year} {2012})\BibitemShut {NoStop}%
\bibitem [{\citenamefont {Lenk}\ \emph {et~al.}(2011)\citenamefont {Lenk},
  \citenamefont {Ulrichs}, \citenamefont {Garbs},\ and\ \citenamefont
  {Münzenberg}}]{mag_blocks}%
  \BibitemOpen
  \bibfield  {author} {\bibinfo {author} {\bibfnamefont {B.}~\bibnamefont
  {Lenk}}, \bibinfo {author} {\bibfnamefont {H.}~\bibnamefont {Ulrichs}},
  \bibinfo {author} {\bibfnamefont {F.}~\bibnamefont {Garbs}}, \ and\ \bibinfo
  {author} {\bibfnamefont {M.}~\bibnamefont {Münzenberg}},\ }\href {\doibase
  https://doi.org/10.1016/j.physrep.2011.06.003} {\bibfield  {journal}
  {\bibinfo  {journal} {Physics Reports}\ }\textbf {\bibinfo {volume} {507}},\
  \bibinfo {pages} {107 } (\bibinfo {year} {2011})}\BibitemShut {NoStop}%
\bibitem [{\citenamefont {Kruglyak}\ \emph {et~al.}(2010)\citenamefont
  {Kruglyak}, \citenamefont {Demokritov},\ and\ \citenamefont
  {Grundler}}]{kruglyak2010mag}%
  \BibitemOpen
  \bibfield  {author} {\bibinfo {author} {\bibfnamefont {V.}~\bibnamefont
  {Kruglyak}}, \bibinfo {author} {\bibfnamefont {S.}~\bibnamefont
  {Demokritov}}, \ and\ \bibinfo {author} {\bibfnamefont {D.}~\bibnamefont
  {Grundler}},\ }\href@noop {} {\bibfield  {journal} {\bibinfo  {journal}
  {Journal of Physics D: Applied Physics}\ }\textbf {\bibinfo {volume} {43}},\
  \bibinfo {pages} {264001} (\bibinfo {year} {2010})}\BibitemShut {NoStop}%
\bibitem [{\citenamefont {Gubbiotti}(2019)}]{gubbiotti_3Dmagnonics}%
  \BibitemOpen
  \bibfield  {author} {\bibinfo {author} {\bibfnamefont {G.}~\bibnamefont
  {Gubbiotti}},\ }\href {https://books.google.sk/books?id=FfxJxgEACAAJ} {\emph
  {\bibinfo {title} {Three-Dimensional Magnonics}}}\ (\bibinfo  {publisher}
  {Jenny Stanford Publishing},\ \bibinfo {year} {2019})\BibitemShut {NoStop}%
\bibitem [{\citenamefont {Demokritov}(2018)}]{Demokritov_magn}%
  \BibitemOpen
  \bibfield  {author} {\bibinfo {author} {\bibfnamefont {S.~O.}\ \bibnamefont
  {Demokritov}},\ }\enquote {\bibinfo {title} {Magnons},}\ in\ \href {\doibase
  10.1007/978-3-319-97334-0_10} {\emph {\bibinfo {booktitle} {Topology in
  Magnetism}}},\ \bibinfo {editor} {edited by\ \bibinfo {editor} {\bibfnamefont
  {J.}~\bibnamefont {Zang}}, \bibinfo {editor} {\bibfnamefont {V.}~\bibnamefont
  {Cros}}, \ and\ \bibinfo {editor} {\bibfnamefont {A.}~\bibnamefont
  {Hoffmann}}}\ (\bibinfo  {publisher} {Springer International Publishing},\
  \bibinfo {address} {Cham},\ \bibinfo {year} {2018})\ pp.\ \bibinfo {pages}
  {299--334}\BibitemShut {NoStop}%
\bibitem [{\citenamefont {Prabhakar}\ and\ \citenamefont
  {Stancil}(2009)}]{SWbook}%
  \BibitemOpen
  \bibfield  {author} {\bibinfo {author} {\bibfnamefont {A.}~\bibnamefont
  {Prabhakar}}\ and\ \bibinfo {author} {\bibfnamefont {D.~D.}\ \bibnamefont
  {Stancil}},\ }\href@noop {} {\emph {\bibinfo {title} {{Spin waves: Theory and
  applications}}}},\ Vol.~\bibinfo {volume} {5}\ (\bibinfo  {publisher}
  {Springer},\ \bibinfo {year} {2009})\BibitemShut {NoStop}%
\bibitem [{\citenamefont {Association}\ \emph {et~al.}(2015)\citenamefont
  {Association} \emph {et~al.}}]{asi2015itrs}%
  \BibitemOpen
  \bibfield  {author} {\bibinfo {author} {\bibfnamefont {A.}~\bibnamefont
  {Association}} \emph {et~al.},\ }\href@noop {} {\  (\bibinfo {year}
  {2015})}\BibitemShut {NoStop}%
\bibitem [{\citenamefont {Davies}\ \emph {et~al.}(2015)\citenamefont {Davies},
  \citenamefont {Francis}, \citenamefont {Sadovnikov}, \citenamefont
  {Chertopalov}, \citenamefont {Bryan}, \citenamefont {Grishin}, \citenamefont
  {Allwood}, \citenamefont {Sharaevskii}, \citenamefont {Nikitov},\ and\
  \citenamefont {Kruglyak}}]{Davies_networks}%
  \BibitemOpen
  \bibfield  {author} {\bibinfo {author} {\bibfnamefont {C.~S.}\ \bibnamefont
  {Davies}}, \bibinfo {author} {\bibfnamefont {A.}~\bibnamefont {Francis}},
  \bibinfo {author} {\bibfnamefont {A.~V.}\ \bibnamefont {Sadovnikov}},
  \bibinfo {author} {\bibfnamefont {S.~V.}\ \bibnamefont {Chertopalov}},
  \bibinfo {author} {\bibfnamefont {M.~T.}\ \bibnamefont {Bryan}}, \bibinfo
  {author} {\bibfnamefont {S.~V.}\ \bibnamefont {Grishin}}, \bibinfo {author}
  {\bibfnamefont {D.~A.}\ \bibnamefont {Allwood}}, \bibinfo {author}
  {\bibfnamefont {Y.~P.}\ \bibnamefont {Sharaevskii}}, \bibinfo {author}
  {\bibfnamefont {S.~A.}\ \bibnamefont {Nikitov}}, \ and\ \bibinfo {author}
  {\bibfnamefont {V.~V.}\ \bibnamefont {Kruglyak}},\ }\href {\doibase
  10.1103/PhysRevB.92.020408} {\bibfield  {journal} {\bibinfo  {journal} {Phys.
  Rev. B}\ }\textbf {\bibinfo {volume} {92}},\ \bibinfo {pages} {020408}
  (\bibinfo {year} {2015})}\BibitemShut {NoStop}%
\bibitem [{\citenamefont {Beginin}\ \emph {et~al.}(2018)\citenamefont
  {Beginin}, \citenamefont {Sadovnikov}, \citenamefont {Sharaevskaya},
  \citenamefont {Stognij},\ and\ \citenamefont {Nikitov}}]{beginin_networks}%
  \BibitemOpen
  \bibfield  {author} {\bibinfo {author} {\bibfnamefont {E.}~\bibnamefont
  {Beginin}}, \bibinfo {author} {\bibfnamefont {A.}~\bibnamefont {Sadovnikov}},
  \bibinfo {author} {\bibfnamefont {A.~Y.}\ \bibnamefont {Sharaevskaya}},
  \bibinfo {author} {\bibfnamefont {A.}~\bibnamefont {Stognij}}, \ and\
  \bibinfo {author} {\bibfnamefont {S.}~\bibnamefont {Nikitov}},\ }\href@noop
  {} {\bibfield  {journal} {\bibinfo  {journal} {Applied Physics Letters}\
  }\textbf {\bibinfo {volume} {112}},\ \bibinfo {pages} {122404} (\bibinfo
  {year} {2018})}\BibitemShut {NoStop}%
\bibitem [{\citenamefont {Sadovnikov}\ \emph {et~al.}(2016)\citenamefont
  {Sadovnikov}, \citenamefont {Beginin}, \citenamefont {Odincov}, \citenamefont
  {Sheshukova}, \citenamefont {Sharaevskii}, \citenamefont {Stognij},\ and\
  \citenamefont {Nikitov}}]{sadovnikov_networks}%
  \BibitemOpen
  \bibfield  {author} {\bibinfo {author} {\bibfnamefont {A.}~\bibnamefont
  {Sadovnikov}}, \bibinfo {author} {\bibfnamefont {E.}~\bibnamefont {Beginin}},
  \bibinfo {author} {\bibfnamefont {S.}~\bibnamefont {Odincov}}, \bibinfo
  {author} {\bibfnamefont {S.}~\bibnamefont {Sheshukova}}, \bibinfo {author}
  {\bibfnamefont {Y.~P.}\ \bibnamefont {Sharaevskii}}, \bibinfo {author}
  {\bibfnamefont {A.}~\bibnamefont {Stognij}}, \ and\ \bibinfo {author}
  {\bibfnamefont {S.}~\bibnamefont {Nikitov}},\ }\href@noop {} {\bibfield
  {journal} {\bibinfo  {journal} {Applied Physics Letters}\ }\textbf {\bibinfo
  {volume} {108}},\ \bibinfo {pages} {172411} (\bibinfo {year}
  {2016})}\BibitemShut {NoStop}%
\bibitem [{\citenamefont {Sadovnikov}\ \emph
  {et~al.}(2019{\natexlab{a}})\citenamefont {Sadovnikov}, \citenamefont
  {Beginin}, \citenamefont {Sheshukova}, \citenamefont {Sharaevskii},
  \citenamefont {Stognij}, \citenamefont {Novitski}, \citenamefont {Sakharov},
  \citenamefont {Khivintsev},\ and\ \citenamefont {Nikitov}}]{semicond_magn}%
  \BibitemOpen
  \bibfield  {author} {\bibinfo {author} {\bibfnamefont {A.~V.}\ \bibnamefont
  {Sadovnikov}}, \bibinfo {author} {\bibfnamefont {E.~N.}\ \bibnamefont
  {Beginin}}, \bibinfo {author} {\bibfnamefont {S.~E.}\ \bibnamefont
  {Sheshukova}}, \bibinfo {author} {\bibfnamefont {Y.~P.}\ \bibnamefont
  {Sharaevskii}}, \bibinfo {author} {\bibfnamefont {A.~I.}\ \bibnamefont
  {Stognij}}, \bibinfo {author} {\bibfnamefont {N.~N.}\ \bibnamefont
  {Novitski}}, \bibinfo {author} {\bibfnamefont {V.~K.}\ \bibnamefont
  {Sakharov}}, \bibinfo {author} {\bibfnamefont {Y.~V.}\ \bibnamefont
  {Khivintsev}}, \ and\ \bibinfo {author} {\bibfnamefont {S.~A.}\ \bibnamefont
  {Nikitov}},\ }\href {\doibase 10.1103/PhysRevB.99.054424} {\bibfield
  {journal} {\bibinfo  {journal} {Phys. Rev. B}\ }\textbf {\bibinfo {volume}
  {99}},\ \bibinfo {pages} {054424} (\bibinfo {year}
  {2019}{\natexlab{a}})}\BibitemShut {NoStop}%
\bibitem [{\citenamefont {Stognij}\ \emph
  {et~al.}(2015{\natexlab{a}})\citenamefont {Stognij}, \citenamefont {Lutsev},
  \citenamefont {Bursian},\ and\ \citenamefont {Novitskii}}]{stognij_Si}%
  \BibitemOpen
  \bibfield  {author} {\bibinfo {author} {\bibfnamefont {A.}~\bibnamefont
  {Stognij}}, \bibinfo {author} {\bibfnamefont {L.}~\bibnamefont {Lutsev}},
  \bibinfo {author} {\bibfnamefont {V.}~\bibnamefont {Bursian}}, \ and\
  \bibinfo {author} {\bibfnamefont {N.}~\bibnamefont {Novitskii}},\ }\href@noop
  {} {\bibfield  {journal} {\bibinfo  {journal} {Journal of Applied Physics}\
  }\textbf {\bibinfo {volume} {118}},\ \bibinfo {pages} {023905} (\bibinfo
  {year} {2015}{\natexlab{a}})}\BibitemShut {NoStop}%
\bibitem [{\citenamefont {Stognij}\ \emph
  {et~al.}(2015{\natexlab{b}})\citenamefont {Stognij}, \citenamefont {Lutsev},
  \citenamefont {Novitskii}, \citenamefont {Bespalov}, \citenamefont
  {Golikova}, \citenamefont {Ketsko}, \citenamefont {Gieniusz},\ and\
  \citenamefont {Maziewski}}]{Stognij_GaN}%
  \BibitemOpen
  \bibfield  {author} {\bibinfo {author} {\bibfnamefont {A.}~\bibnamefont
  {Stognij}}, \bibinfo {author} {\bibfnamefont {L.}~\bibnamefont {Lutsev}},
  \bibinfo {author} {\bibfnamefont {N.}~\bibnamefont {Novitskii}}, \bibinfo
  {author} {\bibfnamefont {A.}~\bibnamefont {Bespalov}}, \bibinfo {author}
  {\bibfnamefont {O.}~\bibnamefont {Golikova}}, \bibinfo {author}
  {\bibfnamefont {V.}~\bibnamefont {Ketsko}}, \bibinfo {author} {\bibfnamefont
  {R.}~\bibnamefont {Gieniusz}}, \ and\ \bibinfo {author} {\bibfnamefont
  {A.}~\bibnamefont {Maziewski}},\ }\href {\doibase
  10.1088/0022-3727/48/48/485002} {\bibfield  {journal} {\bibinfo  {journal}
  {Journal of Physics D: Applied Physics}\ }\textbf {\bibinfo {volume} {48}},\
  \bibinfo {pages} {485002} (\bibinfo {year} {2015}{\natexlab{b}})}\BibitemShut
  {NoStop}%
\bibitem [{\citenamefont {Lutsev}\ \emph {et~al.}(2018)\citenamefont {Lutsev},
  \citenamefont {Stognij}, \citenamefont {Novitskii}, \citenamefont {Bursian},
  \citenamefont {Maziewski},\ and\ \citenamefont {Gieniusz}}]{Lutsev_2DEgas}%
  \BibitemOpen
  \bibfield  {author} {\bibinfo {author} {\bibfnamefont {L.~V.}\ \bibnamefont
  {Lutsev}}, \bibinfo {author} {\bibfnamefont {A.~I.}\ \bibnamefont {Stognij}},
  \bibinfo {author} {\bibfnamefont {N.~N.}\ \bibnamefont {Novitskii}}, \bibinfo
  {author} {\bibfnamefont {V.~E.}\ \bibnamefont {Bursian}}, \bibinfo {author}
  {\bibfnamefont {A.}~\bibnamefont {Maziewski}}, \ and\ \bibinfo {author}
  {\bibfnamefont {R.}~\bibnamefont {Gieniusz}},\ }\href {\doibase
  10.1088/1361-6463/aad41b} {\bibfield  {journal} {\bibinfo  {journal} {Journal
  of Physics D: Applied Physics}\ }\textbf {\bibinfo {volume} {51}},\ \bibinfo
  {pages} {355002} (\bibinfo {year} {2018})}\BibitemShut {NoStop}%
\bibitem [{\citenamefont {Sadovnikov}\ \emph
  {et~al.}(2019{\natexlab{b}})\citenamefont {Sadovnikov}, \citenamefont
  {Beginin}, \citenamefont {Sheshukova}, \citenamefont {Sharaevskii},
  \citenamefont {Stognij}, \citenamefont {Novitski}, \citenamefont {Sakharov},
  \citenamefont {Khivintsev},\ and\ \citenamefont {Nikitov}}]{Sadovnikov_GaAs}%
  \BibitemOpen
  \bibfield  {author} {\bibinfo {author} {\bibfnamefont {A.~V.}\ \bibnamefont
  {Sadovnikov}}, \bibinfo {author} {\bibfnamefont {E.~N.}\ \bibnamefont
  {Beginin}}, \bibinfo {author} {\bibfnamefont {S.~E.}\ \bibnamefont
  {Sheshukova}}, \bibinfo {author} {\bibfnamefont {Y.~P.}\ \bibnamefont
  {Sharaevskii}}, \bibinfo {author} {\bibfnamefont {A.~I.}\ \bibnamefont
  {Stognij}}, \bibinfo {author} {\bibfnamefont {N.~N.}\ \bibnamefont
  {Novitski}}, \bibinfo {author} {\bibfnamefont {V.~K.}\ \bibnamefont
  {Sakharov}}, \bibinfo {author} {\bibfnamefont {Y.~V.}\ \bibnamefont
  {Khivintsev}}, \ and\ \bibinfo {author} {\bibfnamefont {S.~A.}\ \bibnamefont
  {Nikitov}},\ }\href@noop {} {\bibfield  {journal} {\bibinfo  {journal} {Phys.
  Rev. B}\ }\textbf {\bibinfo {volume} {99}} (\bibinfo {year}
  {2019}{\natexlab{b}})}\BibitemShut {NoStop}%
\bibitem [{\citenamefont {Krawczyk}\ and\ \citenamefont
  {Grundler}(2014)}]{Krawczyk2014MC}%
  \BibitemOpen
  \bibfield  {author} {\bibinfo {author} {\bibfnamefont {M.}~\bibnamefont
  {Krawczyk}}\ and\ \bibinfo {author} {\bibfnamefont {D.}~\bibnamefont
  {Grundler}},\ }\href {\doibase 10.1088/0953-8984/26/12/123202} {\bibfield
  {journal} {\bibinfo  {journal} {Journal of Physics: Condensed Matter}\
  }\textbf {\bibinfo {volume} {26}},\ \bibinfo {pages} {123202} (\bibinfo
  {year} {2014})}\BibitemShut {NoStop}%
\bibitem [{\citenamefont {Chumak}\ \emph {et~al.}(2017)\citenamefont {Chumak},
  \citenamefont {Serga},\ and\ \citenamefont {Hillebrands}}]{chumak2017MC}%
  \BibitemOpen
  \bibfield  {author} {\bibinfo {author} {\bibfnamefont {A.}~\bibnamefont
  {Chumak}}, \bibinfo {author} {\bibfnamefont {A.}~\bibnamefont {Serga}}, \
  and\ \bibinfo {author} {\bibfnamefont {B.}~\bibnamefont {Hillebrands}},\
  }\href@noop {} {\bibfield  {journal} {\bibinfo  {journal} {Journal of Physics
  D: Applied Physics}\ }\textbf {\bibinfo {volume} {50}},\ \bibinfo {pages}
  {244001} (\bibinfo {year} {2017})}\BibitemShut {NoStop}%
\bibitem [{\citenamefont {Chumak}\ \emph {et~al.}(2009)\citenamefont {Chumak},
  \citenamefont {Neumann}, \citenamefont {Serga}, \citenamefont {Hillebrands},\
  and\ \citenamefont {Kostylev}}]{Chumak2009_current}%
  \BibitemOpen
  \bibfield  {author} {\bibinfo {author} {\bibfnamefont {A.~V.}\ \bibnamefont
  {Chumak}}, \bibinfo {author} {\bibfnamefont {T.}~\bibnamefont {Neumann}},
  \bibinfo {author} {\bibfnamefont {A.~A.}\ \bibnamefont {Serga}}, \bibinfo
  {author} {\bibfnamefont {B.}~\bibnamefont {Hillebrands}}, \ and\ \bibinfo
  {author} {\bibfnamefont {M.~P.}\ \bibnamefont {Kostylev}},\ }\href {\doibase
  10.1088/0022-3727/42/20/205005} {\bibfield  {journal} {\bibinfo  {journal}
  {Journal of Physics D: Applied Physics}\ }\textbf {\bibinfo {volume} {42}},\
  \bibinfo {pages} {205005} (\bibinfo {year} {2009})}\BibitemShut {NoStop}%
\bibitem [{\citenamefont {Nikitin}\ \emph {et~al.}(2015)\citenamefont
  {Nikitin}, \citenamefont {Ustinov}, \citenamefont {Semenov}, \citenamefont
  {Chumak}, \citenamefont {Serga}, \citenamefont {Vasyuchka}, \citenamefont
  {L{\"a}hderanta}, \citenamefont {Kalinikos},\ and\ \citenamefont
  {Hillebrands}}]{nikitin2015_current}%
  \BibitemOpen
  \bibfield  {author} {\bibinfo {author} {\bibfnamefont {A.~A.}\ \bibnamefont
  {Nikitin}}, \bibinfo {author} {\bibfnamefont {A.~B.}\ \bibnamefont
  {Ustinov}}, \bibinfo {author} {\bibfnamefont {A.~A.}\ \bibnamefont
  {Semenov}}, \bibinfo {author} {\bibfnamefont {A.~V.}\ \bibnamefont {Chumak}},
  \bibinfo {author} {\bibfnamefont {A.~A.}\ \bibnamefont {Serga}}, \bibinfo
  {author} {\bibfnamefont {V.~I.}\ \bibnamefont {Vasyuchka}}, \bibinfo {author}
  {\bibfnamefont {E.}~\bibnamefont {L{\"a}hderanta}}, \bibinfo {author}
  {\bibfnamefont {B.~A.}\ \bibnamefont {Kalinikos}}, \ and\ \bibinfo {author}
  {\bibfnamefont {B.}~\bibnamefont {Hillebrands}},\ }\href@noop {} {\bibfield
  {journal} {\bibinfo  {journal} {Applied Physics Letters}\ }\textbf {\bibinfo
  {volume} {106}},\ \bibinfo {pages} {102405} (\bibinfo {year}
  {2015})}\BibitemShut {NoStop}%
\bibitem [{\citenamefont {Kryshtal}\ and\ \citenamefont
  {Medved}(2012)}]{kryshtal2012surface}%
  \BibitemOpen
  \bibfield  {author} {\bibinfo {author} {\bibfnamefont {R.}~\bibnamefont
  {Kryshtal}}\ and\ \bibinfo {author} {\bibfnamefont {A.}~\bibnamefont
  {Medved}},\ }\href@noop {} {\bibfield  {journal} {\bibinfo  {journal}
  {Applied Physics Letters}\ }\textbf {\bibinfo {volume} {100}},\ \bibinfo
  {pages} {192410} (\bibinfo {year} {2012})}\BibitemShut {NoStop}%
\bibitem [{\citenamefont {Sadovnikov}\ \emph
  {et~al.}(2019{\natexlab{c}})\citenamefont {Sadovnikov}, \citenamefont
  {Grachev}, \citenamefont {Serdobintsev}, \citenamefont {Sheshukova},
  \citenamefont {Yankin},\ and\ \citenamefont
  {Nikitov}}]{sadovnikov2019magnon}%
  \BibitemOpen
  \bibfield  {author} {\bibinfo {author} {\bibfnamefont {A.~V.}\ \bibnamefont
  {Sadovnikov}}, \bibinfo {author} {\bibfnamefont {A.~A.}\ \bibnamefont
  {Grachev}}, \bibinfo {author} {\bibfnamefont {A.~A.}\ \bibnamefont
  {Serdobintsev}}, \bibinfo {author} {\bibfnamefont {S.~E.}\ \bibnamefont
  {Sheshukova}}, \bibinfo {author} {\bibfnamefont {S.~S.}\ \bibnamefont
  {Yankin}}, \ and\ \bibinfo {author} {\bibfnamefont {S.~A.}\ \bibnamefont
  {Nikitov}},\ }\href@noop {} {\bibfield  {journal} {\bibinfo  {journal} {IEEE
  Magnetics Letters}\ }\textbf {\bibinfo {volume} {10}},\ \bibinfo {pages} {1}
  (\bibinfo {year} {2019}{\natexlab{c}})}\BibitemShut {NoStop}%
\bibitem [{\citenamefont {Tikhonov}\ \emph {et~al.}(2016)\citenamefont
  {Tikhonov}, \citenamefont {Litvinenko}, \citenamefont {Sadovnikov},\ and\
  \citenamefont {Nikitov}}]{tikhonov2016brillouin}%
  \BibitemOpen
  \bibfield  {author} {\bibinfo {author} {\bibfnamefont {V.}~\bibnamefont
  {Tikhonov}}, \bibinfo {author} {\bibfnamefont {A.}~\bibnamefont
  {Litvinenko}}, \bibinfo {author} {\bibfnamefont {A.}~\bibnamefont
  {Sadovnikov}}, \ and\ \bibinfo {author} {\bibfnamefont {S.}~\bibnamefont
  {Nikitov}},\ }\href@noop {} {\bibfield  {journal} {\bibinfo  {journal}
  {Bulletin of the Russian Academy of Sciences: Physics}\ }\textbf {\bibinfo
  {volume} {80}},\ \bibinfo {pages} {1242} (\bibinfo {year}
  {2016})}\BibitemShut {NoStop}%
\bibitem [{\citenamefont {Litvinenko}\ \emph {et~al.}(2021)\citenamefont
  {Litvinenko}, \citenamefont {Khymyn}, \citenamefont {Tyberkevych},
  \citenamefont {Tikhonov}, \citenamefont {Slavin},\ and\ \citenamefont
  {Nikitov}}]{litvinenko2021tunable}%
  \BibitemOpen
  \bibfield  {author} {\bibinfo {author} {\bibfnamefont {A.}~\bibnamefont
  {Litvinenko}}, \bibinfo {author} {\bibfnamefont {R.}~\bibnamefont {Khymyn}},
  \bibinfo {author} {\bibfnamefont {V.}~\bibnamefont {Tyberkevych}}, \bibinfo
  {author} {\bibfnamefont {V.}~\bibnamefont {Tikhonov}}, \bibinfo {author}
  {\bibfnamefont {A.}~\bibnamefont {Slavin}}, \ and\ \bibinfo {author}
  {\bibfnamefont {S.}~\bibnamefont {Nikitov}},\ }\href@noop {} {\bibfield
  {journal} {\bibinfo  {journal} {Physical Review Applied}\ }\textbf {\bibinfo
  {volume} {15}},\ \bibinfo {pages} {034057} (\bibinfo {year}
  {2021})}\BibitemShut {NoStop}%
\bibitem [{\citenamefont {Litvinenko}\ \emph {et~al.}(2018)\citenamefont
  {Litvinenko}, \citenamefont {Grishin}, \citenamefont {Sharaevskii},
  \citenamefont {Tikhonov},\ and\ \citenamefont
  {Nikitov}}]{litvinenko2018chaotic}%
  \BibitemOpen
  \bibfield  {author} {\bibinfo {author} {\bibfnamefont {A.~N.}\ \bibnamefont
  {Litvinenko}}, \bibinfo {author} {\bibfnamefont {S.~V.}\ \bibnamefont
  {Grishin}}, \bibinfo {author} {\bibfnamefont {Y.~P.}\ \bibnamefont
  {Sharaevskii}}, \bibinfo {author} {\bibfnamefont {V.~V.}\ \bibnamefont
  {Tikhonov}}, \ and\ \bibinfo {author} {\bibfnamefont {S.~A.}\ \bibnamefont
  {Nikitov}},\ }\href@noop {} {\bibfield  {journal} {\bibinfo  {journal}
  {Technical Physics Letters}\ }\textbf {\bibinfo {volume} {44}},\ \bibinfo
  {pages} {263} (\bibinfo {year} {2018})}\BibitemShut {NoStop}%
\bibitem [{\citenamefont {Ustinov}\ \emph {et~al.}(2019)\citenamefont
  {Ustinov}, \citenamefont {Drozdovskii}, \citenamefont {Nikitin},
  \citenamefont {Semenov}, \citenamefont {Bozhko}, \citenamefont {Serga},
  \citenamefont {Hillebrands}, \citenamefont {L{\"a}hderanta},\ and\
  \citenamefont {Kalinikos}}]{FE_Voltage_dynamic}%
  \BibitemOpen
  \bibfield  {author} {\bibinfo {author} {\bibfnamefont {A.~B.}\ \bibnamefont
  {Ustinov}}, \bibinfo {author} {\bibfnamefont {A.~V.}\ \bibnamefont
  {Drozdovskii}}, \bibinfo {author} {\bibfnamefont {A.~A.}\ \bibnamefont
  {Nikitin}}, \bibinfo {author} {\bibfnamefont {A.~A.}\ \bibnamefont
  {Semenov}}, \bibinfo {author} {\bibfnamefont {D.~A.}\ \bibnamefont {Bozhko}},
  \bibinfo {author} {\bibfnamefont {A.~A.}\ \bibnamefont {Serga}}, \bibinfo
  {author} {\bibfnamefont {B.}~\bibnamefont {Hillebrands}}, \bibinfo {author}
  {\bibfnamefont {E.}~\bibnamefont {L{\"a}hderanta}}, \ and\ \bibinfo {author}
  {\bibfnamefont {B.~A.}\ \bibnamefont {Kalinikos}},\ }\href@noop {} {\bibfield
   {journal} {\bibinfo  {journal} {Communications Physics}\ }\textbf {\bibinfo
  {volume} {2}},\ \bibinfo {pages} {1} (\bibinfo {year} {2019})}\BibitemShut
  {NoStop}%
\bibitem [{\citenamefont {Wang}\ \emph {et~al.}(2017)\citenamefont {Wang},
  \citenamefont {Chumak}, \citenamefont {Jin}, \citenamefont {Zhang},
  \citenamefont {Hillebrands},\ and\ \citenamefont
  {Zhong}}]{Voltage_anisotripy}%
  \BibitemOpen
  \bibfield  {author} {\bibinfo {author} {\bibfnamefont {Q.}~\bibnamefont
  {Wang}}, \bibinfo {author} {\bibfnamefont {A.~V.}\ \bibnamefont {Chumak}},
  \bibinfo {author} {\bibfnamefont {L.}~\bibnamefont {Jin}}, \bibinfo {author}
  {\bibfnamefont {H.}~\bibnamefont {Zhang}}, \bibinfo {author} {\bibfnamefont
  {B.}~\bibnamefont {Hillebrands}}, \ and\ \bibinfo {author} {\bibfnamefont
  {Z.}~\bibnamefont {Zhong}},\ }\href {\doibase 10.1103/PhysRevB.95.134433}
  {\bibfield  {journal} {\bibinfo  {journal} {Phys. Rev. B}\ }\textbf {\bibinfo
  {volume} {95}},\ \bibinfo {pages} {134433} (\bibinfo {year}
  {2017})}\BibitemShut {NoStop}%
\bibitem [{\citenamefont {xiong Li}\ \emph {et~al.}(2015)\citenamefont {xiong
  Li}, \citenamefont {guang Wang}, \citenamefont {wei Wang}, \citenamefont
  {zhuang Nie}, \citenamefont {Tang},\ and\ \citenamefont {hua
  Guo}}]{MC_domain_walls}%
  \BibitemOpen
  \bibfield  {author} {\bibinfo {author} {\bibfnamefont {Z.}~\bibnamefont
  {xiong Li}}, \bibinfo {author} {\bibfnamefont {X.}~\bibnamefont {guang
  Wang}}, \bibinfo {author} {\bibfnamefont {D.}~\bibnamefont {wei Wang}},
  \bibinfo {author} {\bibfnamefont {Y.}~\bibnamefont {zhuang Nie}}, \bibinfo
  {author} {\bibfnamefont {W.}~\bibnamefont {Tang}}, \ and\ \bibinfo {author}
  {\bibfnamefont {G.}~\bibnamefont {hua Guo}},\ }\href {\doibase
  https://doi.org/10.1016/j.jmmm.2015.04.012} {\bibfield  {journal} {\bibinfo
  {journal} {Journal of Magnetism and Magnetic Materials}\ }\textbf {\bibinfo
  {volume} {388}},\ \bibinfo {pages} {10 } (\bibinfo {year}
  {2015})}\BibitemShut {NoStop}%
\bibitem [{\citenamefont {Banerjee}\ \emph {et~al.}(2017)\citenamefont
  {Banerjee}, \citenamefont {Gruszecki}, \citenamefont {Klos}, \citenamefont
  {Hellwig}, \citenamefont {Krawczyk},\ and\ \citenamefont
  {Barman}}]{domain_walls_experiment}%
  \BibitemOpen
  \bibfield  {author} {\bibinfo {author} {\bibfnamefont {C.}~\bibnamefont
  {Banerjee}}, \bibinfo {author} {\bibfnamefont {P.}~\bibnamefont {Gruszecki}},
  \bibinfo {author} {\bibfnamefont {J.~W.}\ \bibnamefont {Klos}}, \bibinfo
  {author} {\bibfnamefont {O.}~\bibnamefont {Hellwig}}, \bibinfo {author}
  {\bibfnamefont {M.}~\bibnamefont {Krawczyk}}, \ and\ \bibinfo {author}
  {\bibfnamefont {A.}~\bibnamefont {Barman}},\ }\href {\doibase
  10.1103/PhysRevB.96.024421} {\bibfield  {journal} {\bibinfo  {journal} {Phys.
  Rev. B}\ }\textbf {\bibinfo {volume} {96}},\ \bibinfo {pages} {024421}
  (\bibinfo {year} {2017})}\BibitemShut {NoStop}%
\bibitem [{\citenamefont {Ma}\ \emph {et~al.}(2015)\citenamefont {Ma},
  \citenamefont {Zhou}, \citenamefont {Braun},\ and\ \citenamefont
  {Lew}}]{skyrmions_MC}%
  \BibitemOpen
  \bibfield  {author} {\bibinfo {author} {\bibfnamefont {F.}~\bibnamefont
  {Ma}}, \bibinfo {author} {\bibfnamefont {Y.}~\bibnamefont {Zhou}}, \bibinfo
  {author} {\bibfnamefont {H.-B.}\ \bibnamefont {Braun}}, \ and\ \bibinfo
  {author} {\bibfnamefont {W.~S.}\ \bibnamefont {Lew}},\ }\href@noop {}
  {\bibfield  {journal} {\bibinfo  {journal} {Nano Letters}\ }\textbf {\bibinfo
  {volume} {15}},\ \bibinfo {pages} {4029} (\bibinfo {year}
  {2015})}\BibitemShut {NoStop}%
\bibitem [{\citenamefont {Dobrovolskiy}\ \emph {et~al.}(2019)\citenamefont
  {Dobrovolskiy}, \citenamefont {Sachser}, \citenamefont {Br{\"a}cher},
  \citenamefont {B{\"o}ttcher}, \citenamefont {Kruglyak}, \citenamefont {Vovk},
  \citenamefont {Shklovskij}, \citenamefont {Huth}, \citenamefont
  {Hillebrands},\ and\ \citenamefont {Chumak}}]{MC_fluxon}%
  \BibitemOpen
  \bibfield  {author} {\bibinfo {author} {\bibfnamefont {O.}~\bibnamefont
  {Dobrovolskiy}}, \bibinfo {author} {\bibfnamefont {R.}~\bibnamefont
  {Sachser}}, \bibinfo {author} {\bibfnamefont {T.}~\bibnamefont
  {Br{\"a}cher}}, \bibinfo {author} {\bibfnamefont {T.}~\bibnamefont
  {B{\"o}ttcher}}, \bibinfo {author} {\bibfnamefont {V.}~\bibnamefont
  {Kruglyak}}, \bibinfo {author} {\bibfnamefont {R.}~\bibnamefont {Vovk}},
  \bibinfo {author} {\bibfnamefont {V.}~\bibnamefont {Shklovskij}}, \bibinfo
  {author} {\bibfnamefont {M.}~\bibnamefont {Huth}}, \bibinfo {author}
  {\bibfnamefont {B.}~\bibnamefont {Hillebrands}}, \ and\ \bibinfo {author}
  {\bibfnamefont {A.}~\bibnamefont {Chumak}},\ }\href@noop {} {\bibfield
  {journal} {\bibinfo  {journal} {Nature Physics}\ }\textbf {\bibinfo {volume}
  {15}},\ \bibinfo {pages} {477} (\bibinfo {year} {2019})}\BibitemShut
  {NoStop}%
\bibitem [{\citenamefont {Chumak}\ \emph {et~al.}(2010)\citenamefont {Chumak},
  \citenamefont {Dhagat}, \citenamefont {Jander}, \citenamefont {Serga},\ and\
  \citenamefont {Hillebrands}}]{moving_MC_dopler}%
  \BibitemOpen
  \bibfield  {author} {\bibinfo {author} {\bibfnamefont {A.}~\bibnamefont
  {Chumak}}, \bibinfo {author} {\bibfnamefont {P.}~\bibnamefont {Dhagat}},
  \bibinfo {author} {\bibfnamefont {A.}~\bibnamefont {Jander}}, \bibinfo
  {author} {\bibfnamefont {A.}~\bibnamefont {Serga}}, \ and\ \bibinfo {author}
  {\bibfnamefont {B.}~\bibnamefont {Hillebrands}},\ }\href@noop {} {\bibfield
  {journal} {\bibinfo  {journal} {Physical Review B}\ }\textbf {\bibinfo
  {volume} {81}},\ \bibinfo {pages} {140404} (\bibinfo {year}
  {2010})}\BibitemShut {NoStop}%
\bibitem [{\citenamefont {Fetisov}\ and\ \citenamefont
  {Makovkin}(1996)}]{fetisov1996_heat}%
  \BibitemOpen
  \bibfield  {author} {\bibinfo {author} {\bibfnamefont {Y.}~\bibnamefont
  {Fetisov}}\ and\ \bibinfo {author} {\bibfnamefont {A.}~\bibnamefont
  {Makovkin}},\ }\href@noop {} {\bibfield  {journal} {\bibinfo  {journal}
  {Journal of applied physics}\ }\textbf {\bibinfo {volume} {79}},\ \bibinfo
  {pages} {5721} (\bibinfo {year} {1996})}\BibitemShut {NoStop}%
\bibitem [{\citenamefont {Obry}\ \emph {et~al.}(2012)\citenamefont {Obry},
  \citenamefont {Vasyuchka}, \citenamefont {Chumak}, \citenamefont {Serga},\
  and\ \citenamefont {Hillebrands}}]{heatMc2012}%
  \BibitemOpen
  \bibfield  {author} {\bibinfo {author} {\bibfnamefont {B.}~\bibnamefont
  {Obry}}, \bibinfo {author} {\bibfnamefont {V.~I.}\ \bibnamefont {Vasyuchka}},
  \bibinfo {author} {\bibfnamefont {A.~V.}\ \bibnamefont {Chumak}}, \bibinfo
  {author} {\bibfnamefont {A.~A.}\ \bibnamefont {Serga}}, \ and\ \bibinfo
  {author} {\bibfnamefont {B.}~\bibnamefont {Hillebrands}},\ }\href@noop {}
  {\bibfield  {journal} {\bibinfo  {journal} {Applied Physics Letters}\
  }\textbf {\bibinfo {volume} {101}},\ \bibinfo {pages} {192406} (\bibinfo
  {year} {2012})}\BibitemShut {NoStop}%
\bibitem [{\citenamefont {Vogel}\ \emph {et~al.}(2015)\citenamefont {Vogel},
  \citenamefont {Chumak}, \citenamefont {Waller}, \citenamefont {Langner},
  \citenamefont {Vasyuchka}, \citenamefont {Hillebrands},\ and\ \citenamefont
  {Von~Freymann}}]{heatMC2015}%
  \BibitemOpen
  \bibfield  {author} {\bibinfo {author} {\bibfnamefont {M.}~\bibnamefont
  {Vogel}}, \bibinfo {author} {\bibfnamefont {A.~V.}\ \bibnamefont {Chumak}},
  \bibinfo {author} {\bibfnamefont {E.~H.}\ \bibnamefont {Waller}}, \bibinfo
  {author} {\bibfnamefont {T.}~\bibnamefont {Langner}}, \bibinfo {author}
  {\bibfnamefont {V.~I.}\ \bibnamefont {Vasyuchka}}, \bibinfo {author}
  {\bibfnamefont {B.}~\bibnamefont {Hillebrands}}, \ and\ \bibinfo {author}
  {\bibfnamefont {G.}~\bibnamefont {Von~Freymann}},\ }\href@noop {} {\bibfield
  {journal} {\bibinfo  {journal} {Nature Physics}\ }\textbf {\bibinfo {volume}
  {11}},\ \bibinfo {pages} {487} (\bibinfo {year} {2015})}\BibitemShut
  {NoStop}%
\bibitem [{\citenamefont {Barman}\ \emph
  {et~al.}(2021{\natexlab{b}})\citenamefont {Barman}, \citenamefont
  {Gubbiotti}, \citenamefont {Ladak}, \citenamefont {Adeyeye}, \citenamefont
  {Krawczyk}, \citenamefont {Gr{\"a}fe}, \citenamefont {Adelmann},
  \citenamefont {Cotofana}, \citenamefont {Naeemi}, \citenamefont {Vasyuchka}
  \emph {et~al.}}]{magnonicsroadmap}%
  \BibitemOpen
  \bibfield  {author} {\bibinfo {author} {\bibfnamefont {A.}~\bibnamefont
  {Barman}}, \bibinfo {author} {\bibfnamefont {G.}~\bibnamefont {Gubbiotti}},
  \bibinfo {author} {\bibfnamefont {S.}~\bibnamefont {Ladak}}, \bibinfo
  {author} {\bibfnamefont {A.~O.}\ \bibnamefont {Adeyeye}}, \bibinfo {author}
  {\bibfnamefont {M.}~\bibnamefont {Krawczyk}}, \bibinfo {author}
  {\bibfnamefont {J.}~\bibnamefont {Gr{\"a}fe}}, \bibinfo {author}
  {\bibfnamefont {C.}~\bibnamefont {Adelmann}}, \bibinfo {author}
  {\bibfnamefont {S.}~\bibnamefont {Cotofana}}, \bibinfo {author}
  {\bibfnamefont {A.}~\bibnamefont {Naeemi}}, \bibinfo {author} {\bibfnamefont
  {V.~I.}\ \bibnamefont {Vasyuchka}},  \emph {et~al.},\ }\href@noop {}
  {\bibfield  {journal} {\bibinfo  {journal} {Journal of Physics: Condensed
  Matter}\ }\textbf {\bibinfo {volume} {33}},\ \bibinfo {pages} {413001}
  (\bibinfo {year} {2021}{\natexlab{b}})}\BibitemShut {NoStop}%
\bibitem [{\citenamefont {Serga}\ \emph {et~al.}(2010)\citenamefont {Serga},
  \citenamefont {Chumak},\ and\ \citenamefont {Hillebrands}}]{serga2010yig}%
  \BibitemOpen
  \bibfield  {author} {\bibinfo {author} {\bibfnamefont {A.}~\bibnamefont
  {Serga}}, \bibinfo {author} {\bibfnamefont {A.}~\bibnamefont {Chumak}}, \
  and\ \bibinfo {author} {\bibfnamefont {B.}~\bibnamefont {Hillebrands}},\
  }\href@noop {} {\bibfield  {journal} {\bibinfo  {journal} {Journal of Physics
  D: Applied Physics}\ }\textbf {\bibinfo {volume} {43}},\ \bibinfo {pages}
  {264002} (\bibinfo {year} {2010})}\BibitemShut {NoStop}%
\bibitem [{\citenamefont {Seshadri}(1970)}]{seshadri1970surface}%
  \BibitemOpen
  \bibfield  {author} {\bibinfo {author} {\bibfnamefont {S.}~\bibnamefont
  {Seshadri}},\ }\href@noop {} {\bibfield  {journal} {\bibinfo  {journal}
  {Proceedings of the IEEE}\ }\textbf {\bibinfo {volume} {58}},\ \bibinfo
  {pages} {506} (\bibinfo {year} {1970})}\BibitemShut {NoStop}%
\bibitem [{\citenamefont {Kawasaki}\ \emph {et~al.}(1974)\citenamefont
  {Kawasaki}, \citenamefont {Takagi},\ and\ \citenamefont
  {Umeno}}]{kawasaki1974interaction}%
  \BibitemOpen
  \bibfield  {author} {\bibinfo {author} {\bibfnamefont {K.}~\bibnamefont
  {Kawasaki}}, \bibinfo {author} {\bibfnamefont {H.}~\bibnamefont {Takagi}}, \
  and\ \bibinfo {author} {\bibfnamefont {M.}~\bibnamefont {Umeno}},\
  }\href@noop {} {\bibfield  {journal} {\bibinfo  {journal} {IEEE Transactions
  on Microwave Theory and Techniques}\ }\textbf {\bibinfo {volume} {22}},\
  \bibinfo {pages} {918} (\bibinfo {year} {1974})}\BibitemShut {NoStop}%
\bibitem [{\citenamefont {Stancil}(1986)}]{stancil1986phenomenological}%
  \BibitemOpen
  \bibfield  {author} {\bibinfo {author} {\bibfnamefont {D.~D.}\ \bibnamefont
  {Stancil}},\ }\href@noop {} {\bibfield  {journal} {\bibinfo  {journal}
  {Journal of applied physics}\ }\textbf {\bibinfo {volume} {59}},\ \bibinfo
  {pages} {218} (\bibinfo {year} {1986})}\BibitemShut {NoStop}%
\bibitem [{\citenamefont {Kindyak}(1995)}]{kindyak1995magnetostatic_screening}%
  \BibitemOpen
  \bibfield  {author} {\bibinfo {author} {\bibfnamefont {A.}~\bibnamefont
  {Kindyak}},\ }\href@noop {} {\bibfield  {journal} {\bibinfo  {journal}
  {Materials Letters}\ }\textbf {\bibinfo {volume} {24}},\ \bibinfo {pages}
  {359} (\bibinfo {year} {1995})}\BibitemShut {NoStop}%
\bibitem [{\citenamefont {Almeida}\ and\ \citenamefont
  {Mills}(1996)}]{almeida1996eddy}%
  \BibitemOpen
  \bibfield  {author} {\bibinfo {author} {\bibfnamefont {N.}~\bibnamefont
  {Almeida}}\ and\ \bibinfo {author} {\bibfnamefont {D.}~\bibnamefont
  {Mills}},\ }\href@noop {} {\bibfield  {journal} {\bibinfo  {journal}
  {Physical Review B}\ }\textbf {\bibinfo {volume} {53}},\ \bibinfo {pages}
  {12232} (\bibinfo {year} {1996})}\BibitemShut {NoStop}%
\bibitem [{\citenamefont {Fetisov}\ \emph {et~al.}(1996)\citenamefont
  {Fetisov}, \citenamefont {Makovkin},\ and\ \citenamefont
  {Studenov}}]{fetisov1996optically_semicond}%
  \BibitemOpen
  \bibfield  {author} {\bibinfo {author} {\bibfnamefont {Y.}~\bibnamefont
  {Fetisov}}, \bibinfo {author} {\bibfnamefont {A.}~\bibnamefont {Makovkin}}, \
  and\ \bibinfo {author} {\bibfnamefont {V.}~\bibnamefont {Studenov}},\ }in\
  \href@noop {} {\emph {\bibinfo {booktitle} {International Topical Meeting on
  Microwave Photonics. MWP'96 Technical Digest. Satellite Workshop (Cat. No.
  96TH8153)}}}\ (\bibinfo {organization} {IEEE},\ \bibinfo {year} {1996})\ pp.\
  \bibinfo {pages} {37--40}\BibitemShut {NoStop}%
\bibitem [{\citenamefont {Kindyak}(1999)}]{kindyak1999nonlinear}%
  \BibitemOpen
  \bibfield  {author} {\bibinfo {author} {\bibfnamefont {A.}~\bibnamefont
  {Kindyak}},\ }\href@noop {} {\bibfield  {journal} {\bibinfo  {journal}
  {Technical Physics}\ }\textbf {\bibinfo {volume} {44}},\ \bibinfo {pages}
  {715} (\bibinfo {year} {1999})}\BibitemShut {NoStop}%
\bibitem [{\citenamefont {Kindyak}(1996)}]{Kindyak}%
  \BibitemOpen
  \bibfield  {author} {\bibinfo {author} {\bibfnamefont {A.~S.}\ \bibnamefont
  {Kindyak}},\ }\href@noop {} {\bibfield  {journal} {\bibinfo  {journal}
  {Journal of communications technology \& electronics}\ }\textbf {\bibinfo
  {volume} {41}} (\bibinfo {year} {1996})}\BibitemShut {NoStop}%
\bibitem [{\citenamefont {Beginin}\ \emph {et~al.}(2012)\citenamefont
  {Beginin}, \citenamefont {Filimonov}, \citenamefont {Pavlov}, \citenamefont
  {Vysotskii},\ and\ \citenamefont {Nikitov}}]{Beginin_metal_MC}%
  \BibitemOpen
  \bibfield  {author} {\bibinfo {author} {\bibfnamefont {E.~N.}\ \bibnamefont
  {Beginin}}, \bibinfo {author} {\bibfnamefont {Y.~A.}\ \bibnamefont
  {Filimonov}}, \bibinfo {author} {\bibfnamefont {E.~S.}\ \bibnamefont
  {Pavlov}}, \bibinfo {author} {\bibfnamefont {S.~L.}\ \bibnamefont
  {Vysotskii}}, \ and\ \bibinfo {author} {\bibfnamefont {S.~A.}\ \bibnamefont
  {Nikitov}},\ }\href {\doibase 10.1063/1.4730374} {\bibfield  {journal}
  {\bibinfo  {journal} {Applied Physics Letters}\ }\textbf {\bibinfo {volume}
  {100}},\ \bibinfo {pages} {252412} (\bibinfo {year} {2012})},\ \Eprint
  {http://arxiv.org/abs/https://doi.org/10.1063/1.4730374}
  {https://doi.org/10.1063/1.4730374} \BibitemShut {NoStop}%
\bibitem [{\citenamefont {Mruczkiewicz}\ \emph {et~al.}(2017)\citenamefont
  {Mruczkiewicz}, \citenamefont {Graczyk}, \citenamefont {Lupo}, \citenamefont
  {Adeyeye}, \citenamefont {Gubbiotti},\ and\ \citenamefont
  {Krawczyk}}]{MC_stripes}%
  \BibitemOpen
  \bibfield  {author} {\bibinfo {author} {\bibfnamefont {M.}~\bibnamefont
  {Mruczkiewicz}}, \bibinfo {author} {\bibfnamefont {P.}~\bibnamefont
  {Graczyk}}, \bibinfo {author} {\bibfnamefont {P.}~\bibnamefont {Lupo}},
  \bibinfo {author} {\bibfnamefont {A.}~\bibnamefont {Adeyeye}}, \bibinfo
  {author} {\bibfnamefont {G.}~\bibnamefont {Gubbiotti}}, \ and\ \bibinfo
  {author} {\bibfnamefont {M.}~\bibnamefont {Krawczyk}},\ }\href {\doibase
  10.1103/PhysRevB.96.104411} {\bibfield  {journal} {\bibinfo  {journal} {Phys.
  Rev. B}\ }\textbf {\bibinfo {volume} {96}},\ \bibinfo {pages} {104411}
  (\bibinfo {year} {2017})}\BibitemShut {NoStop}%
\bibitem [{\citenamefont {Mruczkiewicz}\ and\ \citenamefont
  {Krawczyk}(2014)}]{mruczkiewicz2014nonreci_Me}%
  \BibitemOpen
  \bibfield  {author} {\bibinfo {author} {\bibfnamefont {M.}~\bibnamefont
  {Mruczkiewicz}}\ and\ \bibinfo {author} {\bibfnamefont {M.}~\bibnamefont
  {Krawczyk}},\ }\href@noop {} {\bibfield  {journal} {\bibinfo  {journal}
  {Journal of Applied Physics}\ }\textbf {\bibinfo {volume} {115}},\ \bibinfo
  {pages} {113909} (\bibinfo {year} {2014})}\BibitemShut {NoStop}%
\bibitem [{\citenamefont {Mruczkiewicz}\ \emph {et~al.}(2014)\citenamefont
  {Mruczkiewicz}, \citenamefont {Pavlov}, \citenamefont {Vysotsky},
  \citenamefont {Krawczyk}, \citenamefont {Filimonov},\ and\ \citenamefont
  {Nikitov}}]{MC_nonreciprocity}%
  \BibitemOpen
  \bibfield  {author} {\bibinfo {author} {\bibfnamefont {M.}~\bibnamefont
  {Mruczkiewicz}}, \bibinfo {author} {\bibfnamefont {E.}~\bibnamefont
  {Pavlov}}, \bibinfo {author} {\bibfnamefont {S.}~\bibnamefont {Vysotsky}},
  \bibinfo {author} {\bibfnamefont {M.}~\bibnamefont {Krawczyk}}, \bibinfo
  {author} {\bibfnamefont {Y.~A.}\ \bibnamefont {Filimonov}}, \ and\ \bibinfo
  {author} {\bibfnamefont {S.}~\bibnamefont {Nikitov}},\ }\href@noop {}
  {\bibfield  {journal} {\bibinfo  {journal} {Physical Review B}\ }\textbf
  {\bibinfo {volume} {90}},\ \bibinfo {pages} {174416} (\bibinfo {year}
  {2014})}\BibitemShut {NoStop}%
\bibitem [{\citenamefont {Korchagin}\ \emph {et~al.}(2021)\citenamefont
  {Korchagin}, \citenamefont {Pleshakova}, \citenamefont {Alexandrova},
  \citenamefont {Dolgov}, \citenamefont {Dogadina}, \citenamefont
  {Serdechnyy},\ and\ \citenamefont {Bublikov}}]{korchagin2021mathematical}%
  \BibitemOpen
  \bibfield  {author} {\bibinfo {author} {\bibfnamefont {S.}~\bibnamefont
  {Korchagin}}, \bibinfo {author} {\bibfnamefont {E.}~\bibnamefont
  {Pleshakova}}, \bibinfo {author} {\bibfnamefont {I.}~\bibnamefont
  {Alexandrova}}, \bibinfo {author} {\bibfnamefont {V.}~\bibnamefont {Dolgov}},
  \bibinfo {author} {\bibfnamefont {E.}~\bibnamefont {Dogadina}}, \bibinfo
  {author} {\bibfnamefont {D.}~\bibnamefont {Serdechnyy}}, \ and\ \bibinfo
  {author} {\bibfnamefont {K.}~\bibnamefont {Bublikov}},\ }\href@noop {}
  {\bibfield  {journal} {\bibinfo  {journal} {Mathematics}\ }\textbf {\bibinfo
  {volume} {9}},\ \bibinfo {pages} {2948} (\bibinfo {year} {2021})}\BibitemShut
  {NoStop}%
\bibitem [{\citenamefont {Cherepanov}\ \emph {et~al.}(1993)\citenamefont
  {Cherepanov}, \citenamefont {Kolokolov},\ and\ \citenamefont
  {L'vov}}]{cherepanovYIG}%
  \BibitemOpen
  \bibfield  {author} {\bibinfo {author} {\bibfnamefont {V.}~\bibnamefont
  {Cherepanov}}, \bibinfo {author} {\bibfnamefont {I.}~\bibnamefont
  {Kolokolov}}, \ and\ \bibinfo {author} {\bibfnamefont {V.}~\bibnamefont
  {L'vov}},\ }\href@noop {} {\bibfield  {journal} {\bibinfo  {journal} {Physics
  reports}\ }\textbf {\bibinfo {volume} {229}},\ \bibinfo {pages} {81}
  (\bibinfo {year} {1993})}\BibitemShut {NoStop}%
\bibitem [{\citenamefont {Stafe}\ \emph {et~al.}(2013)\citenamefont {Stafe},
  \citenamefont {Marcu},\ and\ \citenamefont {Puscas}}]{laserablationboock}%
  \BibitemOpen
  \bibfield  {author} {\bibinfo {author} {\bibfnamefont {M.}~\bibnamefont
  {Stafe}}, \bibinfo {author} {\bibfnamefont {A.}~\bibnamefont {Marcu}}, \ and\
  \bibinfo {author} {\bibfnamefont {N.~N.}\ \bibnamefont {Puscas}},\
  }\href@noop {} {\emph {\bibinfo {title} {Pulsed laser ablation of solids:
  basics, theory and applications}}},\ Vol.~\bibinfo {volume} {53}\ (\bibinfo
  {publisher} {Springer Science \& Business Media},\ \bibinfo {year}
  {2013})\BibitemShut {NoStop}%
\bibitem [{\citenamefont {Beginin}\ \emph {et~al.}(2013)\citenamefont
  {Beginin}, \citenamefont {Sadovnikov}, \citenamefont {Sharaevsky},\ and\
  \citenamefont {Nikitov}}]{beginin2013spatiotemporal}%
  \BibitemOpen
  \bibfield  {author} {\bibinfo {author} {\bibfnamefont {E.}~\bibnamefont
  {Beginin}}, \bibinfo {author} {\bibfnamefont {A.}~\bibnamefont {Sadovnikov}},
  \bibinfo {author} {\bibfnamefont {Y.~P.}\ \bibnamefont {Sharaevsky}}, \ and\
  \bibinfo {author} {\bibfnamefont {S.}~\bibnamefont {Nikitov}},\ }\href@noop
  {} {\bibfield  {journal} {\bibinfo  {journal} {Bulletin of the Russian
  Academy of Sciences: Physics}\ }\textbf {\bibinfo {volume} {77}},\ \bibinfo
  {pages} {1429} (\bibinfo {year} {2013})}\BibitemShut {NoStop}%
\bibitem [{\citenamefont {Beginin}\ \emph {et~al.}(2014)\citenamefont
  {Beginin}, \citenamefont {Sadovnikov}, \citenamefont {Sharaevsky},\ and\
  \citenamefont {Nikitov}}]{beginin2014multimode}%
  \BibitemOpen
  \bibfield  {author} {\bibinfo {author} {\bibfnamefont {E.~N.}\ \bibnamefont
  {Beginin}}, \bibinfo {author} {\bibfnamefont {A.~V.}\ \bibnamefont
  {Sadovnikov}}, \bibinfo {author} {\bibfnamefont {Y.~P.}\ \bibnamefont
  {Sharaevsky}}, \ and\ \bibinfo {author} {\bibfnamefont {S.~A.}\ \bibnamefont
  {Nikitov}},\ }in\ \href@noop {} {\emph {\bibinfo {booktitle} {Solid State
  Phenomena}}},\ Vol.\ \bibinfo {volume} {215}\ (\bibinfo {organization} {Trans
  Tech Publ},\ \bibinfo {year} {2014})\ pp.\ \bibinfo {pages}
  {389--393}\BibitemShut {NoStop}%
\bibitem [{\citenamefont {Sotoodeh}\ \emph {et~al.}(2000)\citenamefont
  {Sotoodeh}, \citenamefont {Khalid},\ and\ \citenamefont
  {Rezazadeh}}]{semicond_empirical}%
  \BibitemOpen
  \bibfield  {author} {\bibinfo {author} {\bibfnamefont {M.}~\bibnamefont
  {Sotoodeh}}, \bibinfo {author} {\bibfnamefont {A.}~\bibnamefont {Khalid}}, \
  and\ \bibinfo {author} {\bibfnamefont {A.}~\bibnamefont {Rezazadeh}},\
  }\href@noop {} {\bibfield  {journal} {\bibinfo  {journal} {Journal of applied
  physics}\ }\textbf {\bibinfo {volume} {87}},\ \bibinfo {pages} {2890}
  (\bibinfo {year} {2000})}\BibitemShut {NoStop}%
\bibitem [{\citenamefont {Lovejoy}\ \emph {et~al.}(1995)\citenamefont
  {Lovejoy}, \citenamefont {Melloch},\ and\ \citenamefont
  {Lundstrom}}]{GaAs_mu_T_depandence}%
  \BibitemOpen
  \bibfield  {author} {\bibinfo {author} {\bibfnamefont {M.~L.}\ \bibnamefont
  {Lovejoy}}, \bibinfo {author} {\bibfnamefont {M.~R.}\ \bibnamefont
  {Melloch}}, \ and\ \bibinfo {author} {\bibfnamefont {M.~S.}\ \bibnamefont
  {Lundstrom}},\ }\href@noop {} {\bibfield  {journal} {\bibinfo  {journal}
  {Applied physics letters}\ }\textbf {\bibinfo {volume} {67}},\ \bibinfo
  {pages} {1101} (\bibinfo {year} {1995})}\BibitemShut {NoStop}%
\bibitem [{\citenamefont {Damon}\ and\ \citenamefont
  {Eshbach}(1961)}]{damon1961}%
  \BibitemOpen
  \bibfield  {author} {\bibinfo {author} {\bibfnamefont {R.~W.}\ \bibnamefont
  {Damon}}\ and\ \bibinfo {author} {\bibfnamefont {J.}~\bibnamefont
  {Eshbach}},\ }\href@noop {} {\bibfield  {journal} {\bibinfo  {journal}
  {Journal of Physics and Chemistry of Solids}\ }\textbf {\bibinfo {volume}
  {19}},\ \bibinfo {pages} {308} (\bibinfo {year} {1961})}\BibitemShut
  {NoStop}%
\bibitem [{\citenamefont {Bajpai}(1985)}]{bajpai1985}%
  \BibitemOpen
  \bibfield  {author} {\bibinfo {author} {\bibfnamefont {S.}~\bibnamefont
  {Bajpai}},\ }\href@noop {} {\bibfield  {journal} {\bibinfo  {journal}
  {Journal of applied physics}\ }\textbf {\bibinfo {volume} {58}},\ \bibinfo
  {pages} {910} (\bibinfo {year} {1985})}\BibitemShut {NoStop}%
\bibitem [{\citenamefont {A.G.}\ and\ \citenamefont
  {G.A.}(1996)}]{gurevichmelkov}%
  \BibitemOpen
  \bibfield  {author} {\bibinfo {author} {\bibfnamefont {G.}~\bibnamefont
  {A.G.}}\ and\ \bibinfo {author} {\bibfnamefont {M.}~\bibnamefont {G.A.}},\
  }\href {https://books.google.sk/books?id=YgQtSvFIvFQC} {\emph {\bibinfo
  {title} {Magnetization Oscillations and Waves}}}\ (\bibinfo  {publisher}
  {Taylor \& Francis},\ \bibinfo {year} {1996})\BibitemShut {NoStop}%
\bibitem [{\citenamefont {Blakemore}(1982)}]{blakemore1982semiconducting}%
  \BibitemOpen
  \bibfield  {author} {\bibinfo {author} {\bibfnamefont {J.}~\bibnamefont
  {Blakemore}},\ }\href@noop {} {\bibfield  {journal} {\bibinfo  {journal}
  {Journal of Applied Physics}\ }\textbf {\bibinfo {volume} {53}},\ \bibinfo
  {pages} {R123} (\bibinfo {year} {1982})}\BibitemShut {NoStop}%
\bibitem [{\citenamefont {Raymond}\ \emph {et~al.}(1979)\citenamefont
  {Raymond}, \citenamefont {Robert},\ and\ \citenamefont
  {Bernard}}]{raymond1979electron}%
  \BibitemOpen
  \bibfield  {author} {\bibinfo {author} {\bibfnamefont {A.}~\bibnamefont
  {Raymond}}, \bibinfo {author} {\bibfnamefont {J.}~\bibnamefont {Robert}}, \
  and\ \bibinfo {author} {\bibfnamefont {C.}~\bibnamefont {Bernard}},\
  }\href@noop {} {\bibfield  {journal} {\bibinfo  {journal} {Journal of Physics
  C: Solid State Physics}\ }\textbf {\bibinfo {volume} {12}},\ \bibinfo {pages}
  {2289} (\bibinfo {year} {1979})}\BibitemShut {NoStop}%
\bibitem [{\citenamefont {Demokritov}\ \emph
  {et~al.}(2001{\natexlab{a}})\citenamefont {Demokritov}, \citenamefont
  {Hillebrands},\ and\ \citenamefont {Slavin}}]{demokritov2001}%
  \BibitemOpen
  \bibfield  {author} {\bibinfo {author} {\bibfnamefont {S.~O.}\ \bibnamefont
  {Demokritov}}, \bibinfo {author} {\bibfnamefont {B.}~\bibnamefont
  {Hillebrands}}, \ and\ \bibinfo {author} {\bibfnamefont {A.~N.}\ \bibnamefont
  {Slavin}},\ }\href@noop {} {\bibfield  {journal} {\bibinfo  {journal}
  {Physics Reports}\ }\textbf {\bibinfo {volume} {348}},\ \bibinfo {pages}
  {441} (\bibinfo {year} {2001}{\natexlab{a}})}\BibitemShut {NoStop}%
\bibitem [{\citenamefont {Demidov}\ \emph {et~al.}(2008)\citenamefont
  {Demidov}, \citenamefont {Dzyapko}, \citenamefont {Demokritov}, \citenamefont
  {Melkov},\ and\ \citenamefont {Slavin}}]{Demidov2008}%
  \BibitemOpen
  \bibfield  {author} {\bibinfo {author} {\bibfnamefont {V.}~\bibnamefont
  {Demidov}}, \bibinfo {author} {\bibfnamefont {O.}~\bibnamefont {Dzyapko}},
  \bibinfo {author} {\bibfnamefont {S.}~\bibnamefont {Demokritov}}, \bibinfo
  {author} {\bibfnamefont {G.}~\bibnamefont {Melkov}}, \ and\ \bibinfo {author}
  {\bibfnamefont {A.}~\bibnamefont {Slavin}},\ }\href@noop {} {\bibfield
  {journal} {\bibinfo  {journal} {Physical review letters}\ }\textbf {\bibinfo
  {volume} {100}},\ \bibinfo {pages} {047205} (\bibinfo {year}
  {2008})}\BibitemShut {NoStop}%
\bibitem [{\citenamefont {Yeh}(1979)}]{yeh1979electromagnetic}%
  \BibitemOpen
  \bibfield  {author} {\bibinfo {author} {\bibfnamefont {P.}~\bibnamefont
  {Yeh}},\ }\href@noop {} {\bibfield  {journal} {\bibinfo  {journal} {Josa}\
  }\textbf {\bibinfo {volume} {69}},\ \bibinfo {pages} {742} (\bibinfo {year}
  {1979})}\BibitemShut {NoStop}%
\bibitem [{\citenamefont {Mruczkiewicz}\ \emph {et~al.}(2013)\citenamefont
  {Mruczkiewicz}, \citenamefont {Krawczyk}, \citenamefont {Gubbiotti},
  \citenamefont {Tacchi}, \citenamefont {Filimonov}, \citenamefont {Kalyabin},
  \citenamefont {Lisenkov},\ and\ \citenamefont
  {Nikitov}}]{mruczkiewicz2013nonreciprocity}%
  \BibitemOpen
  \bibfield  {author} {\bibinfo {author} {\bibfnamefont {M.}~\bibnamefont
  {Mruczkiewicz}}, \bibinfo {author} {\bibfnamefont {M.}~\bibnamefont
  {Krawczyk}}, \bibinfo {author} {\bibfnamefont {G.}~\bibnamefont {Gubbiotti}},
  \bibinfo {author} {\bibfnamefont {S.}~\bibnamefont {Tacchi}}, \bibinfo
  {author} {\bibfnamefont {Y.~A.}\ \bibnamefont {Filimonov}}, \bibinfo {author}
  {\bibfnamefont {D.}~\bibnamefont {Kalyabin}}, \bibinfo {author}
  {\bibfnamefont {I.}~\bibnamefont {Lisenkov}}, \ and\ \bibinfo {author}
  {\bibfnamefont {S.}~\bibnamefont {Nikitov}},\ }\href@noop {} {\bibfield
  {journal} {\bibinfo  {journal} {New Journal of Physics}\ }\textbf {\bibinfo
  {volume} {15}},\ \bibinfo {pages} {113023} (\bibinfo {year}
  {2013})}\BibitemShut {NoStop}%
\bibitem [{\citenamefont {van~den Berg}(1991)}]{van1991magnetostatic}%
  \BibitemOpen
  \bibfield  {author} {\bibinfo {author} {\bibfnamefont {H.}~\bibnamefont
  {van~den Berg}},\ }\href@noop {} {\bibfield  {journal} {\bibinfo  {journal}
  {IEEE transactions on magnetics}\ }\textbf {\bibinfo {volume} {27}},\
  \bibinfo {pages} {5480} (\bibinfo {year} {1991})}\BibitemShut {NoStop}%
\bibitem [{\citenamefont {O’keeffe}\ and\ \citenamefont
  {Patterson}(1978)}]{Okeeffe}%
  \BibitemOpen
  \bibfield  {author} {\bibinfo {author} {\bibfnamefont {T.}~\bibnamefont
  {O’keeffe}}\ and\ \bibinfo {author} {\bibfnamefont {R.}~\bibnamefont
  {Patterson}},\ }\href@noop {} {\bibfield  {journal} {\bibinfo  {journal}
  {Journal of Applied Physics}\ }\textbf {\bibinfo {volume} {49}},\ \bibinfo
  {pages} {4886} (\bibinfo {year} {1978})}\BibitemShut {NoStop}%
\bibitem [{\citenamefont {Demokritov}\ \emph
  {et~al.}(2001{\natexlab{b}})\citenamefont {Demokritov}, \citenamefont
  {Hillebrands},\ and\ \citenamefont {Slavin}}]{demokritov2001brillouin}%
  \BibitemOpen
  \bibfield  {author} {\bibinfo {author} {\bibfnamefont {S.~O.}\ \bibnamefont
  {Demokritov}}, \bibinfo {author} {\bibfnamefont {B.}~\bibnamefont
  {Hillebrands}}, \ and\ \bibinfo {author} {\bibfnamefont {A.~N.}\ \bibnamefont
  {Slavin}},\ }\href@noop {} {\bibfield  {journal} {\bibinfo  {journal}
  {Physics Reports}\ }\textbf {\bibinfo {volume} {348}},\ \bibinfo {pages}
  {441} (\bibinfo {year} {2001}{\natexlab{b}})}\BibitemShut {NoStop}%
\bibitem [{\citenamefont {Sze}\ \emph {et~al.}(2021)\citenamefont {Sze},
  \citenamefont {Li},\ and\ \citenamefont {Ng}}]{sze2021physics}%
  \BibitemOpen
  \bibfield  {author} {\bibinfo {author} {\bibfnamefont {S.~M.}\ \bibnamefont
  {Sze}}, \bibinfo {author} {\bibfnamefont {Y.}~\bibnamefont {Li}}, \ and\
  \bibinfo {author} {\bibfnamefont {K.~K.}\ \bibnamefont {Ng}},\ }\href@noop {}
  {\emph {\bibinfo {title} {{Physics of semiconductor devices}}}}\ (\bibinfo
  {publisher} {John wiley \& sons},\ \bibinfo {year} {2021})\BibitemShut
  {NoStop}%
\bibitem [{\citenamefont {Pohorelec}\ \emph {et~al.}(2020)\citenamefont
  {Pohorelec}, \citenamefont {{\v{T}}apajna}, \citenamefont
  {Gregu{\v{s}}ov{\'a}}, \citenamefont {Gucmann}, \citenamefont
  {Hasen{\"o}hrl}, \citenamefont {Ha{\v{s}}{\v{c}}{\'i}k}, \citenamefont
  {Stoklas}, \citenamefont {Seifertov{\'a}}, \citenamefont {P{\'e}cz},
  \citenamefont {Tóth} \emph {et~al.}}]{pohorelec2020investigation}%
  \BibitemOpen
  \bibfield  {author} {\bibinfo {author} {\bibfnamefont {O.}~\bibnamefont
  {Pohorelec}}, \bibinfo {author} {\bibfnamefont {M.}~\bibnamefont
  {{\v{T}}apajna}}, \bibinfo {author} {\bibfnamefont {D.}~\bibnamefont
  {Gregu{\v{s}}ov{\'a}}}, \bibinfo {author} {\bibfnamefont {F.}~\bibnamefont
  {Gucmann}}, \bibinfo {author} {\bibfnamefont {S.}~\bibnamefont
  {Hasen{\"o}hrl}}, \bibinfo {author} {\bibfnamefont {{\v{S}}.}~\bibnamefont
  {Ha{\v{s}}{\v{c}}{\'i}k}}, \bibinfo {author} {\bibfnamefont {R.}~\bibnamefont
  {Stoklas}}, \bibinfo {author} {\bibfnamefont {A.}~\bibnamefont
  {Seifertov{\'a}}}, \bibinfo {author} {\bibfnamefont {B.}~\bibnamefont
  {P{\'e}cz}}, \bibinfo {author} {\bibfnamefont {L.}~\bibnamefont {Tóth}},
  \emph {et~al.},\ }\href@noop {} {\bibfield  {journal} {\bibinfo  {journal}
  {Applied Surface Science}\ }\textbf {\bibinfo {volume} {528}},\ \bibinfo
  {pages} {146824} (\bibinfo {year} {2020})}\BibitemShut {NoStop}%
\bibitem [{\citenamefont {{COMSOL Multiphysics}}\ and\ \citenamefont {{COMSOL
  Multiphysics Modeling Guide}}(2017)}]{RFguide}%
  \BibitemOpen
  \bibfield  {author} {\bibinfo {author} {\bibnamefont {{COMSOL
  Multiphysics}}}\ and\ \bibinfo {author} {\bibnamefont {{COMSOL Multiphysics
  Modeling Guide}}},\ }\href@noop {} {\bibfield  {journal} {\bibinfo  {journal}
  {Stockholm, Sweden: COMSOL AB}\ } (\bibinfo {year} {2017})}\BibitemShut
  {NoStop}%
\bibitem [{\citenamefont {Sadovnikov}\ \emph
  {et~al.}(2015{\natexlab{a}})\citenamefont {Sadovnikov}, \citenamefont
  {Beginin}, \citenamefont {Bublikov}, \citenamefont {Grishin}, \citenamefont
  {Sheshukova}, \citenamefont {Sharaevskii},\ and\ \citenamefont
  {Nikitov}}]{sadovnikov2015brillouin}%
  \BibitemOpen
  \bibfield  {author} {\bibinfo {author} {\bibfnamefont {A.}~\bibnamefont
  {Sadovnikov}}, \bibinfo {author} {\bibfnamefont {E.}~\bibnamefont {Beginin}},
  \bibinfo {author} {\bibfnamefont {K.}~\bibnamefont {Bublikov}}, \bibinfo
  {author} {\bibfnamefont {S.}~\bibnamefont {Grishin}}, \bibinfo {author}
  {\bibfnamefont {S.}~\bibnamefont {Sheshukova}}, \bibinfo {author}
  {\bibfnamefont {Y.~P.}\ \bibnamefont {Sharaevskii}}, \ and\ \bibinfo {author}
  {\bibfnamefont {S.}~\bibnamefont {Nikitov}},\ }\href@noop {} {\bibfield
  {journal} {\bibinfo  {journal} {Journal of Applied Physics}\ }\textbf
  {\bibinfo {volume} {118}},\ \bibinfo {pages} {203906} (\bibinfo {year}
  {2015}{\natexlab{a}})}\BibitemShut {NoStop}%
\bibitem [{\citenamefont {Sadovnikov}\ \emph
  {et~al.}(2015{\natexlab{b}})\citenamefont {Sadovnikov}, \citenamefont
  {Bublikov}, \citenamefont {Beginin}, \citenamefont {Sheshukova},
  \citenamefont {Sharaevskii},\ and\ \citenamefont
  {Nikitov}}]{sadovnikov2015nonreciprocal}%
  \BibitemOpen
  \bibfield  {author} {\bibinfo {author} {\bibfnamefont {A.~V.}\ \bibnamefont
  {Sadovnikov}}, \bibinfo {author} {\bibfnamefont {K.}~\bibnamefont
  {Bublikov}}, \bibinfo {author} {\bibfnamefont {E.~N.}\ \bibnamefont
  {Beginin}}, \bibinfo {author} {\bibfnamefont {S.~E.}\ \bibnamefont
  {Sheshukova}}, \bibinfo {author} {\bibfnamefont {Y.~P.}\ \bibnamefont
  {Sharaevskii}}, \ and\ \bibinfo {author} {\bibfnamefont {S.~A.}\ \bibnamefont
  {Nikitov}},\ }\href@noop {} {\bibfield  {journal} {\bibinfo  {journal} {JETP
  letters}\ }\textbf {\bibinfo {volume} {102}},\ \bibinfo {pages} {142}
  (\bibinfo {year} {2015}{\natexlab{b}})}\BibitemShut {NoStop}%
\bibitem [{\citenamefont {Gurevich}(1973)}]{gurevichRussian}%
  \BibitemOpen
  \bibfield  {author} {\bibinfo {author} {\bibfnamefont {A.}~\bibnamefont
  {Gurevich}},\ }\href@noop {} {\enquote {\bibinfo {title} {Magnetic resonance
  in ferrites and antiferromagnets},}\ } (\bibinfo {year} {1973})\BibitemShut
  {NoStop}%
\bibitem [{\citenamefont {Bass}\ and\ \citenamefont
  {Bulgakov}(1997)}]{bass1997kinetic}%
  \BibitemOpen
  \bibfield  {author} {\bibinfo {author} {\bibfnamefont {F.}~\bibnamefont
  {Bass}}\ and\ \bibinfo {author} {\bibfnamefont {A.}~\bibnamefont
  {Bulgakov}},\ }\href@noop {} {\emph {\bibinfo {title} {Kinetic and
  Electrodynamic Phenomena in Classical and Quantum Semiconductor
  Superlattices}}}\ (\bibinfo  {publisher} {Nova Science Publishers},\ \bibinfo
  {year} {1997})\BibitemShut {NoStop}%
\bibitem [{\citenamefont {Kindyak}\ \emph {et~al.}(2002)\citenamefont
  {Kindyak}, \citenamefont {Boardman},\ and\ \citenamefont
  {Kindyak}}]{kindyak2002surface}%
  \BibitemOpen
  \bibfield  {author} {\bibinfo {author} {\bibfnamefont {A.}~\bibnamefont
  {Kindyak}}, \bibinfo {author} {\bibfnamefont {A.}~\bibnamefont {Boardman}}, \
  and\ \bibinfo {author} {\bibfnamefont {V.}~\bibnamefont {Kindyak}},\
  }\href@noop {} {\bibfield  {journal} {\bibinfo  {journal} {Journal of
  magnetism and magnetic materials}\ }\textbf {\bibinfo {volume} {253}},\
  \bibinfo {pages} {8} (\bibinfo {year} {2002})}\BibitemShut {NoStop}%
\bibitem [{\citenamefont {Eliseeva}\ and\ \citenamefont
  {Sementsov}(2005)}]{eliseeva2005electromagnetic}%
  \BibitemOpen
  \bibfield  {author} {\bibinfo {author} {\bibfnamefont {S.}~\bibnamefont
  {Eliseeva}}\ and\ \bibinfo {author} {\bibfnamefont {D.}~\bibnamefont
  {Sementsov}},\ }\href@noop {} {\bibfield  {journal} {\bibinfo  {journal}
  {Technical physics}\ }\textbf {\bibinfo {volume} {50}},\ \bibinfo {pages}
  {924} (\bibinfo {year} {2005})}\BibitemShut {NoStop}%
\bibitem [{\citenamefont {Eliseeva}\ \emph {et~al.}(2008)\citenamefont
  {Eliseeva}, \citenamefont {Sementsov},\ and\ \citenamefont
  {Stepanov}}]{eliseeva2008dispersion}%
  \BibitemOpen
  \bibfield  {author} {\bibinfo {author} {\bibfnamefont {S.}~\bibnamefont
  {Eliseeva}}, \bibinfo {author} {\bibfnamefont {D.}~\bibnamefont {Sementsov}},
  \ and\ \bibinfo {author} {\bibfnamefont {M.}~\bibnamefont {Stepanov}},\
  }\href@noop {} {\bibfield  {journal} {\bibinfo  {journal} {Technical
  Physics}\ }\textbf {\bibinfo {volume} {53}},\ \bibinfo {pages} {1319}
  (\bibinfo {year} {2008})}\BibitemShut {NoStop}%
\bibitem [{\citenamefont {Eliseeva}\ \emph {et~al.}(2010)\citenamefont
  {Eliseeva}, \citenamefont {Sannikov},\ and\ \citenamefont
  {Sementsov}}]{eliseeva2010anisotropy}%
  \BibitemOpen
  \bibfield  {author} {\bibinfo {author} {\bibfnamefont {S.}~\bibnamefont
  {Eliseeva}}, \bibinfo {author} {\bibfnamefont {D.}~\bibnamefont {Sannikov}},
  \ and\ \bibinfo {author} {\bibfnamefont {D.}~\bibnamefont {Sementsov}},\
  }\href@noop {} {\bibfield  {journal} {\bibinfo  {journal} {Journal of
  magnetism and magnetic materials}\ }\textbf {\bibinfo {volume} {322}},\
  \bibinfo {pages} {3807} (\bibinfo {year} {2010})}\BibitemShut {NoStop}%
\end{thebibliography}%

\end{document}